\documentclass[11pt, openright]{memoir}
\usepackage{Preamble}
\begin{document}
\pagenumbering{Roman}
\thispagestyle{empty}
\newcommand{\form}[1]{\scalebox{1.087}{\boldmath{#1}}}
\sffamily
\begin{textblock}{160}(-0.3,-27)
\vspace{-\parskip}
\begin{center}
\includegraphics[width=10mm]{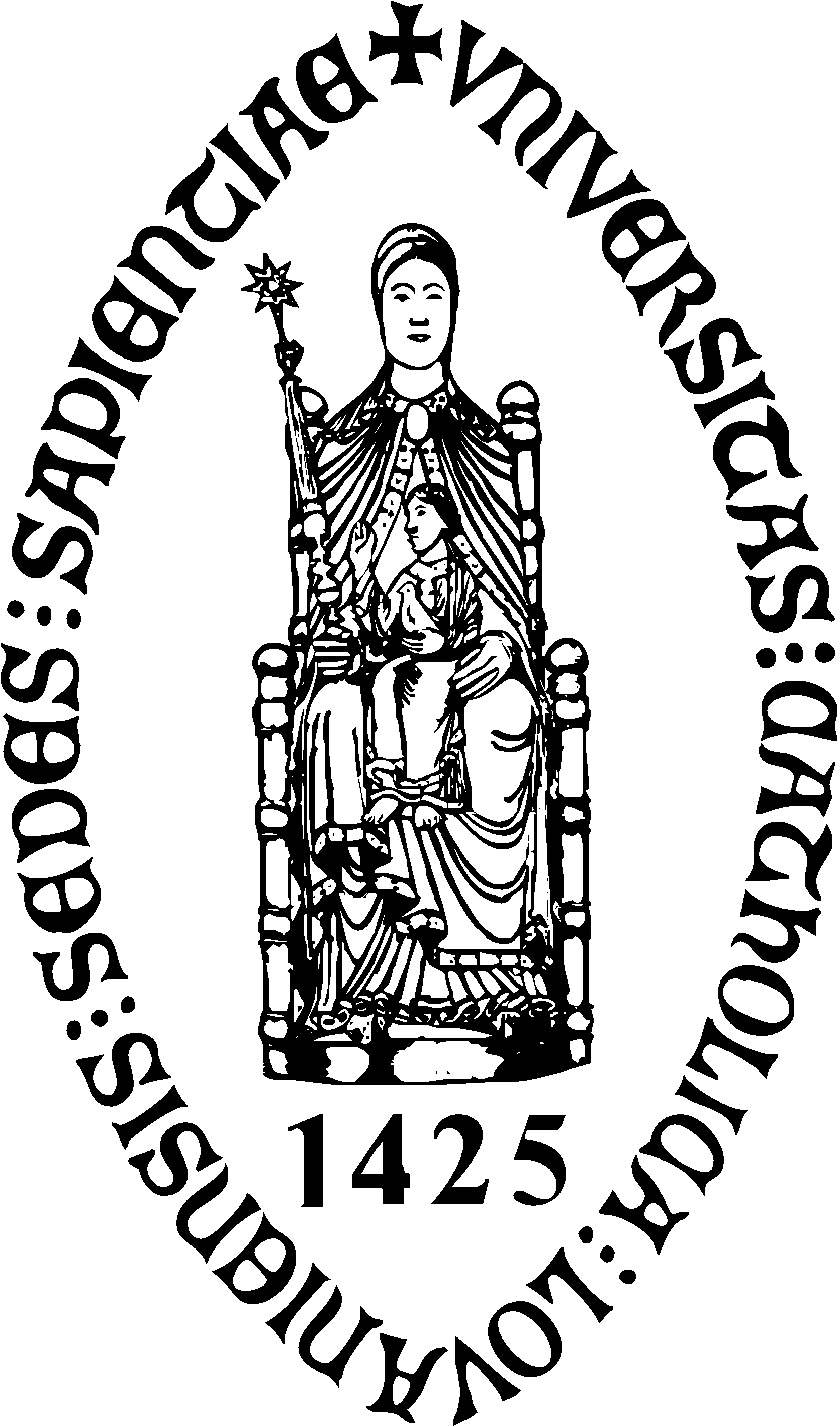}
\end{center}
\end{textblock}
\begin{textblock}{140}(72.5,-2)
\colorbox{gray}{\hspace{9mm}\ \parbox[c][21.3truemm]{83mm}{%
\fontsize{13}{15} \textcolor{white}{\textbf{FACULTY OF SCIENCE}}\\[0.25ex]
\fontsize{11}{13} \textcolor{white}{Department \textsl{Physics and Astronomy}}\\[0.25ex]
\fontsize{11}{13} \textcolor{white}{Section \textsl{Theoretical Physics}}
}}
\end{textblock}
\begin{textblock}{70}(4.3,-2)\textblockcolour{}\includegraphics*[height=23.43truemm]{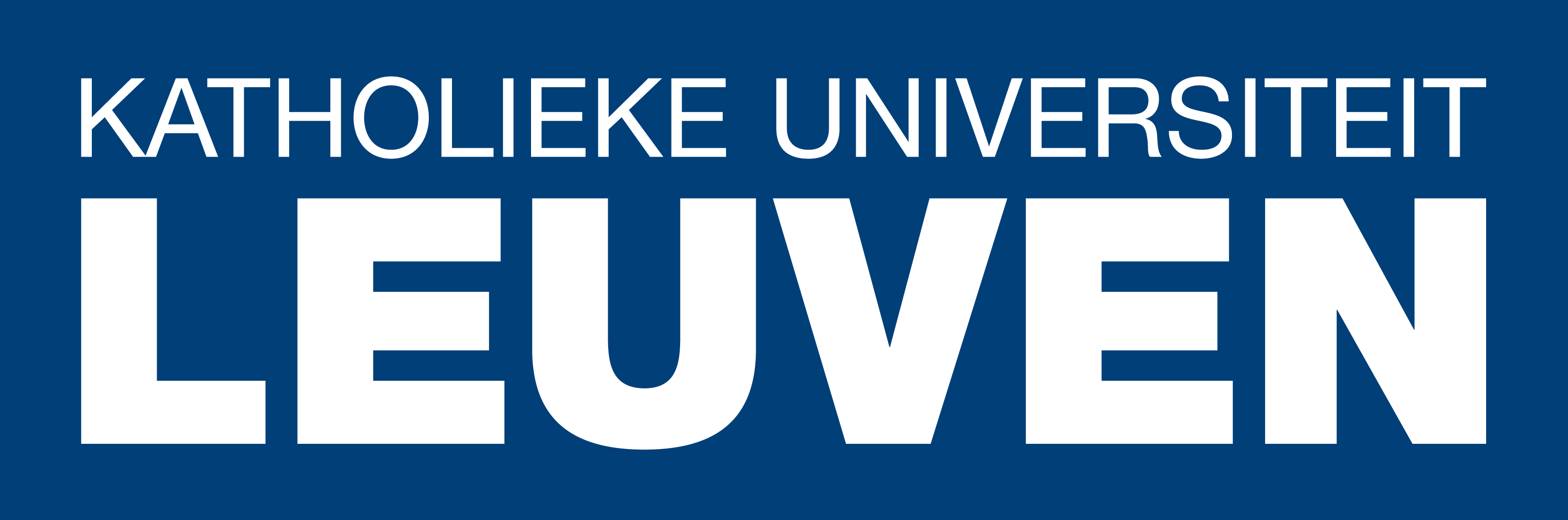}
\end{textblock}
\begin{textblock}{160}(6,81)
\textblockcolour{}
\vspace{-\parskip}
\begin{center}
\fontsize{16}{18} \textbf{On the Quantum-to-Classical Transition of Primordial Perturbations}\\[3mm]
\end{center}
\end{textblock}
\begin{textblock}{160}(6,125)
\textblockcolour{}
\vspace{-\parskip}
\begin{center}
\fontsize{12}{14} by
\end{center}
\end{textblock}
\begin{textblock}{160}(7,158)
\textblockcolour{}
\vspace{-\parskip}
\begin{center}
\fontsize{12}{14}
{Wouter RYSSENS}
\end{center}
\end{textblock}
\begin{textblock}{160}(17.5,212)
\textblockcolour{}
\vspace{-\parskip}
\begin{tabular}{@{}p{9.7cm}@{}l@{}}
 Promotor: dr.~W.~Struyve       & Dissertation presented in \\[0.25ex]
  & fulfillment of the requirements \\[0.25ex]
  & for the degree of Master of  \\[0.25ex]
                                     & Physics\\
\end{tabular}
\end{textblock}
\begin{textblock}{160}(7.5,238)
\textblockcolour{}
\vspace{-\parskip}
\begin{center}
Academic Year 2011-2012
\end{center}
\end{textblock}
\vfill
\newpage

\thispagestyle{empty}
\null
\vfill
\copyright \hbox{} Copyright by KU Leuven
\newline
\newline
Without written permission of the promotors and the authors it is forbidden to reproduce or adapt in any form or by any means any part of this publication. Requests for obtaining the right to reproduce or utilize parts of this publication should be addressed to KU Leuven, Faculteit Wetenschappen, Geel Huis, Kasteelpark Arenberg 11, 3001 Leuven (Heverlee), Telephone +32 16 32 14 01.
\newline
\newline
A written permission of the promotor is also required to use the methods, products, schematics and programs described in this work for industrial or commercial use, and for submitting this publication in scientific contests.
\setcounter{tocdepth}{1}
\setcounter{page}{1}
\tableofcontents*
\clearpage
\pagenumbering{arabic}
\setcounter{page}{1}
\addcontentsline{toc}{chapter}{Introduction}
\chapter*{Introduction}\label{Intro}\markboth{Introduction}{Introduction}
At the time of its discovery, the cosmic microwave background generated extensive interest, culminating in half a Nobel prize for the discoverers Penzias and Wilson. This `relic radiation' was found to consist of photons that have been travelling for billions of years, effectively starting at the time of decoupling and not stopping until they were captured in our detectors.
\par
\begin{figure}[h]
 \begin{center}
 \includegraphics[scale = 0.26]{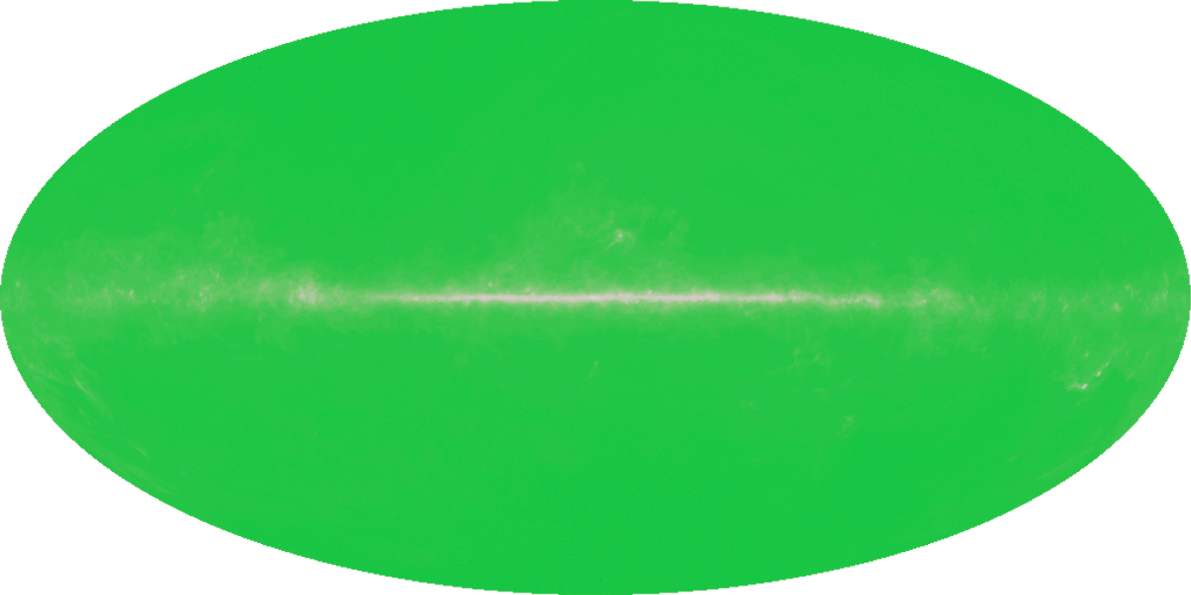}
\caption{Simulated temperature map of the sky as it could have been measured by Penzias and Wilson. The central white band is due to contamination from our Milky Way. Figure courtesy of NASA.}\label{WMAPHomogeen}
 \end{center}
\end{figure}

 In fig. \ref{WMAPHomogeen} we see a simulated image of the temperature distribution of the sky, as it could have been measured by Penzias and Wilson. Apart from contamination by our own Milky Way (the central white `band'), we see that the temperature distribution is extremely homogeneous. This can be regarded as the most important experimental confirmation made by the discovery of the background radiation: it confirms the \emph{cosmological principle}. This is a very important principle in cosmology, stating that the universe is \emph{homogeneous} and \emph{isotropic}. Technically, this means that there are no preferred directions in space and that relevant observables as temperature and energy density are the same everywhere at a given instant of time. Of course, this principle applies only if we focus on sufficiently big scales, meaning scales that are far larger than typical supercluster size, typically above $10^2$ Mpc \cite{CosmologyCursus}. The evidence provided by the cosmic microwave background is compelling, illustrating that the universe was extremely homogeneous in the early part of it's history. It should be noted that the cosmological principle is supported by some other observations, such as the distribution of radio galaxies.
\par
\begin{figure}[h]
 \begin{center}
 \includegraphics[scale = 0.5]{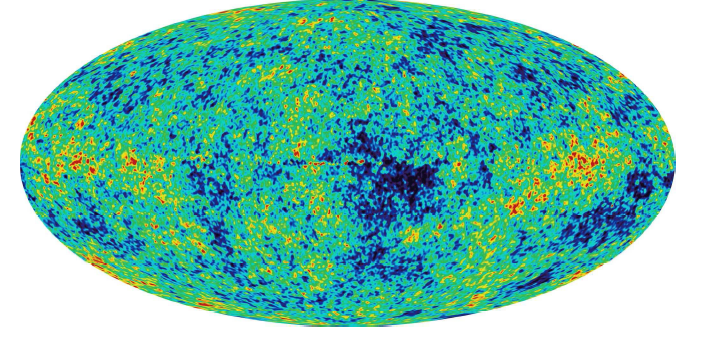}
 \end{center}
\caption{Anisotropy in the microwave background as measured by three years of the WMAP experiment. Figure courtesy of NASA.}\label{WMAPInHomogeen}
\end{figure}

 When we accept the cosmological principle, we have to acknowledge the fact that the universe is not homogeneous and isotropic on small scales. The Milky Way is not spherically symmetric and the mass distribution inside our solar system is not homogeneous. In particular, the cosmic microwave background is not perfectly homogeneous. Deviations from the mean temperature are very small, but could eventually be detected. The three-year results of the WMAP mission are displayed on figure \ref{WMAPInHomogeen}, displaying the deviations from the mean temperature in function of the direction on the sky. The fluctuations are of order $10^{-4}$ with respect to the mean temperature and can be characterised in terms of their power spectrum.
\par These tiny fluctuations in the background radiation mean that there were tiny fluctuations in the energy density of the early universe. The anisotropies in the CMB provide us with a snapshot of this period. The general idea is then that these initial perturbations will amplify in time due to the attractive nature of gravity, in turn leading to the small-scale structure of the universe as we observe it today. The evolution of these perturbations after the recombination and decoupling epoch and the formation of structure in the universe is a very active field of research, with famous examples as Jeans theory and the Millennium Simulation. While it is not entirely understood how the structure has formed exactly, it is usually thought that these inhomogeneities are the direct cause of structure formation.
\newpage
\par In this thesis, we will not be concerned with the evolution of these inhomogeneities after decoupling, but with their origin. In the 1980's the theory of inflation was formulated, in response to a set of problems cosmologists faced. The theory of inflation puts forward that the universe went through a period of accelerated expansion, (partially) solving what was know as the flatness problem, the homogeneity problem and the magnetic monopoles problem. It turned out that a quantised version of this theory was able to explain the origin and the power spectrum of the observed inhomogeneities. This property of inflation is in these days regarded as a far more compelling reason to attach credibility to the theory of inflation than the original problems.
\par The idea of inflation is rather simple. We take the very early universe as a starting point, before the suspected time at which inflation is suspected to have started. The whole universe is assumed to be isotropic at this point in time, meaning that the energy distribution was homogeneous. But on very small scales quantum mechanical effects become important and, because of the Heisenberg uncertainty principle, this homogeneity is uncertain. This phenomenon is commonly called `quantum fluctuations', but it is dangerous to engage in trying to sketch a picture of this, therefore we will not try to give the reader a more `visual' idea of these fluctuations. At that moment the universe enters the epoch of inflation and all length scales expand enormously. The quantum effects get transferred to much larger scales and get 'frozen' into fluctuations in the energy density. Eventually, at the moment of decoupling, scattered photons will get gravitationally redshifted by the energy density. The difference in mass distribution gives
rise to a difference in redshift which gives rise to the difference in temperature we observe. %
\par The picture sketched above already has one fundamental problem. The idea is that a quantum phenomenon gives rise to a classical phenomenon, quantum fluctuations giving rise to classical energy density perturbations. While quantum theory has been around since the early 20th century, we still do not fully understand how quantum theory reduces to the classical world we observe every day. This problem is especially severe in the cosmological context since the events we wish to consider are generally far removed in both space and time from us as observers. Several mechanisms have been proposed to solve the classicality problem in the context of primordial perturbations. So far, no proposition is completely compelling and the scientific community is divided.
\par We will conclude this introduction of the problem by citing some authorities in the field on the subject of these perturbations. Weinberg writes in \cite{Weinberg:2008zzc}
\begin{quote}
 `\textit{ [\ldots] the field configurations must become locked into one of an ensemble of classical configurations, with ensemble averages given by the quantum expectation
values calculated [\ldots]. It is not apparent just how this happens \ldots}'
\end{quote}
In \cite{Lyth:2009zz} it is stated that
\begin{quote}
`\textit{There remains the usual interpretation problem, arising whenever we talk about measurement in quantum theory. We would like to know why eigenstates of the field operator have been selected by Nature, since after all they provide only one from an infinity of possible ways of expanding the state vector. And we would like to be assured that one of the eigenstates was indeed selected in the early Universe, since otherwise it would seem that the pattern of say the cosmic microwave anisotropy may not have existed before it was observed.}'
\end{quote}
As a final example we cite Mukhanov, one of the founders of the theory of quantum inflationary perturbations, in \cite{MukhanovBook}
\begin{quote}
`\textit{If these galaxies originated from initial quantum fluctuations, a natural question arises: how does a galaxy, e.g.\ Andromeda, find itself at a particular place if the initial vacuum state was translational-invariant with no preferred position in space? Quantum mechanical unitary evolution does not destroy translational invariance and hence the answer to this question must lie in the transition from quantum fluctuations to classical inhomogeneities. Decoherence is a necessary condition for the emergence of classical inhomogeneities and can easily be justified for amplified cosmological perturbations. However, decoherence is not sufficient to explain the breaking of translational invariance. It can be shown that as a result of unitary evolution we obtain a state which is a superposition of many macroscopically different states, each corresponding to a particular realization of galaxy distribution.[\ldots] Such a state is a close cosmic analog of the Schr\"odinger cat.}'
\end{quote}
\newpage
\paragraph{Organisation of this Thesis}
In part \ref{Exploring} we will construct the theoretical basis to be able to formulate the problem as specifically as possible. We will start in chapter \ref{Inflation} by investigating the nature and origin of the cosmic microwave background. The simplest model of inflation, known as slow-roll inflation, will be introduced. In chapter \ref{Perturbations} we will then look at perturbations of the metric and the inflaton field to the homogeneous inflationary theory. At the end of this chapter, we will promote the theory to a quantum theory. The framework is then in place to discuss the nature of the inhomogeneities in chapter \ref{Observation}. First the observed inhomogeneities will be characterised, followed by a brief exposition of how the perturbations from chapter \ref{Perturbations} are supposed to give rise to the inhomogeneities in the sky. The questions we will abord in this thesis can then be explicitly stated in section \ref{PROBLEM}.
\par Part \ref{Usual} of this thesis will focus on the usual solutions presented in the literature. The discussion will be opened in chapter \ref{QtoC} where we will treat the general question of the classical limit, using two exemplary physical systems. In chapter \ref{Squeezing} we will critically review the proposed solution known as `squeezing'. Following that, we will explore the possibilities of decoherence in chapter \ref{Decoherence}. All solutions considered in this part of the thesis are proposed solutions that remain within the confines of quantum theory. We mention that there are several other schemes to explain the observed classicality within and outside standard quantum mechanics, the examples we have studied are far from exhaustive. %
\par The final part of this thesis, part \ref{PWCase}, will be concerned with pilot-wave theory. We will introduce the framework in chapter \ref{PilotWave} and immediately apply it to the examples used in chapter \ref{QtoC}, studying what the classical limit entails in de Broglie-Bohm theory. This will be applied to the case of the primordial perturbations in chapter \ref{PilotWavePert}, where we will find that pilot-wave theory presents a rather natural solution to the problem. We will continue by presenting some simulations in this framework in chapter \ref{Quantifying}, thereby providing more insight in the actual dynamics of the perturbations in the context of pilot-wave mechanics. We will argue from these simulations that one can expect the mechanism that is provided reaches classical behavior fast enough to explain observations. In chapter \ref{Test}, we will look at some other implications of pilot-wave theory in cosmology: it is possible that the difference between pilot-wave theory and standard quantum mechanics becomes observable in that context.

\part{Inhomogeneities in the Sky: Exploring the Problem}\label{Exploring}
\chapter{The Cosmic Microwave Background and Inflation}\label{Inflation}When Penzias and Wilson reported an `all-pervading' noise in their radio experiments at Bell labs in New Jersey, they paved the road for most of the observational cosmology of today. This background radiaton that pervades the earth from all directions is measured to have a temperature of $2.725K$, giving it a mean wavelength in the order of micrometers. Therefore, it is commonly called the \emph{Cosmic Microwave Background} or the CMB in short. It is among the best sources of information we have on the early universe, and while it is the most accurately measured black-body spectrum in human history, there still are experiments under way attempting to improve our knowledge and understanding of this phenomenon. Most striking is the homogeneity of this background: the temperature is homogeneous to one part in $10^{-4}$ for every direction in the sky!
\par In this chapter we will briefly look into the origin of the CMB, as it is explained by the hot Big Bang model.\footnote{The CMB can in fact be considered as a most striking experimental confirmation of the hot Big Bang model.} However, the conclusions we can draw on the state of the early universe present us with some extra questions. We will then introduce the (simplest) model of inflation, \emph{slow-roll inflation}, a model that provides an answer for several of this questions.A brief discussion of possible extensions of this simple model will be included. To finish this chapter, we will introduce the observed inhomogeneities in the CMB, the deviations of order $10^{-4}$ that will be the primary focus of this thesis.
\par For the duration of this thesis, we will use natural units, meaning that $\hbar = c= 1$. The only exception is chapter \ref{QtoC} where the explicit presence of the factors of $\hbar$ is enlightening.
\section{Origin of the Cosmic Microwave Background}
We will begin our account by assuming the cosmological principle, stating that the universe is (in first approximation) isotropic and homogeneous on large scales. This assumption is supported very well by our observations of the homogeneity of the CMB. In addition, it is supported by the measurement of the celestial distribution of radio galaxies and the isotropic nature of Hubble's law.
\par Under the assumption of the cosmological principle, the universe must be a space of constant curvature on large scales. There are thus only three very distinct possibilities. Either the universe has positive, negative or zero curvature, usually denoted by $K = 1,-1,0$ respectively. The metric dictated by general relativity is then the \emph{Friedmann-Lema\^itre-Robertson-Walker line element}, or FLRW metric. It can be found in most textbooks on general relativity, as for instance \cite{dInverno:1992rk}.
\begin{equation}\label{FRWMetric}
 ds^2 = dt^2 - a^{2}(t)\frac{d\vec{x}^2}{\left[1 + \frac{1}{4}K\vec{x}^2\right]^2}
\end{equation}
Here $a(t)$ is called the scale factor. Note that it can only depend on time, since any spatial dependence would violate the cosmological principle. The only aspects of the metric that are not fixed by the cosmological principle are $a(t)$ and $K$.
\par The Einsteins equations are, in a universe without cosmological constant,
\begin{equation}
  R_{\nu}^{\mu} - \frac{1}{2}Rg_{\nu}^{\mu} = 8 \pi G T_{\nu}^{\mu},
\end{equation}
where $R_{\nu}^{\mu}$ is the contracted curvature tensor, $R$ is the Ricci scalar and $g_{\mu \nu}$ is the FLRW metric from eq.\eqref{FRWMetric}. To solve for $a(t)$ and $K$, we also need a suitable energy momentum tensor $T_{\mu \nu}$. One usually assumes that the universe is filled with a perfect fluid and the energy-momentum tensor $T_{\mu \nu}$ is given by
\begin{equation}
 T_{\mu \nu} = \left(\rho + p\right)U_{\mu}U_{\nu} - p g_{\mu \nu},
\end{equation}
where $\rho$ and $p$ denote the density and the pressure of the fluid respectively, while $U_{\mu}$ denotes the velocity field four-vector of the liquid. This set of equations is usually supplemented by an equation of state for the fluid $p(\rho)$.
\par This set-up is well known and leads to two independent relations, called the \emph{Friedmann equations}\cite{LiddleBook}
\begin{align}\label{Friedmann}
 \left(\frac{\dot{a}}{a}\right)^2 &= \frac{8\pi G}{3}\rho - \frac{K}{a^2},\\\label{FluidEquation}
\dot{\rho} &= - 3 \frac{\dot{a}}{a}\left(\rho  + p\right),
\end{align}
where dots indicate derivatives with respect to time. An important quantity is clearly $\frac{\dot{a}}{a}$. It is called \emph{Hubble's parameter} and is usually denoted by $H$.
\par We now make some simplifying assumptions. First, the universe we observe today seems to be flat, so we take $K = 0$. In addition, we take the fluid to consist of a mixture of matter and radiation. Perfect matter is a fluid without pressure, while radiation has the following simple equation of state
\begin{equation}
p = \frac{\rho}{3}.
\end{equation}
We can solve eq.\ \eqref{FluidEquation} for each component separately.
\begin{align}\label{DensityBehavior}
\rho_{matter} &\sim \frac{1}{a^3}, & \rho_{radiation} \sim \frac{1}{a^4}.
\end{align}
For an expanding universe, the volume increases as $a^3$, so the matter result comes as no surprise. Radiation, in addition to diluting, redshifts in an expanding universe, which results in an extra loss of energy and in the extra factor $a^{-1}$. Solving equation \eqref{Friedmann} for a mixture of matter and radiation can be done, but is very messy. To make things easier we imagine one of both components dominating. The solutions for $a(t)$ are respectively given by
\begin{align}\label{FLRWOplossingen}
 a_{matter}(t) &= \left(\frac{t}{t_0}\right)^{2/3}, & a_{radiation}(t) =  \left(\frac{t}{t_0}\right)^{1/2}.
\end{align}
where the constant $t_0$ is chosen in such a way that the present value of $a$ is $1$.
\par We can now, very schematically, paint a picture of the different eras the universe evolved through. We start our evolution in a very small universe that we imagine to be dominated by radiation. Since the wavelengths of photons behave as $\lambda \sim a$, their energy will decrease as $a^{-1}$. Following this reasoning, the initial stages of the universe must have been very hot. In fact the temperature was so high that the available protons and electrons didn't combine into hydrogen. Any hydrogen system that formed would almost instantly be ionised by the omnipresent photons, who carried a lot more energy than the $13.6$ eV required. As a consequence, the mean free path length for the photons was very small. It is at this time that the photons were able to establish thermal equilibrium with their surroundings, forming a black-body spectrum.
\par In this regime, the universe expanded as $a \sim t^{1/2}$ and thus the temperature behaved as $t^{-1/2}$. Since radiation density decreases a lot faster with time than matter energy density, eqs.\ \eqref{DensityBehavior}, our approximation of radiation dominating the expansion gets worse and worse. At a certain time, the situation will be inverted: the matter energy density will be more important. We thus transfer to a regime where matter dominates, meaning $a \sim t^{2/3}$.
\par While the universe expands, the photons redshift to lower and lower frequencies. Eventually the energy of the average photon drops to the point where it can no longer ionize hydrogen. Consequently, the protons and electrons combine into (mostly) hydrogen. This is called the period of \emph{recombination}. The photons have less free electrons to scatter off and their mean free path increases.
\par Soon after, as the temperature decreases further, the photons lose the capability to excite hydrogen to any of its energy eigenstates and all the electrons drop to the ground state of hydrogen. The probability for the photons to interact with the hydrogen atoms becomes vanishingly small and the mean free path for a thermal photon becomes very large. This is commonly called the \emph{period of decoupling}. The approximate temperature of the universe during this period was $3 \cdot 10^{3}K$\cite{LiddleBook}. The photons we see now in the night sky are originating from a sphere of radius $ct$, where $t$ is the time elapsed between decoupling and our observations. This sphere is called the surface of last scattering.
\par We already established that the photons were in thermal equilibrium during the early stages of the universe. The initial energy distribution over the frequency domain of the photons $\epsilon_i(f)$ was at that time
\begin{equation}
 \epsilon_i(f) =  4 \frac{f^3}{\exp \left(\frac{2\pi f}{k_B T_{eq}}\right) - 1}.
\end{equation}
However, we know that $f \sim a^{-1}$. The left hand side of the equation is an energy density: it has to scale as $a^{-3}$ for a given $f$. This is exactly what the numerator of the right hand side scales like. The denominator, we can take as constant in time, provided we take $T_{eq} \sim a^{-1}$. The evolution of the universe preserves the form of the initial black-body spectrum, but transfers it to increasingly lower equilibrium temperatures.
\par This explains very roughly where the photons we observe in the night sky come from and why they essentially behave as being emitted by a black body at a temperature of $2.725K$. Models other than the hot Big Bang model have trouble explaining the black body spectrum. Because of this, the presence of the CMB can be thought of as a `smoking gun' for the hot Big Bang model.
\clearpage
\section{Problems of Hot Big Bang Cosmology}\label{Problems}
Hot big bang cosmology is, of course, not complete. In this section we will zoom in on three observations that are usually cited as the reason to implement inflation.
\paragraph{Flatness Problem}
Turning back to the more general Friedmann equation \eqref{Friedmann}, one can derive the following solution for the curvature $K$
\begin{equation}
 K = \left(\rho(t) - \frac{3H^2}{8\pi}\right)\frac{8}{3}\pi a^2.
\end{equation}
Taking the value of the Hubble parameter now to be $H_0$, the quantity $\frac{3H_0^2}{8\pi}$ is called the \emph{critical density}. One immediately sees that the ratio of the present-day density $\rho_0$ and the critical density $\rho_c$ determines the type of universe we live in: positive, negative or zero curvature. Rewriting this equation in terms of this ratio $\Omega$
\begin{equation}\label{CritDensEvo}
 |\Omega(t) - 1 | = \frac{|K|}{a^2 H^2}.
\end{equation}
Supposing that there is enough fluid density present to dominate over any curvature and/or cosmological constant terms, we will take $a(t)$ as one of the solutions in eqs.\ \eqref{FLRWOplossingen}. We then have as time evolution
\begin{align}
 |\Omega(t) - 1| &\sim t, & \text{(Radiation)}\\
&\sim  t^{2/3 }.& \text{(Matter)}
\end{align}
For both cases, we see that $\Omega(t)$ evolves away from $1$, provided that $\Omega(t) \not = 1$. This means that a slight deviation from the critical density will be greatly amplified in time.
\par Current observations of type IA supernovae (`standard candles') and the CMB place the current ratio $\Omega_0$ at $1$, with an experimental uncertainty of the order of $1\%$. So the value of $|\Omega_0 - 1|$ can be at most $0.01$. But this is a quantity that has strictly increased with time. To account for the present value, the inital value of $|\Omega_i - 1|$ would have to be very small. If one uses the time evolution of the universe, as described above, and evolves back to Planck time the condition on $\Omega(t_{Planck})$ becomes \cite{MukhanovBook}
\begin{equation}
 |\Omega_i - 1| < 10^{-56}
\end{equation}
This is a very stringent condition on the initial condition of our universe indeed!
\paragraph{Horizon Problem}
Looking at the CMB, we see that the temperature distribution across the sky is very homogeneous. Some rough estimates then lead us to what is called the `horizon problem'.
The present scale of homogeneity (that we can observe) is at least the size of the observable universe, on the order of $10^{28}$ cm. Evolving our universe back into time, the homogeneity originated at time $t_i$ in a patch of size at least $10^{28} \frac{a(t_i)}{a(t_0)}$ cm. Now at that time, light only had the time to travel a distance $t_i$. Thus the scale of physical processes, the causal scale, was at most $t_i$. Denoting the homogeneity scale by $l_h$ and the causal scale by $l_c$ and evaluating at Planck time we get\cite{MukhanovBook}
\begin{equation}
 \frac{l_h}{l_c} \sim 10^{28}.
\end{equation}
Thus, at time $t_i$ we have on our hands a homogeneous region, that is immensely bigger than the causal scale. Every possible equilibration process that could have occurred before $t_i$ could only have established homogeneity on scales up to $l_c$. Thus, under standard cosmology, the observation of the homogeneity in the CMB implies that the initial state of the universe was already homogeneous on large scales!
\paragraph{Magnetic Monopole Problem}
Several Grand Unifying Theories, i.e.\ superstring theory, predict a plethora of new particles at the Grand Unification Scale, the scale on which the four fundamental forces are thought to be unified. One of the most infamous of these are magnetic monopoles.  We only note that these monopoles are predicted to be stable and they are extremely heavy, with masses on the order of $10^{16}$GeV.
\par As we have seen, the temperature of the universe decreased as $a^{-1}$. We can conclude that the very early universe was an extremely hot place, and several GUT's predict that a copious amount of monopoles would have been formed. Being very heavy, these monopoles would very fast behave non-relativistically and thus would rapidly behave as pressureless matter. Now the energy density of radiation falls as $a^{-4}$, while the density of the monopoles would fall as $a^{-3}$. The energy density of the monopoles would very fast have dominated the energy density of the radiation.
\par However, present day limits on the existence of monopoles are extremely stringent. The GUT-scenario, however, predicts stable magnetic monopoles, constituting a density far greater than the density of radiation in our present universe. This is called the monopole problem.\footnote{Historically, there was only the `monopole problem'. Following the developments in high-energy physics and GU theories, there are now several hypothetical particles that cause similar problems.}
\section{Inflation}
\paragraph{Answering the problems} In the 80's Guth \cite{Guth:1980zm} proposed a way to solve the problems touched upon in the previous section. The paradigm known as inflation promises to solve the flatness problem, the horizon problem and the magnetic monopole problem by hypothesising a period of extreme accelerated expansion in the early universe. We will treat the answers that inflation offers to these problems.
\par The rough picture goes as follows. An initially small patch in our universe undergoes, during a brief period, severe expansion. We assume that this patch has some initial nonzero curvature and some initial nonzero magnetic monopole density. The monopole density falls off as $a^{-3}$, so if the change in $a$ was big enough, this could bring the monopole density to below present experimental limits.
\par If our patch has some initial nonzero curvature, the ratio of the density and the critical density in that patch is different from $1$. For decelerated expansion, we have seen that in this case $\Omega$ evolves away from $1$. However, if we assume an accelerating inflationary phase we get for the scale factor $a$
\begin{equation}
 \frac{d}{dt}(\dot{a}) = \frac{d}{dt}(aH) > 0.
\end{equation}
Thus, using eq.\ \eqref{CritDensEvo}, we find that $\Omega$ will evolve towards $1$, instead of away from it. Inflation thus has the benefit of smoothing out the curvature.
\par Inflation also solves the horizon problem in a conceptually easy way. Patches of sky that seem to be causally disconnected today were in fact much closer together before the period of inflation. Thermal equilibrium would thus have been established before inflation, in a causally connected universe. Then the patches were causally removed from each other, by enormously enlarging the universe, accounting for the seemingly causally disconnected patches.
\par For inflation to account for observations, we need at least a factor of $e^{N}$ where $N$ is somewhere in the range between $10-70$. The number $N$ is called the number of e-folds the universe has expanded. As typical value, one can safely use $N \approx 60$\cite{Lyth:2009zz}. In orders of magnitude: $e^{60} \sim 10^{26}$.
\clearpage
\paragraph{Problems?}
We close this section with some side remarks on the problems we presented in the last section. One must notice that none of these is a problem in the strict sense of speaking. The magnetic monopole problem is only a problem as soon as we introduce magnetic monopoles, which so far have absolutely not been observed. The flatness problem and horizon problem can both be solved, within a non-inflationary universe, by fine-tuning the initial condition of our universe. If we stay within hot big bang cosmology, these problems merely represent restrictions on the possible initial state of the universe. Such fine tuning is widely regarded as dissatisfactory, but in principle, it is not inconceivable that our universe started out with the special initial conditions required to form our present-day universe. One could question the motivations for the theory of inflation, as Penrose does in \cite{Penrose}. His arguments are the following.
\par The monopole problem, while certainly a problem for people working on superstring theory or other Grand Unified Theories, does not present itself in the context of standard cosmology, for the simple reason that the Standard Model does not incorporate magnetic monopoles or other exotic particles.
\par Given a certain amount of inflation, say $60$ e-folds, there certainly are manifolds that are not smooth on a scale of $e^{-60}$. Penrose takes as example the Mandelbrot fractal, a structure that does not smooth out when enlarged. One could argue that this is kind of a pathological example, but it is clear that there are a lot of possible initial spacetime structures that are not smoothed out by $60$ e-folds of inflation. In a way, we can say that the inflation paradigm limits the amount of required finetuning for the flatness problem, but does not eliminate it entirely.
\par The phrasing of the horizon problem also merits some extra attention. The observed homogeneity of the CMB can also be explained in terms of the second law of thermodynamics. Let us think of two causally disconnected regions as two isolating boxes containing some (very large) amount of gas molecules. The number of possibilities for the temperature in both boxes to be equal by far exceeds the number of possibilities for the boxes  to have different temperatures. Thus, in the same way both gas containers do not need thermal equilibration, two causally disconnected regions in the universe do not need an equilibration to have the same temperature.
\par These arguments are not trying to disprove the inflationary paradigm in any way. They are meant to be sideremarks to the discussion in section \ref{Problems}.
\paragraph{Implementing Slow-Roll Inflation}
The rest of this section will be devoted to the question how the hypothesis of inflation can be realised within the framework of general relativity.  Our treatment will be brief and loosely based on \cite{Brandenberger:1997yf}, \cite{Lyth:2009zz} and \cite{Brandenberger:2002wm}.
\par Looking back at the Friedmann equations, eq.\ \eqref{Friedmann} and eq.\ \eqref{FluidEquation} for a flat universe
\begin{align}
 \left(\frac{\dot{a}}{a}\right)^2 &= \frac{8\pi G}{3}\rho, & \dot{\rho} = - 3 \frac{\dot{a}}{a}\left(\rho  + p\right).
\end{align}
It is clear that, when $\dot{\rho} = 0$, a possible solution for $a(t)$ is an increasing exponential. %
\par We will construct a fluid satisfying $\rho \approx - p$ in the following way. The action for a minimally coupled massless scalar field $\varphi$, that we will call the \emph{inflaton}, in a potential $V(\varphi)$ is
\begin{equation}\label{InflatonAction}
S = \frac{1}{2}\int d^4 x \sqrt{-g} \left[g^{\mu \nu}\partial_{\mu}\varphi \partial_{\nu}\varphi - V(\varphi)\right].
\end{equation}
From the action, we can extract the energy momentum tensor
\begin{equation}
  T_{\mu \nu} = \partial_{\mu} \varphi \partial_{\nu}\varphi - \frac{1}{2}g_{\mu \nu} g_{\rho \sigma}\partial^{\rho} \varphi \partial^{\sigma}\varphi + g_{\mu \nu}V(\varphi).
\end{equation}
We read off the pressure and the energy density
\begin{align}
 \rho &= \frac{1}{2} \dot{\varphi}^2 - \frac{1}{2}\left(\nabla \varphi\right)^2 + V(\varphi), & p = \frac{1}{2} \dot{\varphi}^2 - \frac{1}{6}\left(\nabla \varphi\right)^2 - V(\varphi).
\end{align}
So, for a homogeneous field that slowly varies in time, the potential $V(\varphi)$ will dominate both the pressure and the energy density, realising $p \approx - \rho$.
\par The Friedmann equations become for this fluid
\begin{align}
 H^2 &= \frac{8\pi G}{3} \left(\frac{1}{2} \dot{\varphi}^2 - \frac{1}{2} a^{-2}\left(\nabla\varphi\right)^2 + V(\varphi)\right)\\
\ddot{\varphi} &= - 3H\dot{\varphi} + a^{-2}\Delta \varphi - \frac{\partial V}{\partial \varphi}
\end{align}
We will be looking for homogeneous solutions for $\varphi$, meaning that the space derivatives on $\varphi$ are well approximated by zero\footnote{Even if this is a bad approximation at the onset of inflation, the factor $a^{-2}$ will certainly quickly suppress the influence of the spatial inhomogeneities.}. Secondly, we will assume that the potential dominates any kinetic energy that is present. Mathematically these two approximations are written
\begin{align}\label{SlowRoll}
 \frac{1}{2}\dot{\varphi}^2 &\ll V(\varphi), &  a^{-2}\left(\nabla\varphi\right)^{2} \ll V(\varphi).
\end{align}
Under the slow-roll and homogeneity assumptions, the Friedmann equations become approximately
\begin{align}\label{SlowFried}
 H^2 &\approx \frac{8\pi G}{3} V(\varphi), & \ddot{\varphi} \approx - 3H\dot{\varphi} - \frac{\partial V}{\partial \varphi}.
\end{align}
So we immediately see that $H$ will be approximately constant. Now, if we allow $\ddot{\varphi}$ to be big, we cannot maintain the slow-roll approximation for long. We will thus look for solutions to these equations in an overdamped regime, meaning that we neglect $\ddot{\varphi}$. The condition for this to be a good approximation is
\begin{equation}\label{SecondSlowroll}
 \ddot{\varphi} \ll 3H\dot{\varphi}.
\end{equation}
This assumption, together with eqs.\ \eqref{SlowRoll}, are the assumptions of slow-roll inflation. These assumptions sketch the picture of a scalar field in a very flat potential, slowly rolling down the hill to some minimum of that potential. We can recast these assumptions to be assumptions on the potential $V$ only\cite{Lyth:2009zz}
\begin{align}\label{Vcond}
 \frac{\partial V}{\partial\varphi}^2 &\ll V(\varphi)^2,  & \left|\frac{\partial^2 V}{\partial \varphi^2}\right| \ll |V(\varphi)|.
\end{align}
A scalar field, moving in a potential that satisfies these constraints, will give rise to a constant term in the right hand side of eq.\ \eqref{SlowFried}. The final result will be a scale factor that increases exponentially $a \sim \exp(Ht)$. Thus we have found a model that incorporates accelerating expansion of the universe, solving the problems we described in section \ref{Problems}.
\par To conclude our construction of the simplest inflationary scenario, we remark on the inflaton field. It is a hypothetical field that is thought to source inflation. It moves in the potential $V(\phi)$, but aside from eqs.\ \eqref{Vcond}, there are no restrictions on this potential. There is a wide variety of possibilities for the potential and there is no reason to suspect the inflaton field is one of the fields present in the Standard Model. Candidates for the inflaton field may be sought in supersymmetric theories or string theory.
\paragraph{Ending inflation}
If inflation wishes to explain the present universe, there has to be a way for it to end. Let's look at the second slow-roll condition, eq.\ \eqref{SecondSlowroll}. This assumption cannot be maintained forever, since the scalar field rolling down the potential will eventually pick-up some speed. Alternatively, one can look at this in terms of eqs.\ \eqref{Vcond}. We need $V(\varphi_0) = 0$ in our present universe. $V(\varphi_0) \not = 0$ would lead to a large cosmological constant in the Einstein equations, something that is not supported by observations today. It is thus clear that the assumption in the second equation of eqs.\ \eqref{Vcond} cannot hold forever.
\par The breakdown of the second of the equations in eqs.\ \eqref{Vcond} is usually cited as the moment slow-roll inflation ends. Eq. \eqref{SlowFried} will then describe an oscillator that is no longer in an overdamped regime. The inflaton field $\varphi$ will start oscillating, while the term $3H \dot{\varphi}$ will damp the amplitude of the oscillation.
\par Usually, one also supposes that the inflaton is coupled to the particles of the Standard Model, allowing the inflaton field to decay into different combinations of elementary particles. Thus, one modifies eq.\ \eqref{SlowFried} by an extra term on the right hand side proportional to $- \Gamma \dot{\varphi}$. Here $\Gamma$ is the total decay rate of the inflaton field into different kinds of particles. This extra term will result in extra damping of the oscillatory motion. During the inflationary phase, one presumes it to be dominated by the term $H\dot{\varphi}$. When inflation ends, $H$ will decrease and the decay term will take over, thereby increasing the damping of the amplitude of the oscillation.
\par This allows for a natural way to repopulate the universe. Since inflation expanded the universe enormously, it will also have cooled enormously. In addition, any initial density of particles present in the Standard Model will have been diluted. The photons and other elementary particles formed by a decaying inflaton field are then a very natural way to restore the temperature in the universe and repopulate it with the particles we see today. This process is called the process of \emph{reheating}.\footnote{This explanation is a very basic one. Normally one should also include the effect of parametric resonances in the oscillator equation eq.\ \ref{SlowFried}, but this is outside the scope of this thesis. For more details, see for example \cite{Brandenberger:2002wm}.}
\par It is however important to note that the reheating process is not understood very well. Since the nature of the inflaton field is unknown, we also know nothing of its decay rates and decay products. We will for the rest of this thesis neglect any couplings of the inflaton field to other particles, since we will mostly be considering the inflationary period itself. %
\section{Inhomogeneities}
As said before, the CMB is homogeneous to one part in $10^{-4}$. That is a remarkable fact and good evidence in support of the cosmological principles. However, it also implies that there are inhomogeneities in the CMB of order $10^{-4}$. In chapter \ref{Observation}, we will go more into detail on characterising the observed anisotropies.
\par Intuitively it is very clear that these inhomogeneities can be explained by inhomogeneities in the matter distribution at the surface of last scattering. How these matter density perturbations exactly lead to the observed anisotropies can be explained by the \emph{nonintegrated Sachs-Wolfe effect}. Very roughly, the scattered photons are gravitationally redshifted at the surface of last scattering. The amount of redshift depends on the distribution of matter. For small density perturbations, general relativity relates the shift in frequency $\delta f$ of a photon travelling from $\vec{r}$ to $\vec{r} + d\vec{r}$ to the difference of the $00$ component of the perturbed metric between those places. It is in good approximation given by
\begin{equation}\label{SachsWolfe}
 \ \frac{\delta f}{f} \approx \frac{\delta \left(\sqrt{g_{00}}\right)}{\sqrt{g_{00}}}.
\end{equation}
The actual situation is obviously more complicated, since this effect does not only play a role at the surface of last scattering. When this effect is also taken into account for the whole journey of the photons from the surface of last scattering to earth, the effect is called the \emph{integrated Sachs-Wolfe} effect. We will not delve deeper into that, since it is outside the scope of this thesis.
\par What we will note however, is that hot big bang cosmology can not account for the presence of these density perturbations. Given initial perturbations in the density of the early universe, hot big bang cosmology can evolve them in time and preserve them for us to observe. It cannot tell us how these perturbations came to be.
\par In contrast, inflation can account for the generation of these inhomogeneities. Roughly speaking one considers the early universe before inflation and looks at vacuum of the inflaton field that is present. If we treat the inflaton vacuum quantum mechanically there are the Heisenberg uncertainties to be considered. In the standard situation, `quantum fluctuations' of the vacuum only occur on very small length scales. However, inflation amplifies all length scales enormously, and transfers these vacuum fluctuations to observable scales. According to inflationary theory, the anisotropies we observe today are direct results from the quantum fluctuations that were present at the onset of inflation. Even more striking, inflationary theory can predict the observed spectrum of the anisotropies in the sky!
\par The explanation offered here is of course a sketch, but there is something amiss here. How exactly can the quantum mechanical fluctuations give rise to the density perturbations that behave classically? In chapter \ref{Perturbations} we will first develop the formalism needed to describe the density perturbations, while we will go more in depth on the origin of the inhomogeneities in chapter \ref{Observation}.
\section{Generalisations}
The scenario sketched above is only one of many possibilities and is commonly known as chaotic inflation. It constitutes the most simple model for inflation, and we will use it as basis for the rest of this thesis. For completeness sake, we list here some possible generalisations. One can of course start by considering more than one inflaton field. Every inflaton field induces its own inflation era. There is no upper limit to the amount of inflaton fields one can have.
\par It is also possible to construct potentials of two scalar fields $\chi$ and $\varphi$ such that the contribution to the energy density is mostly due to a very massive $\chi$, for $\varphi$ bigger than some critical value $\varphi_c$. One then usually takes $\varphi$ to be a slowly rolling field, and when $\varphi$ drops beneath $\varphi_c$ the potential of $\chi$ acquires a minimum at $\chi = 0$. Then $\chi$ begins to move and inflation ends as the potential vanishes. In this way, one scalar field takes the role of the inflaton while another one takes the role of potential.
\par The exponential behavior for the scale factor $a(t)$ presented above is not the only possible behavior. For example, Lucchin \cite{Lucchin}
proves that power-law inflation, with $a \sim t^p$ for $p > 1$, can also prove a satisfactory explanation for the flatness and horizon problem. To account for the correct spectrum of anisotropies, Lucchin shows that $p = 2$ or $p \geq 10$.

\chapter{The Theory of Cosmological Perturbations}\label{Perturbations}
The aim of this chapter is to develop the theory of cosmological perturbations. We will look at the machinery that will allow us to talk about the impact of the primordial quantum fluctuations that are supposed to be the source of the inhomogeneities in the CMB.
\par To do that, we will first look at the classical picture. Supposing some initial inhomogeneities, what do the Einstein equations tell us about their time evolution? We will analyse deviations of the FLRW-metric and inhomogeneities of the inflaton, in order to find the relevant physical variables. This will turn out to be a single scalar field. Second, we will quantise the action of the fluctuations by imposing the canonical commutation relations. Thirdly, we take a look at what happens when we write the theory using creation and annihilation operators. We will mostly follow the treatment and notation from \cite{MukhanovBook} and \cite{Brandenberger:1993zc}, but we will also use elements from \cite{Brandenberger:2003vk}, \cite{CosmologyCursus} and \cite{Bertschinger:2001is}.
\section{The Classical Theory}\label{ClassTheory}
\paragraph{Metric Perturbations}
As in the preceding chapter, we will consider a flat FLRW-universe, using the following metric
\begin{equation}
ds^2 =  dt^2 -a^2(t) d\vec{x}^2.
\end{equation}
In what follows, we will find it easier to work in a different coordinate system. We define conformal time $\eta$ as $d \eta = a^{-1}(t)dt$, so that the FLRW metric takes the form
\begin{equation}\label{FLRWmetric}
 ds^2 = a(\eta)^2 (d\eta^2 - d\vec{x}^2).
\end{equation}
When $t$ ranges between $-\infty$ and $\infty$, $\eta$ ranges from $-\infty$ to $0$. We are now interested in what happens to small deviations of this FLRW-metric. We identify
\begin{equation}
g_{\mu \nu} = g^{0}_{\mu \nu} + \delta g_{\mu \nu},
\end{equation}
where $g^{0}_{\alpha \beta}$ denotes the background FLRW-metric. Since we obviously want $\delta g_{\alpha \beta}$ to transform as a tensor quantity, we will decompose it into parts, according to their transformation properties. Thus we acquire \textit{scalar, vector} and \textit{tensor} contributions to $g_{\alpha \beta}$.
\par The time-time component $\delta g_{00}$ behaves as a scalar and so we write
\begin{equation}
\delta g_{00} = 2 a^2 \phi,
\end{equation}
where $\phi$ is some scalar function. The quantities $\delta g_{0i}$, where $i$ ranges over spatial indices, are vectors. Any vector can be decomposed as the gradient of a scalar $B$ and a three-vector $\overline{S}$ with zero divergence.
\begin{equation}
\delta g_{0i} = a^2 (\partial_i B + S_i)
\end{equation}
The quantities $\delta g_{ij}$ should now behave as a tensor, and thus can be written as \cite{MukhanovBook}
\begin{equation}
\delta g_{ij} = a^2(2 \psi \delta_{ij} + \partial_i \partial_j E + \partial_j F_i + \partial_i F_j + h_{ij}),
\end{equation}
where $\psi$ and $E$ are scalar functions, $\overline{F}$ has zero divergence and the three-tensor $h_{ij}$ satisfies the analogous constraints
\begin{align}
h_{ij}\delta^{ij} &= 0, & \partial_i h_{kj}\delta^{ki} &= 0.
\end{align}
Briefly counting the independent degrees of freedom, we see that the scalar functions $\psi, E$ and $B$ bring three degrees of freedom. $\overline{S}$ and $\overline{F}$ both contribute two degrees of freedom, since they each satisfy a constraint. $h_{ij}$ then, as a symmetric three-tensor satisfying four constraints, contributes two degrees of freedom. So we have ten degrees of freedom, exactly the number of independent components of the symmetric tensor $\delta g_{\mu \nu}$.
\par Collecting all the terms, we can now write the metric of our universe if only scalar, vector or tensor perturbations are present. In that order
\begin{align}
\label{ScalarMetric}
ds^2 &= a^2 \left[(1 + 2 \phi)d\eta^2  + 2 \partial_i B dx^i d\eta  - \big((1-2\psi) \delta_{ij} - 2 \partial_i\partial_j E\big)dx^idx^j\right],\\
ds^2 &= a^2 \left[d\eta^2 + 2 S_i dx^id\eta - \left(\delta_{ij} - \partial_j F_{i} - \partial_i F_{j} \right)dx^i dx^j\right],\\
ds^2 &= a^2 \left[d\eta^2 -(\delta_{ij} - h_{ij})dx^idx^j \right].
\end{align}
It is now important to look at the behaviour of this metric perturbations under general coordinate transformations, since some of the ten degrees of freedom represent our freedom in the choice of coordinates. One can show that the simplest gauge invariant quantities we can construct are
\begin{eqnarray}\label{GIMetrPert1}
\Phi &=& \phi - \frac{1}{a}\left[a (B - E')\right]',\\\label{GIMetrPert2}
\Psi &=& \psi + \frac{a'}{a}\left[ B - E'\right],\\\label{GIMetrPert3}
V_i &=& S_i - F_i',
\end{eqnarray}
where primes denote derivatives with respect to conformal time. Note that $h_{ij}$ is already gauge invariant. Thus, we have seven remaining degrees of freedom and three degrees of freedom can be attributed to our choice of coordinate system.
\paragraph{Einstein Equations}
The aim is not only to perturb the metric around a flat FLRW universe, but also to consider perturbations of the energy momentum tensor. The Einstein equations are
\begin{equation}
 R_{\nu}^{\mu} - \frac{1}{2}\delta^{\mu}_{\nu} = 8 \pi G T_{\nu}^{\mu}.
\end{equation}
We split the Einsteintensor $G_{\nu}^{\mu} \approx G_{\nu}^{\mu (0)} + \delta G_{\nu}^{\mu}$, where $\delta G_{\nu}^{\mu}$ is linear in the metric perturbations. Dividing the energy-momentum tensor in a similar way we find the linearised Einstein equations
\begin{equation}
 \delta G_{\nu}^{\mu} = 8 \pi G \delta T_{\nu}^{\mu}.
\end{equation}
As before, we need to identify the gauge invariant degrees of freedom first. One can show that the simplest gauge invariant quantities are
\begin{eqnarray}\label{GIEMtensor}
 \overline{\delta G^{0 }_0} &=& \delta G^{0 }_0 - [\delta G^{0 (0)}_0]'(B - E'),\\
\overline{\delta G^{0 }_i} &=& \delta G^{0 }_i - \left(\delta G^{0 (0)}_0 - \frac{1}{3}\delta G^{k (0)}_k\right)\partial_i (B - E'),\\
\overline{\delta G^{i }_j}  &=& \delta G^{i }_j - \left(\delta G^{i (0)}_j\right)'(B - E'),
\end{eqnarray}
and similar for $\overline{\delta T^{\mu }_\nu}$. Thus we find
\begin{equation}\label{EinsteinGauge}
 \overline{\delta G_{\nu}^{\mu}} = 8 \pi G \overline{\delta T_{\nu}^{\mu}}.
\end{equation}
As before, one could now decompose $ \overline{\delta T_{\nu}^{\mu}}$ into the different contributions: scalar, vector and tensor perturbations. From the linear nature of eq.\ \eqref{EinsteinGauge}, we can already expect that the metric perturbations only couple to their corresponding energy-momentum perturbations; i.e. scalar metric perturbations couple to scalar energy-momentum perturbations and analogously for the vector and tensor perturbations.
\par Physically, scalar perturbations correspond to density perturbations, while vector perturbations are related to the rotational motion of our cosmic liquids and tensor modes are associated with gravitational waves. However, vector perturbations do not grow in time and tensor perturbations contribute far less to the energy density and pressure\cite{ Brandenberger:1993zc}\footnote{Vector and tensor perturbations are less important for cosmic structure formation, but they are important for observations of the CMB\cite{Bertschinger:2001is}. The Planck mission\cite{PlanckMissie} will in fact look for \textit{B-mode polarisation} in the CMB, a signature of vector and tensor perturbations.}. A typical ratio of tensor contributions to scalar contributions is in the range of $0.2$ to $0.3$ \cite{MukhanovBook}. In what follows, we will restrict ourselves to scalar perturbations as they are the ones that contribute most to the density variations in our universe, e.g.\ we will be concerned with the linearised Einstein equations using the metric of eq.\ \eqref{ScalarMetric}.
\paragraph{Introducing the Inflaton Field}
Now we suppose that the universe is filled with a minimally coupled real scalar field $\varphi$, the inflaton field that is subject to a potential $V(\varphi)$, with action given by eq.\ \eqref{InflatonAction}.  The inflaton then has an energy-momentum tensor
\begin{equation}\label{EMtensor}
 T_{\mu \nu} = \partial_{\mu} \varphi \partial_{\nu}\varphi - \frac{1}{2}g_{\mu \nu}\partial^{\mu} \varphi \partial^{\nu}\varphi + g_{\mu \nu}V(\varphi).
\end{equation}
As before, see chapter \ref{Inflation}, one can regard this scalar field as a perfect fluid.\footnote{In fact one can do everything that follows for different fluids, as is done in \cite{MukhanovBook} and \cite{Albrecht:1992kf}. One retains then an extra parameter in $c_s$, the \textit{speed of sound}, that depends on the equation of state of the liquid. For the canonical scalar field that we use, $c_s$ is equal to the speed of light.}
\par Let's now perturb this inflaton field around it's mean value: $\varphi(\eta, \overline{x}) = \varphi_0(\eta) + \delta \varphi(\eta, \overline{x})$. The equations of motion to linear order in the perturbations become
\begin{align}\label{BackgroundEOM1}
 \varphi_0'' &=  - 3H\varphi_0' - a^2 \partial_{\varphi}V,\\\label{BackgroundEOM2}
 \delta \varphi'' &= - 3H \delta \varphi ' + \Delta(\delta \varphi + \varphi_0'(B- E')) - a^2\partial_{\varphi}\partial_{\varphi}V \delta \varphi+ \varphi_0'(3\psi + \phi)' - 2a^2 \partial_{\varphi}V \phi.
\end{align}
We notice that the background variable $\varphi$ still obeys the equations of motion from the previous chapter, to first order. From the analogues of eq.\ \eqref{GIEMtensor} for $\overline{\delta T^{i }_j}$ we find that the relevant gauge invariant quantity in these equations is
\begin{equation}\label{GIPert}
 \overline{\delta \varphi} = \delta \varphi + \varphi'(B - E').
\end{equation}
The gauge invariant equation of motion for the scalar perturbations becomes
\begin{equation}\label{GIPertEOM}
 \overline{\delta \varphi}'' = - 2H \overline{\delta \varphi} ' + \Delta \overline{\delta \varphi} - a^2\partial_{\varphi}\partial_{\varphi}V \overline{\delta \varphi} + \varphi_0'(3\Psi + \Phi)' - 2a^2 \partial_{\varphi}V \Phi.
\end{equation}
It is important to note that this last equation was derived using eq.\ \eqref{BackgroundEOM1} and \eqref{BackgroundEOM2}. %
In \cite{MukhanovBook} one proceeds to show that the gauge invariant quantities $\Phi$ and $\Psi$ need to be equal. The metric from eq.\ \eqref{ScalarMetric} then reduces to
\begin{equation}\label{ReducedMetric}
 ds^2 = a^2 \left[(1 + 2 \Phi)d\eta^2   - (1-2\Phi) d\overline{x}^2\right].
\end{equation}
Notice that this is very reminiscent of the Newtonian limit of General Relativity: with $a = 1$ we can identify $\Phi$ with the Newtonian potential. %
The variable $\Phi$ usually goes by the name of \emph{Bardeen potential}. It is the only physical degree of freedom left for the metric perturbations.
\paragraph{Equations of Motion for a Slow-Roll Theory}\label{EOMSlowRoll}
We are now in a position to actually solve the linearised Einstein equations, eq.\ \eqref{EinsteinGauge}, by substituting equation \eqref{EMtensor} and calculating the gauge invariant perturbation of the Einstein Tensor $\overline{\delta G_{\nu}^{\mu}}$ with metric given by eq.\ \eqref{ReducedMetric}. This involves a lot of tedious algebra that we will omit here. The result is \cite{Brandenberger:1993zc}
\begin{align}\label{ClassicalEOM}
  \nu''&= \left(\Delta  + \frac{z''}{z}\right)\nu, \\
  \nu &= a \left( \overline{\delta \varphi} + \frac{\varphi_0'}{H} \Phi \right),\\ \label{MukhSas}
  z &= a\frac{\varphi_0'}{H},
\end{align}
where $\nu$ is called the \emph{Mukhanov-Sasaki} variable. While we omitted the algebra involved, it is important to note that the background equations for $\varphi_0$  and $g_{\mu \nu}$, eq.\ \eqref{GIPertEOM} and eq.\ \eqref{FLRWmetric}, were of critical importance in obtaining this result.
\par It may seem strange that we retain only one physical degree of freedom at the end of this story. However, the metric fluctuations contain only one degree of freedom, as do the inflaton perturbations. These degrees of freedom satisfy a constraint: the Einstein equations. It is in this way that only one degree of freedom remains: the Mukhanov-Sasaki variable. We can already note that $\nu$ is not the physical variable: the gauge invariant perturbation is $\overline{\delta \varphi} + \frac{\varphi_0'}{H} \Phi$. $\nu$ includes an extra scaling and we conclude that the physical variable is $a^{-1}\nu$.
\par In the case of slow-roll inflation we can further simplify the equations of motion. Using the approximations in eqs.\ \eqref{SlowRoll} and treating $H$ as a constant we can show that
\begin{equation}\label{ZalsA}
 \frac{z''}{z} \approx \frac{a''}{a}  + \frac{2H \varphi''}{\varphi'} + \frac{\varphi'''}{\varphi'}.
\end{equation}
For a perfect slowly rolling field, the second and third term are zero. In more general slow-roll inflation, we can safely ignore the second and third term. Thus we find that the equation of motion in the regime of slow-roll inflation is given by
\begin{equation}\label{ClassicalEOMSRInf}
 \nu''=\left(\Delta  + \frac{a''}{a}\right)\nu.
\end{equation}
Given a background $a(\eta)$, a solution to some slow-roll model, one could now compute the classical evolution of perturbations. For this however, we need some initial conditions. Solving this equation is what we will do in the next paragraph.
\paragraph{Solving the Equations of Motion}
Lets first look at eq.\ \eqref{ClassicalEOM} in more detail. It is a generalisation of the Klein-Gordon equation for the scalar field $\nu$ on expanding spacetime. As in standard field theory, we can decompose the Mukhanov-Sasaki variable in Fourier components
\begin{equation}\label{Fourier}
 \nu( \eta, \vec{x}) = \frac{1}{(2\pi)^{3/2}} \int  d^3k   \nu_{\vec{k}}(\eta) e^{i \vec{k}\cdot \vec{x}}.
\end{equation}
And it is now easy to check that the components $\nu_{\vec{k}}$ satisfy the condition
\begin{equation}\label{ClassicalEOMFourier}
 \nu_{\vec{k}}''= \left(- k^2  + \frac{z''}{z}\right)\nu_{\vec{k}}.
\end{equation}
How to make sense of this equation? There are two limits in which the solution is clear. As explained in the previous chapter, we imagine the observable part of the universe as originating from a small volume of space of a typical scale on the order of $H^{-1}$. The physically relevant modes are then those with wavelengths well inside the Hubble radius: $\frac{\lambda}{2 \pi} < H^{-1}$ or equivalently $k > H$. Thus, at the onset of inflation, we imagine that gravitational effects are not yet important for these modes,  as their wavelength is significantly smaller than the Hubble scale. In this regime, we expect $k^2$ to dominate over $\frac{z''}{z}$. Eq. \eqref{ClassicalEOMFourier} reduces (as it should) to the Klein-Gordon equation for a free scalar field on Minkowski spacetime. The solutions to eq.\ \eqref{ClassicalEOMFourier} are plane waves.
Then as inflation progresses, the physical wavelengths grow in time, as discussed in the previous chapter. At a certain moment, the physical wavelength $\lambda_{phys} = a\lambda$ will equal the Hubble radius. On this scale, the expansion of space becomes important for the time evolution of the mode $\nu_{\vec{k}}$. Thus the term $ \frac{z''}{z} \approx \frac{a''}{a}$ becomes more and more important. In the limit where $k^2 \ll \frac{a''}{a}$ the solutions behave essentially as $a$. Thus, for $\eta \to 0$, $\nu \approx C a$, where $C$ is some complex constant. The amplitude of the Mukhanov-Sasaki variable grows rapidly in this regime.
\par The transition between both regimes is commonly called the \emph{Hubble exit} of the mode. The condition for this to occur is that $k^2 = \frac{a''}{a}$. For exponential and power-law inflation this means $|k\eta| = 1$. The regimes are thus characterised by $|k\eta|$: the mode is a free field on Minkowski spacetime when $|k\eta| \gg 1$ and grows in time proportional to the scale factor when $|k\eta| \ll 1$.
\par One can also draw the conclusion that the system described by a single Fourier mode $\nu_{\vec{k}}$ is in fact a time dependent harmonic oscillator. The frequency of the system is then
\begin{equation}\label{Frequency}
 \omega^2 = \left(k^2  - \frac{z''}{z}\right).
\end{equation}
Indeed, when $\omega^2$ is positive, the solutions are plane waves, Re$(\nu_{\vec{k}})$ and Im$(\nu_{\vec{k}})$ oscillate. But when the mode exits the Hubble radius $\omega^2$ changes sign. The system then transforms into an upside down harmonic oscillator and $\nu_{\vec{k}}$ will go `rolling down the hill'. During the rest of this thesis we will use the upside-down harmonic oscillator as a toy model and a useful analogy.
\par Owing to the above analysis, natural initial conditions for the Cauchy problem posed by eq.\ \eqref{ClassicalEOMFourier} are given by a free field on Minkowski space. Thus, denoting the time when inflation begins with $\eta_{ini}$
\begin{align}\label{InitialConditions}
\nu_{\vec{k}}(\eta_{ini}) = \frac{1}{\sqrt{2 k }} e^{-i k \eta_{ini}} && \left. \nu_{\vec{k}}(\eta)'\right|_{\eta = \eta_{ini}}  = -i\sqrt{\frac{k}{2}} e^{-i k \eta_{ini}}.
\end{align}
For exponential and power-law inflation, where the scale factor is proportional to $\eta^{-p}$, one can solve eq.\ \eqref{ClassicalEOMFourier} supplemented by these initial conditions exactly. The solution is given in terms of Hankel functions of the first kind of degree $p+ \frac{1}{2}$ \cite{Martin:2004um}
\begin{equation}\label{exactF}
\nu_{\vec{k}}(\eta) = \sqrt{\frac{\pi}{4k}}e^{- i k \eta_{i}}e^{\pi \left(\frac{1+p}{2}\right)}\sqrt{-k\eta}H^{(1)}_{p+ \frac{1}{2}}(-k\eta).
\end{equation}
For de Sitter-expansion ($p = 1$), this reduces to \cite{CosmologyCursus}
\begin{equation}\label{P1Solution}
 \nu_{\vec{k}}(\eta) = \frac{1}{\sqrt{2k}}\left(1 - \frac{i}{k\eta}\right)e^{-i k \eta}.
\end{equation}
This solution is clearly a plane wave for $|k\eta|$ big, while it becomes proportional to $\frac{-1}{H\eta} = a$ when $|k \eta|$ becomes small. Some additional properties of these solutions to the classical equations of motion for general inflationary models are given in appendix \ref{AppLemma}.
\section{An Overview}\label{Overview}
To promote the outlined theory to a quantum theory, we need an action to quantise. Let's use the opportunity to briefly redo what we did in the previous chapter and section, in a conceptually more transparent way.\footnote{This way is however a lot more computationally intensive.}
\par We started from the Einstein-Hilbert action and the action for a minimally coupled real massless scalar field $\varphi$, as in chapter \ref{Inflation}
\begin{equation}
 S^{0} = -\frac{1}{16\pi G}\int d^4x \sqrt{-g} \left(R + \partial^{\mu} \varphi \partial_{\mu}\varphi - V(\varphi)\right),
\end{equation}
and by varying with respect to the metric $g_{\mu \nu}$ we found the equations of motion, being the Einstein field equations. By applying the cosmological principle we found the FLRW metric and the Friedmann equations. As a result we get homogeneous background solutions $\varphi_{(u)}(\eta)$ and $a_{(u)}(\eta)$, where the subscript $u$ stands for `unperturbed'.
\par Now, we look at perturbations around the solutions to these equations of motion. We write
\begin{align}
g_{(p) \mu \nu} (\eta, \vec{x}) &= g^{0}_{(u)\mu \nu} (\eta) + \delta g_{\mu \nu} (\eta, \vec{x}),& \varphi_{(p)}(\eta, \vec{x}) = \varphi^{0}_{(u)}(\eta) + \delta \varphi(\eta, \vec{x}),
\end{align}
where $g^{0}_{\mu \nu}$ is the FLRW metric, incorporating some scale factor $a_{p}$, where the subscript $p$ now stands for `perturbed'. Using this division between background and perturbation, we identified the gauge invariant parts of the perturbations in equations \eqref{GIMetrPert1}, \eqref{GIMetrPert2} and \eqref{GIMetrPert3}. As before, we restrict ourselves to scalar perturbations. The following step is writing the Einstein-Hilbert action up to second order in the gauge invariant scalar perturbations, denoted by bars.
\begin{equation}\label{FullAction}
S = S^{0}(g^{0}_{(p) \mu \nu}, \varphi^{0}_{(p)}) + \delta S_1 (g^{0}_{(p) \mu \nu}, \varphi^{0}_{(p)},\overline{\delta g_{\mu \nu}},\overline{\delta \varphi} )
 + \delta S_2 (g^{0}_{(p) \mu \nu}, \varphi^{0}_{(p)},\overline{\delta g_{\mu \nu}},\overline{\delta \varphi} )
\end{equation}
Here $\delta S_1$ is the first order perturbation, $\delta S_2$ the second order. For small perturbations, we can assume that the equations of motion for $g_{\mu \nu}^{0}$ and $\varphi^0$ are still solved by the same solutions as before. In other words, we assume that $a_{(u)}(\eta) \approx a_{(p)}(\eta)$ and $\varphi_{(u)}(\eta) \approx \varphi^{0}_{(p)}(\eta)$. Since these are solutions to the equations of motion, we know that $\delta S_1$ vanishes\footnote{This immediately explains why we wrote the action up to second order.}. Thus, by choosing these particular solutions, $S^{0}$ becomes fixed and we are left with $\delta S_2$. In terms of the Mukhanov-Sasaki variable $\delta S_2$ equals \cite{MukhanovBook}
\begin{equation}\label{actie}
 \delta S_2 = \frac{1}{2}\int d^4x \left( \eta^{\mu \nu}\partial_{\mu} \nu \partial_{\nu} \nu + \frac{z''}{z}\nu^2\right),
\end{equation}
where $\eta^{\mu \nu}$ is the Minkowski metric. Varying this action with respect to $\nu$ yields eq.\ \eqref{ClassicalEOM}, or eq.\eqref{ClassicalEOMFourier} in terms of Fourier modes.
\section{The Quantum Theory}\label{QuanTheor}
\paragraph{Semiclassical versus Quantum Theory}
It is the action from eq.\ \eqref{actie} that we will now quantise and promote our theory from a classical to a semi-classical one. The quantised version of this action will not constitute a full quantum theory of cosmological perturbations since the full action is in fact given by eq.\ \eqref{FullAction}. A complete theory would thus quantise the background variables as well as the Mukhanov-Sasaki variable $\nu$. It is clear that quantising the full action is a difficult matter, not only computationally, but also conceptually. The result of such quantisation would be wavefunctions for $g_{\mu \nu}$ and interpreting those is not a trivial matter. By only quantising the action for the perturbations, we avoid such problems for the moment. Later on, in section \ref{dBBCosm}, we will return to the problem of quantising the full action.
\paragraph{Quantisation}\label{Quantisation}
To quantise the action in eq.\ \eqref{actie}, we will first transfer from the Lagrangian to the Hamiltonian formalism. The momentum adjoint to $\nu$ is
\begin{equation}\label{momentum}
\Pi = \nu'.
\end{equation}
And thus the Hamiltonian reads
\begin{equation}
 H =  \frac{1}{2}\int  d\vec{x} \left[ \Pi^2 + \left(\Delta - \frac{z''}{z}\right) \nu^2 \right].
\end{equation}
Inspecting this Hamiltonian, we see that quantising the Mukhanov-Sasaki variable is analogous to quantising a scalar field on flat Minkowski space with a time-dependent mass $m^2 = - \frac{z''}{z}$.
\par To quantise the theory, we want to promote $\nu$ and $\Pi$ to operators $\hat{\nu}$ and $\hat{\Pi}$ and impose the equal-time commutation relations on them.
\begin{align}\label{EqTiComRel}
\left[\hat{\nu}(\eta, \vec{x}), \hat{\nu}(\eta, \vec{y})\right] &= \left[\hat{\Pi}(\eta, \vec{x}), \hat{\Pi}(\eta, \vec{y})\right] = 0,\\
\left[\hat{\nu}(\eta, \vec{x}), \hat{\Pi}(\eta, \vec{y})\right] &= i \delta(\vec{x} - \vec{y}).
\end{align}
Proceeding as in the standard case, we decompose $\hat{\nu}$ into its Fourier components
\begin{equation}
  \hat{\nu}( \eta, \vec{x}) = \frac{1}{(2\pi)^{3/2}} \int  d^3k \quad \hat{\nu}_{\vec{k}}(\eta) e^{i \vec{k}\cdot \vec{x}}.
\end{equation}
Because our field is real, we demand that $\hat{\nu}_{\vec{k}}^{*} = -\hat{\nu}_{\vec{k}}$. The Hamiltonian in terms of Fourier modes, with $\Pi^{*}_{\vec{k}} = \nu_{\vec{k}}'$, then becomes
\begin{equation}\label{QHamiltonian}
\hat{H} = \frac{1}{2}\int d^3k \left[ \Pi_{\vec{k}}\Pi_{\vec{k}}^* + \left(k^2 - \frac{z''}{z}\right)\nu_{\vec{k}}^* \nu_{\vec{k}}\right],
\end{equation}
and we notice that the Fourier modes are all decoupled, as in standard field theory. This decoupling happens only because we are dealing with a perturbative theory: expanding the action from eq.\ \eqref{FullAction} to higher order would yield interaction terms in the field theory of $\nu$, as for example done in \cite{Martineau:2006ki}.
\paragraph{Annihilation and Creation Operators}
It will pay off to look at the theory in its second-quantised form, in other words, in terms of annihilation and creation operators, as for example done in \cite{Lesgourgues:1996jc}.
\par We will start by decomposing the position and momentum operators:
\begin{align}\label{Bogoliubov}
\hat{\nu}_{\vec{k}} &= f_k(\eta) \hat{a}_{\vec{k}} + f_k^{*}(\eta)\hat{a}^{\dagger}_{\vec{k}}, &
\hat{\Pi}_{\vec{k}} = -i \left(g_k(\eta) \hat{a}_{\vec{k}} - g_k^{*}(\eta)\hat{a}^{\dagger}_{\vec{k}}\right).
\end{align}
Such a transformation is called a \emph{Bogoliubov transformation}. For the case of scalar field on flat Minkowski space, the usual decomposition is given by $f_k = 2k^{-1/2}$ and $g_k = 2^{-1/2}\sqrt{k}$. In our case, the functions $f_k,g_k$ will depend on time since we have a time dependent Hamiltonian due to the growth of $a(\eta)$. 
\par Let's further specialise using the action from eq.\ \eqref{actie}. %
Imposing the equal-time commutation relations, eqs.\ \eqref{EqTiComRel}, implies the following commutation relations for $\hat{a}_{\vec{k}}$ and $\hat{a}^{\dagger}_{\vec{k}}$
\begin{align}
\left[\hat{a}_{\vec{k_1}}, \hat{a}_{\vec{k_2}}\right] &= \left[\hat{a}^{\dagger}_{\vec{k_1}}, \hat{a}^{\dagger}_{\vec{k_2}}\right] = 0,\\
\left[\hat{a}^{\dagger}_{\vec{k_1}}, \hat{a}^{\dagger}_{\vec{k_2}}\right] &=\frac{1}{f_k g_k^{*} + g_kf_k^{*}}\delta(\vec{k_1} - \vec{k_2}).
\end{align}
If we want to recover the usual interpretation of $\hat{a}_{\vec{k}}$ and $\hat{a}^{\dagger}_{\vec{k}}$ as annihilation and creation operators, we should impose a condition on the time dependent functions $f_{k}$ and $g_k$
\begin{equation}\label{Wronskian}
 (g_{k}  f^{*}_{\vec{k}} +  g^{*}_{k} f_{k}) = 1.
\end{equation}
Using this condition, we can easily invert eq.\ \eqref{Bogoliubov}
\begin{align}
\hat{a}_{\vec{k}} &= g_k^{*}\hat{\nu}_{\vec{k}} + i f_{k}^{*}\hat{\Pi}_{\vec{k}}, &
\hat{a}^{\dagger}_{\vec{k}} = g_k\hat{\nu}_{\vec{k}} - i f_{k}\hat{\Pi}_{\vec{k}}.
\end{align}
In addition, the Heisenberg equations of motion read
\begin{align}\label{HeisenbergEOM1}
\frac{d\hat{\nu}_{\vec{k}}}{d \eta} &= -i\left[\hat{\nu}_{\vec{k}}, \hat{H} \right] = \hat{\Pi}_{\vec{k}}^{*},\\ \label{HeisenbergEOM2}
\frac{d\hat{\Pi}_{\vec{k}}}{d \eta} &=  -i\left[\hat{\Pi}_{\vec{k}}, \hat{H} \right] = - \left(k^2 - \frac{z''}{z}\right)\hat{\nu}_{\vec{k}}^*.
\end{align}
This implies that $g_k = if'_k$. It is in this context that we can regard eq.\eqref{Wronskian} as a Wronskian condition on the system of differential equations defined by eq.\ \eqref{HeisenbergEOM1} and eq.\ \eqref{HeisenbergEOM2}. It imposes the need for linear independent solutions to these equations. Moreover, since we want our field $\nu_{\vec{k}}$ to behave as a free field on Minkowski space for very large values of $|k\eta|$, we will choose as initial conditions for $f_k$ and $g_k$
\begin{align}
f_k(\eta_{i}) &= \frac{1}{\sqrt{2k}} e^{-i k \eta_{i}}, & g_k(\eta_{i}) = \sqrt{\frac{k}{2}}  e^{-i k \eta_{i}}.
\end{align}
Filling in the Bogoliubov transformation, eq.\ \eqref{Bogoliubov}, we find that $f_{k}$ has to obey the following equation
\begin{equation}
f_{k}'' + \left(k^2 - \frac{z''}{z}\right)f_{k} = 0.
\end{equation}
We notice that this is exactly the classical equation of motion, eq.\ \eqref{ClassicalEOM}!
\par As a consistency check, we can check that the creation operator is indeed time independent. Using the Wronskian condition and noting that $f_{k}$ is a solution to the classical equations of motion we get
\begin{eqnarray}
 \frac{d}{d\eta} \hat{a}^{\dagger}_{\vec{k}} = 0.
\end{eqnarray}
We conclude that both $\hat{a}^{\dagger}_{\vec{k}}$ and $\hat{a}_{\vec{k}}$ are time independent. This means that the operator $\hat{a}^{\dagger}_{\vec{k}}\hat{a}_{\vec{k}}$ will be an invariant of motion. Note that this means that $N$-particle states are robust under time evolution.\footnote{This fact will be used in appendix \ref{WFAppendix} to derive the wavefunctions of similar systems.}
\par For each $\vec{k}$ one can now continue to define a vacuum $|0\rangle$ in the Fock space of $a_{\vec{k}}$ and $a^{\dagger}_{\vec{k}}$ by demanding
\begin{equation}\label{Vac}
a_{\vec{k}} |0\rangle_{\vec{k}} = 0.
\end{equation}
In general, one defines a state $|n\rangle$ by applying $a^{\dagger}_{\vec{k}}$ repeatedly:
\begin{equation}\label{Nparticles}
|n\rangle_{\vec{k}} = \frac{(a^{\dagger}_{\vec{k}})^{n}}{n!} |0\rangle_{\vec{k}}  .
\end{equation}
To finish, let us briefly look at the possible interpretation of the states $|n\rangle_{\vec{k}}$. When $|k\eta|$ is big, our field evolves as if in Minkowski space, and the Bogoliubov transformation, eq.\ \eqref{Bogoliubov}, reflects that. The annihilation and creation operators are the canonical ones, and we can safely interpret the states $|n\rangle_{\vec{k}}$ as a state with $n$ particles (or quanta of energy) with momentum $\vec{k}$.
\par However, when $|k\eta|$ is small, when the mode has exited the Hubble radius, $a^{\dagger}_{\vec{k}}$ and $a_{\vec{k}}$ are far from the canonical creation and annihilation operators of a field living on flat Minkowski space. As noted before, this is due to the time dependence of the frequency of our harmonic oscillator, especially because the potential changes sign when $|k \eta|= 1$. In appendix \ref{AppAlt} we consider a different picture, starting from a different action that is equivalent to the on used here on the classical level. Its interpretation when $|k\eta|$ is small is more straightforward.
\chapter{Inhomogeneities in the CMB: An Inflationary Explanation?}\label{Observation}
In chapter \ref{Inflation}, we discussed the cosmic microwave background, as it is measured by us. It appears to be extremely homogeneous, with deviations from uniformity of the order $10^{-4}$. To explain this homogeneity the inflationary model was developed. However, we now need a mechanism to explain the inhomogeneities and explain their apparent random distribution across the sky.
\par In this chapter we explain the standard account of how quantum fluctuations of the Mukhanov-Sasaki variable, introduced in the previous chapter, give rise to density perturbations in the early universe. At first sight, we will find that the theory explains the observed pattern in the sky. Careful inspection however, will reveal that there is some missing element. There is no mechanism that explains how the \emph{quantum} fluctuations give rise to \emph{classical} density perturbations. At the end of this chapter we will try to formulate this problem as precise as possible, in order to have a solid footing for the discussion on various proposed explanations.
\section{Characterising the CMB Inhomogeneities}
To find a workable theory, we will first need to understand what we see in the sky. The inhomogeneities in the CMB seem random, but what do we mean by \emph{random}?
We can decompose the observed relative temperature variations in spherical harmonics
\begin{equation}\label{SpherHarm}
\frac{\Delta T}{T} = \sum_{lm} a_{lm}Y_{lm}(\theta, \phi),
\end{equation}
where the angles $\theta$ and $\phi$ characterize a certain direction in the sky and $Y_{lm}$ are the spherical harmonics of degree $l$ and order $m$.
A more precise formulation of \emph{random} is now evident: the coefficients $a_{lm}$ are independent random variables, each one distributed according to some probability distribution $F_{lm}$. In addition, we impose that the universe is rotationally invariant, so the fluctuations must be isotropic. For this reason, we restrict \emph{random} to mean that the distributions $F_{lm}$ are independent of direction, and thus not depending on $m$.
\par The CMB can be thought of as a particular realization of the $a_{lm}$. One could imagine ourselves living in a different universe, where the coefficients $a_{lm}$ would be different. If we have a large number of such universes, all the $a_{lm}$ would be distributed according to their distribution $F_{l}$. These reflections already touch the problem upon that we only have \emph{one} universe. A decomposition of the entire sky, would only give us one set of $a_{lm}$. For a single realization of random parameters, it is impossible to deduce anything about their distribution $F_{l}$.
\par In this kind of matter, it is often convenient not to talk about the values of the $a_{lm}$. More manageable quantities are the correlators $\langle a_{lm} a_{l'm'}\rangle$, which we write down as
\begin{equation}\label{correlators}
\langle a_{lm} a_{l'm'} \rangle = P(l)\delta_l^{l'}\delta_{m}^{m'}.
\end{equation}
where we call $P(l)$ the spectrum. Rotation invariance makes it independent of $m$ and the Kronecker deltas make sure that the different $a_{lm}$ are independent variables. One can continue to define higher order correlators, and in fact, one can completely specify the function $F$ in terms of all its correlators.
\par Experimentally, one can find out more about $P(l)$ by dividing the sky into patches. Suppose we want information for some fixed $l$. We then divide the sky into $N_{l}$ patches, in such a way that the size of a typical patch is inversely proportional to $l$, meaning that we take the patch size proportional to the wavelength of the perturbations we consider. For each such patch, we then decompose the available temperature fluctuations and average over $m$. This gives us $N_{l}$ values of $P_{l}$.
\par
\begin{figure}\label{WMAPPower}
\begin{center}
\includegraphics[scale = 0.09]{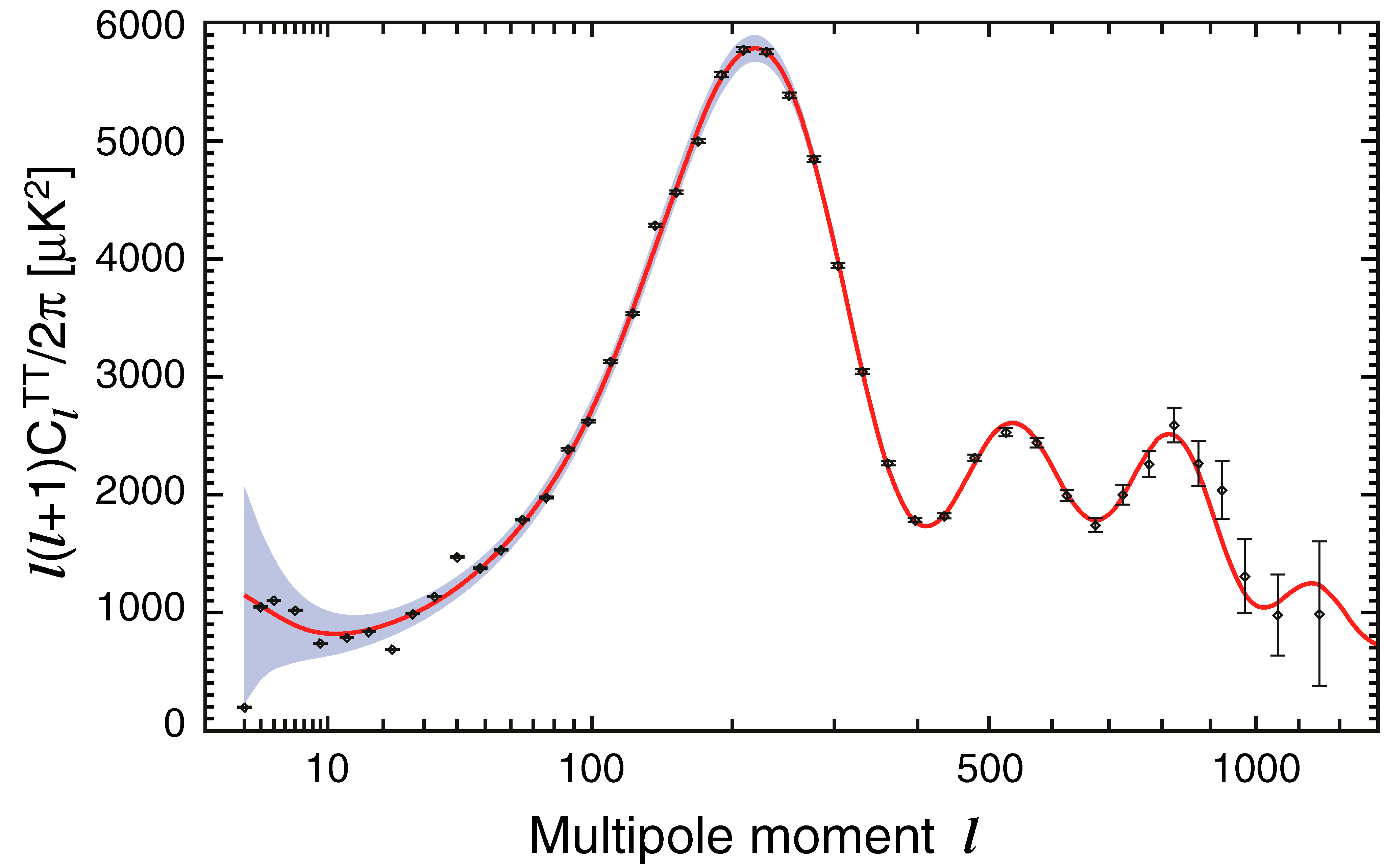}
 \caption{Power spectrum of the CMB anisotropy with experimental errorbars. For large scales, the influence of cosmic variance is suggested by the blue band. The red line is a theoretical model. Figure courtesy of NASA.}
\end{center}
\end{figure}

 The common approach is then to calculate the two-point-correlators, eq. \eqref{correlators}, for each patch. By virtue of the ergodic theorem \cite{Lyth:2009zz}, a spatial average is a good estimator of an average over an ensemble in this case. What it means, is that we can for all practical (read as `averaging') purposes understand each patch as a `distinct universe'. Thus we have $N_{l}$ values for the two point correlator, and we can start statistical analysis. However, one typically works with the dimensionless quantity $ \mathbb{P}(l) = \frac{k^3}{2 \pi^2}P(l)$ which is called the \emph{power spectrum} of the inhomogeneities. The result of the WMAP experiment is usually displayed as in fig. \ref{WMAPPower}.
In this picture, the multipole moments $l$ are plotted on the horizontal axis, while the power spectrum is plotted on the vertical scale. There are many very distinct features on this graph that have been adequately studied during the last decades. The appearance of acoustic peaks for instance and their location and height has been the subject of extensive study. We will not go into the details of this spectrum because the majority of them is explained by acoustic theory in the post-inflationary universe. In short, we are more concerned with the origin of the perturbations than with their evolution through the non-inflationary epochs. A more in depth explanation of the spectrum can be found in \cite{CosmologyCursus}, \cite{Lyth:2009zz}. An animated, but more intuitive approach can be found at \cite{WayneHu}.
\par In the previous paragraph we have (very briefly) studied the experimental observations of the temperature variation across the sky. Thus, any theory about cosmological perturbations should be able to explain a spectrum like the one we see, e.g. in fig. \ref{WMAPPower}. As it turns out, the acoustic theory due to Harrison and Z'eldovich (see chapter \ref{Inflation}) is able to reproduce excellently the observed spectrum, granted that the density perturbations at the end of the inflationary epoch have a Harrison-Z'eldovich spectrum. This power spectrum is easy to describe: $\mathbb{P}_{l}$ is independent of $l$. For this reason it is commonly called a \emph{scale invariant} power spectrum.
\section{Spectrum of $\nu$}\label{ObsSpec}
To see how one gets from a quantum mechanical Hamiltonian, eq. \eqref{QHamiltonian}, to a scale invariant power spectrum, we start by considering the initial state of the universe. Generically one assumes the Mukhanov-Sasaki variable to start evolving from the \emph{Bunch-Davies vacuum}, defined as the product state of the vacuum for all modes
\begin{equation}\label{BunchDavies}
|0\rangle_{BD} = \prod_{\vec{k}} |0\rangle_{\vec{k}}.
\end{equation}
The usual argument is that the Bunch-Davies vacuum is the most symmetric state possible, and thus is the most `natural' initial condition for our universe to be in. A second argument, given in \cite{Sudarsky:2009za}, is that we want to explain the appearance of inhomogeneities in our universe, not assume them from the beginning by choosing an excited state for the Mukhanov-Sasaki variable.
\par To see what happens, one needs to know the expansion of Fourier components of the Mukhanov-Sasaki variable, $\nu_{\vec{k}}$ into spherical components $\nu_{lm}(k, \eta)$. One can show that%
\begin{equation}
\hat{\nu}_{lm}(k, \eta) = k i^{l} \int d\Omega_{k} \hat{\nu}_{\vec{k}}(\eta)  Y^{*}_{lm}\left(\frac{\vec{k}}{k}\right).
\end{equation}
In \cite{Lyth:2009zz}, one continues to show that the correlators for the $\nu_{lm}(k)$ then are given by
\begin{equation}
\langle  \hat{\nu}_{lm}(k, \eta), \hat{\nu}_{l'm'}(k, \eta) \rangle = \delta_{l}^{l'} \delta_{m}^{m'} \langle \hat{\nu}_{\vec{k}}(\eta) \hat{\nu}_{\vec{k'}}(\eta)\rangle ,
\end{equation}
where, as before, the $\nu_{\vec{k}}$ are the Fourier components of $\nu$ and
\begin{equation}
 \langle \hat{\nu}_{\vec{k}}(\eta) \hat{\nu}_{\vec{k'}}(\eta)\rangle = \delta(\vec{k} - \vec{k'}) \frac{2\pi}{k^3} P_{\nu}(k, \eta).
\end{equation}
The factor $P_{\nu}(k)$ is called the spectrum of the Mukhanov-Sasaki variable and can be found by computing the correlators for the Fourier components. Using the commutation relations for the annihilation and creation operators present in $\nu_k$, direct computation gives us the important equation
\begin{equation}\label{MukhPowerSpectrum}
 P_{\nu}(k, \eta) = |f_{k}(\eta)|^2.
\end{equation}
However, we should remember from the previous chapter that $\nu$ isn't the physical variable. It is the rescaling of the fluctuations with a factor $a(\eta)$, thus the physical spectrum is given by
\begin{equation}\label{PS}
 P_{\text{phys}}(k, \eta) = \frac{|f_{k}(\eta)|^2}{a(\eta)^2}.
\end{equation}
For exponential and power-law inflation, the ratio $\frac{|f_{k}|}{a}$ is proportional to $k^{-3/2}$ for late times \cite{Lesgourgues:1996jc}. Taking as example exponential inflation, $p = 1$, $f_{k}$ is given by eq. \eqref{P1Solution} and the power spectrum corresponding to this spectrum becomes
\begin{equation}\label{PowerSpectrumPhys}
 \mathbb{P}_{k} = \frac{k^2 H^2}{4 \pi^2}\left( \eta^2 + \frac{1}{k^2}\right).
\end{equation}
And thus, for any mode that didn't leave the Hubble radius too late ($|k\eta| << 1$), this power spectrum is (approximately) scale invariant. Similar calculations can be done for other $p$ and in fact, a (nearly) scale invariant power spectrum is a very generic prediction for a lot of inflationary models.
\par It is important to note, that it is a valid approximation to neglect the term $\eta^2$ in for all the modes that are discernable across the sky. By this we mean, that all the wavelengths observed by WMAP crossed the Hubble radius early enough\cite{CosmologyCursus}. Let's make a (very crude) estimate. We will take the end of inflation at $t_{e}= 10^{-32} s$ after the big bang, while decoupling happened approximately at $t_{d} = 10^{14} s$ after the big bang\cite{LiddleBook} and we are at present at about $t_p = 10^{17}s$. Taking the Hubble radius at present to be (very) roughly $H^{-1} = 10^{17} s$ and the angular resolution of the WMAP experiment to be of order $1^{\circ}$, we get that the smallest intrinsic wavelength we can discern in the CMB is of order
\begin{equation}
 \lambda_{phys}(t_p) \sim 10^{15}s.
\end{equation}
In a matter dominated universe $a \sim t^{2/3}$ so we have
\begin{equation}
 \lambda_{phys}(t_d) = \lambda_{phys}(t_p)\frac{a(t_d)}{a(t_p)} \sim 10^{13}s.
\end{equation}
Continuing past decoupling, we have $a \sim t^{1/2}$ and we get for the physical wavelength at the end of inflation
\begin{equation}
 \lambda_{phys}(t_e) \sim 10^{-10}s.
\end{equation}
For the Hubble radius during matter and radiation dominance $H^{-1} \sim t$
\begin{equation}
 H^{-1}(t_e) \sim 10^{-49} s.
\end{equation}
So at the end of inflation, the ratio $\frac{\lambda_{phys}}{H^{-1}}$ was of order $10^{39}$! We can thus safely assume that the modes we see in the sky were well outside the Hubble radius.
\par It is also illustrative to note that eq. \eqref{PowerSpectrumPhys} doesn't change much when we consider initial states that are not the Bunch-Davies vacuum. Excited states will contribute extra constant terms to the analogue of eq. \eqref{MukhPowerSpectrum}. Thus, the contribution of these terms to eq.\ \eqref{PowerSpectrumPhys} decreases in time as $a^{-2}$. Given that the energy contained in these excitations is not too far from the ground-state energy, we can safely neglect such terms. This can be regarded as another reason to adopt the Bunch-Davies vacuum as initial state.
\section{Spectrum of the Density Fluctuations}\label{Spectra}
While we succesfully recovered a power spectrum that is (approximately) scale invariant, it is now time to answer the question how fluctuations of the Mukhanov-Sasaki variable result in density perturbations. These density perturbations will then account for the observed inhomogeneities in the CMB, via the Sachs-Wolfe effect (see chapter \ref{Inflation}).
\par The standard account starts by looking for a solution to the Schr\"odinger equation, that corresponds to the ground state of the system: the Bunch-Davies vacuum. That wavefunction can be found in \cite{Martin:2007bw}. In appendix \ref{WFAppendix} we construct the wavefunctions for generic states of $\nu$, and the initial ground state is given by\footnote{Notice that this is the wavefunction for the ground state of the action \eqref{actie}. For the alternative action \eqref{AltAction} one can also construct similar wavefunctions, as for instance found in \cite{Lesgourgues:1998gk} and \cite{PhysRevD.85.083506}.}
\begin{equation}
\Psi_{0}(\eta,\nu) = \prod_{\vec{k}} N(\eta) \exp \left(\frac{i}{2} \frac{\dot{f_k}^{*}}{f_k^{*}}\nu_{\vec{k}}^2\right),
\end{equation}
where $N(\eta)$ is a normalisation constant, that is irrelevant for now. The first thing to notice is that this wavefunction is a Gaussian for each mode $\nu_{\vec{k}}$. The width of this Gaussian can be read off and using the Wronskian condition \ref{Wronskian}
\begin{align}\label{width}
 (\Delta \hat{\nu}_{\vec{k}})^2 &= \left[ \text{Im}\left(\frac{\dot{f}^{*}}{f^{*}}\right)\right]^{-1} = |f_{k}|^2
\end{align}
This means that the physical variable $a^{-1}\hat{\nu}$ has Fourier coefficients with Gaussian wavefunctions of width given by
\begin{equation}\label{PhysWidth}
 (\Delta a^{-1}\hat{\nu}_{\vec{k}})^2 = \left[ \frac{|f_{k}|^2}{a^2}\right].
\end{equation}
Noting that in the slow-roll approximation $V(\phi_0)$ dominates the kinetic energy, the energy density of the inflaton field $\phi$ is given by
\begin{align}
 \rho &= \frac{1}{2}\dot{\phi}^2 + V(\phi) \approx V(\phi).
\end{align}
Thus, the classical first order energy density perturbations are related to the classical(!) Mukhanov-Sasaki variable $\nu$ by
\begin{equation}\label{ClassDens}
 \delta\rho \approx \partial_{\phi}V(\phi)|_{\phi_0} \left( a^{-1}\nu \right),
\end{equation}
where $\phi_0$ refers to the homogeneous background inflaton field. We are more interested in the contrast of density perturbations in momentum space
\begin{equation}\label{ClassicalDensity}
 \left(\frac{\delta\rho}{\rho}\right)_{\vec{k}} \approx \frac{\partial_{\phi}V(\phi)|_{\phi_0}}{V(\phi_0)} \left( a^{-1}\nu_{\vec{k}}\right).
\end{equation}
We see that classical density perturbations are seeded by the classical $\nu_{\vec{k}}$. One now takes the classical $\nu_{\vec{k}}$ as `mimicking' \cite{Padma} the quantum mechanical $\hat{\nu}_{\vec{k}}$. More precisely, one takes the $\nu_{\vec{k}}$ to be classical `random' fields, distributed according to a Gaussian distribution of width given by eq. \eqref{width}. That means that we fill in eq. \eqref{PhysWidth} as an approximation to $a^{-1} \nu_{\vec{k}}$. We get
\begin{align}\label{QDensity}
 \left(\frac{\delta\rho}{\rho}\right)_{\vec{k}} &= \frac{\partial_{\phi}V(\phi)|_{\phi_0}}{V(\phi_0)} a^{-1} \Delta \nu_{\vec{k}}  =  \frac{\partial_{\phi}V(\phi)|_{\phi_0}}{V(\phi_0)}  \frac{|f_k|}{a}.
\end{align}
Taking as example de Sitter inflation, this becomes in the late time limit $|k \eta| << 1$
\begin{eqnarray}
  \left(\frac{\delta\rho}{\rho}\right)_{\vec{k}} &\approx& \frac{\partial_{\phi}V(\phi)|_{\phi_0}}{V(\phi_0)}\frac{H}{\sqrt{2k^3}}.
\end{eqnarray}
We regain a spectrum for $\frac{\delta \rho}{\rho}$ varying as $k^{-3}$. Equivalently we get a (nearly) scale invariant power spectrum $\mathbb{P}_{\rho}$. As said before, power law inflation for different $p$ also leads to a scale invariant power spectrum. Now we have shown that the density perturbations have a scale invariant spectrum, we can let the Wolf-Sachs effect and the acoustic theory of Harrison and Z'eldovich do their job: the result is the power spectrum we see in the cosmic microwave background!
\section{The Problem}\label{PROBLEM}
At first sight, the theory sketched above seems solid. Given that it succeeds in accounting for the observed inhomogeneities in the CMB, this is a veritable triumph for physics! However, this thesis wouldn't be written if there wasn't something to complain about. The replacement of $\nu_{\vec{k}}$ by $\Delta \hat{\nu}_{\vec{k}}$ in eq. \eqref{QDensity} is poorly understood. The essence is that we are replacing quantum mechanical quantities ($\Delta \hat{\nu}_{\vec{k}}$) by classical quantities ($\nu_{\vec{k}}$). Several people in the literature recognize this problem. For instance\footnote{In addition, we remind the reader of the various quotes we presented in the introduction of this thesis.}, Sudarsky writes in \cite{Sudarsky:2009za},
\begin{quote}
 `\textit{..., we must carefully inquire
into the justifications for the identification of these quantum uncertainties and the perturbation spectrum for the seeds of structure that characterize our Universe.}'
\end{quote}
Or even more suggestively, in \cite{Perez:2005gh}:
\begin{quote}
 `\textit{ The observed similarity between the
spectral properties of the (classical) density fluctuations in the early universe and the quantum mechanical fluctuations of the inflaton field after inflation
indeed suggest much more than a coincidence. However, we want to draw attention to the fact that a detailed understanding of the process that leads
from the quantum mechanical to the classical fluctuations is lacking. What is needed to justify the connection between the two spectra is a mechanism
that transforms quantum mechanical uncertainties into classical density fluctuations.}'
\end{quote}
\par There are in fact a lot of theories that couple quantum mechanical and classical quantities, let's say $A$ and $\hat{A}$. In such cases, computations are made in the quantum mechanical regime, using $\hat{A}$. In the end, the quantity $A$ is then computed by identifying it with the expected value of $\hat{A}$. In this way, one does not get a fundamental theory. In many cases however, this does produce an effective theory that is quite good at explaining the experimental results. For example, in solid state physics, the equation of motion for electrons in a crystal are obtained by equating their classical momentum $\vec{P}$ by the expected value of the quantum mechanical operator $\hat{P}$. The result is then interpreted as being Newtons second law for a so-called \emph{effective mass} of the electron.
\par However, there is an important difference for cosmological perturbations. In going from eq. \eqref{ClassicalDensity} to eq. \eqref{QDensity}, we didn't replace $\nu_{\vec{k}}$ by the expected value of $\hat{\nu}_{\vec{k}}$, but by its standard deviation! Notice that the expected value of $\nu_{\vec{k}}$ in the Bunch-Davies vacuum is in fact zero, so that replacing $\nu_{\vec{k}}$ by its expected value would predict no inhomogeneities at all. The claim that there is an analogue between cosmological perturbations and other semi-classical theories, like the effective mass description in solid state physics, is thus unfounded.
\par We are touching an important aspect of the problem, when we note that the expected value of the operator $\nu_{\vec{k}}$ is zero. This is rooted in the fact that our vacuum state is completely symmetric. On the other hand, it is clear that the result of eq. \eqref{QDensity} is an asymmetric state for the density perturbations. Somewhere along the line, some mechanism broke the symmetry present in the vacuum state with as result the particular realisation of density perturbations we can infer from the CMB. It is clear that standard unitary evolution following the Schr\"odinger equation cannot result in any loss of symmetry. There is only one mechanism present that can account for such a symmetry breaking, namely a collapse of the wavefunction. In standard quantum mechanics this type of symmetry breaking is typically induced by a \emph{measurement}.
\par If we call upon a measurement as symmetry breaking mechanism we have to ask the obvious question: who is the observer? It can't be the WMAP probe, or for that matter, any human or device created by humanity, for the simple reason that we ourselves are the product of the concrete realisation of perturbations that we observe. These perturbations are presumed to source the formation of superclusters, clusters, galaxies, stars and ultimately ourselves. Thus the CMB had to be inhomogeneous long before we observed the inhomogeneities in 1992 with the COBE probe. We are clearly not the relevant observers. Moreover, it is not clear what constituted such a measurement, let alone what the measured observable was. %
\par In the next part of this thesis we will look at some standard ideas that claim to solve these problems. Before we move on, we'll list the questions that need to be asked.
\begin{enumerate}[I]
 \item How does one go from an initial homogeneous and isotropic state to a state with inhomogeneities?
 \item Why can we consider the density perturbations to behave essentially classical at the end of inflation?
\end{enumerate}

\part{The Quantum-to-Classical Transition: Usual Solutions}\label{Usual}
\chapter{The Classical Limit of Quantum Mechanics}\label{QtoC}
Recall question number two at the end of last chapter:
\begin{quote}
\textit{`Why can we expect the density perturbations to behave classically at the end of inflation?'}
\end{quote}
This chapter will be devoted to the more general case of this question:
\begin{quote}
 \textit{`Which quantum mechanical systems can we expect to behave classically?'}
\end{quote}
This is a very old question, dating back to a letter Lorentz wrote to Schr\"odinger \cite{LettersWave}. Lorentz wrote to his colleague on the subject of the quantum mechanical theory of the hydrogen atom:
\begin{quote}
  `\textit{If we decide to dissolve the electron completely, so to speak, and to replace this by a system of waves this has both an advantage and a disadvantage. [...] But then how would I understand the phenomena of photo-electricity and the emission of electrons from heated metals? [...] But if we take a wave packet as model for an electron, then by doing so we block the way of restoring matters. For it is indeed asking a lot that a wave packet should condense itself again once it has lost its shape.}'
\end{quote}
In fact, this is still the core of the problem of the classical limit of quantum mechanics: how can we reconcile the spreading of wave packets, inherent in the Schr\"odinger equation, with the image of the world we see around us, full of seemingly well located particles with well defined momenta?
\par Since it is not very clear what \emph{classically} in our question means, we will be focusing on different characterisations of this concept within the framework of standard quantum mechanics. We will treat briefly the role of Heisenbergs uncertainty principle, the WKB approximation and the Wigner function for the toy models of a free particle and the inverted oscillator. For the latter, we will include a treatment of squeezing, a concept not applicable to the free particle.
\par We will look at the importance of the various constants in our model and how they relate to the classical limit. For this reason, we will be keeping all factors of $\hbar$ intact throughout this chapter. The treatment of decoherence will postponed untill chapter \ref{Decoherence}.
\clearpage
\section{Criteria for the Free Particle}\label{FreeParticle}
In this section, we will look at the most simple system in quantum mechanics: the free particle in one dimension. Starting from the classical Hamiltonian, the correspondence principle gives us the quantum Hamiltonian
\begin{equation}\label{FreeHamiltonian}
 \hat{H} = \frac{\hat{p}^2}{2m},
\end{equation}
where $m$ is the mass of our particle and $\hat{p}$ is the momentum conjugate to $\hat{x}$, satisfying the canonical commutation relation
\begin{equation}
 [\hat{x}, \hat{p}] = i \hbar.
\end{equation}
The Schr\"odinger equation reads
\begin{equation}\label{Schrodinger}
i \hbar \frac{\partial}{\partial t}\Psi(x,t) = \hat{H}\Psi(x, t).
\end{equation}Energy eigenfunctions are then given by de Broglie's plane waves, and any solution to Schr\"odingers equation can be thought of as a packet of plane waves. We will consider a Gaussian wavepacket as initial state. It is given by
\begin{equation}\label{wavefunction}
 \phi(p) = (\pi \sigma^2 \hbar^2)^{-1/4}\exp \left(-\frac{(p-p_0)^2}{2 \sigma^2 \hbar^2} - i \frac{p^2}{2m\hbar}t\right).
\end{equation}
\paragraph{Uncertainty}
The most striking aspect of quantum mechanics is perhaps the concept of uncertainty. Heisenberg's uncertainty relations indicate that in standard quantum mechanics there are no such things as particles with welldefined trajectories. Or in other words, if we adhere to the Copenhagen Interpretation of quantum mechanics we cannot think of particles as having a definite position and momentum. In contrast, classical mechanics treats pointlike particles (or pointlike centers of mass) with welldefined momenta.
\par Thus, to regain classical mechanics, one would like the uncertainties in position and momentum to be at least `small'. We of course need something to compare these uncertainties to, and a logical choice is the scale of the experiment we have in question.
\par Working in the Heisenberg picture and using Heisenbergs equation of motion for the operators $\hat{x}$ and $\hat{p}$, one can easily find $\Delta \hat{x}(t)$ and $\Delta \hat{p}(t)$
\begin{align}\label{TE1}
 \Delta \hat{x}(t)^2 &= \Delta \hat{p}(0)^2 \frac{t^2}{m^2} + \left[ \langle \hat{x}(0) \hat{p}(0)\rangle + \langle\hat{p}(0)  \hat{x}(0)\rangle\right]\frac{t}{m} + \Delta \hat{x}(0)^2,\\
\Delta \hat{p}(t)^2 &= \Delta \hat{p}(0)^2.
\end{align}
If we imagine our system to be initially in the state given by eq.\ \eqref{wavefunction}, the initial uncertainties are so that the Heisenberg uncertainty principle is satisfied as an equality
\begin{align}
 \Delta \hat{x}(0)^2 &= \frac{1}{ 2 \sigma^2}, & \Delta \hat{p}(0)^2 = \frac{\hbar^2 \sigma^2}{2}.
\end{align}
The terms linear in time in eq.\ \eqref{TE1} are zero for such a Gaussian packet, and the uncertainty in $\hat{x}$ rises as $t^2$. We see that any initial uncertainty gets amplified as time goes on. This phenomenon is the result of the dispersive aspect of the Schr\"odinger equation.

Let us be more precise, we can say that our wavepacket is not any more of minimum uncertainty once
\begin{align}
\Delta \hat{p}^2(0)\frac{t^2}{m^2}& \approx \Delta \hat{x}^2(0), &\rightarrow&& \frac{\sigma^2 \hbar^2 t^2}{2 m^2} &\approx \frac{1}{2\sigma^2},
\end{align}
or thus when $\frac{\hbar t}{m}$ is comparable to $\sigma^{-2}$. If we want to retain minimum uncertainty as long as possible, we had better have that $\frac{m}{\hbar \sigma}$ be big for a given $\sigma$. This in fact corresponds to our intuition: pretty `heavy' objects remain rather localised in position and momentum space.
\par One last remark is in place: however heavy our particle is, there will always be a time $t$ so that the packet is no longer of minimum uncertainty. Thus, every free particle will eventually be badly localised. However, any realistic physical system that may be described by a free particle on short time scales is far from `free' if we look at longer and longer timescales. As we will see in chapter \ref{Decoherence}, interactions tend to damp the dispersive character of wavefunctions.
\paragraph{WKB approximation}\label{WKBSectie}
Another criterion commonly used to characterise the classical limit is the Wentzel-Kramers-Brillouin (WKB) approximation. One can derive it by rewriting the Schr\"odinger equation. We will first look at the most general case, meaning a particle moving in a potential $V(x)$. This treatment is based on \cite{HollandBook}.
\par Writing $\Psi(x,t)$ as $R(x,t)\exp(iS(x,t))$, for $R(t)$ real and positive and $S(t)$ real, we find that the Schr\"odinger equation for a free particle decouples into
\begin{align}\label{HamiltonJacobi}
\frac{\partial S}{\partial t} &=  - \frac{1}{2m} \left(\frac{\partial S}{\partial x}\right)^2 - V(x) +  \frac{\hbar^2}{2m R}\frac{\partial^2 R}{\partial x^2},\\
\frac{\partial R^2}{\partial t} &= - \frac{\partial}{\partial x} \left( \frac{R^2}{m} \frac{\partial S}{\partial x}\right).
\end{align}
Since $R^2(x,t) = |\Psi(x,t)|^2$, we recognize in the second equation the continuity equation for $\Psi(x,t)$, expressing the conservation of probability. Here, we will not delve into it further. The first equation can be recognized as the Hamilton-Jacobi equation for a particle moving in a potential $V$ and an extra potential given by
\begin{equation}
V_q =- \frac{\hbar^2}{2m R}\frac{\partial^2 R}{\partial x^2}
\end{equation}
 This quantity is often called the \emph{quantum potential} for this reason. If, for some reason, this quantum potential would be small in comparison with the typical energy, then we would regain something that looks like the classical Hamilton-Jacobi equation.
\par One typically continues by assuming a form $\Psi(x,t) = \exp\left(i\frac{\phi(x,t)}{\hbar}\right)$, and assuming that $\phi(x,t)$ can be expanded in powers of $\hbar$.
\begin{equation}\label{PhiExpansie}
\phi(x,t)= \phi_0(x,t) + \hbar \phi_1(x,t) + \dots
\end{equation}
We then look for solutions to the stationary Schr\"odinger equation, for some energy eigenvalue $E$. Since we are looking for the classical limit, we only consider the case $E> V(x)$, since the opposite has no classical analog.
\begin{equation}
\frac{\partial^2}{\partial x^2}\Psi =  -\frac{2m}{\hbar^2}E \Psi
\end{equation}
Filling in eq.\ \eqref{PhiExpansie}, and collecting powers of $\hbar$, we get for the first and second order \cite{HollandBook}
\begin{align}
\phi_0 &= \pm \int dx \sqrt{2m(E- V(x))},\\
\phi_1 &= \frac{i}{4}\log(2m(E-V)).
\end{align}
The third order term is negligible in comparison with the first and second order if
\begin{equation}
\frac{m \hbar \frac{\partial V}{\partial x}}{[2m(E-V)]^{3/2}} \ll 1.
\end{equation}
Interpreting the quantity in the denominator as the classical momentum to the third power, we find that this is in fact a condition on the de Broglie wavelength $\lambda$
\begin{equation}\label{WKBcondition}
 \left|\frac{d}{dx} \lambda \right| \ll 1,
\end{equation}
where $\lambda$ is given by
\begin{equation}\label{dBLambda}
 \lambda = \frac{1}{\sqrt{2m \left(E - V(x) \right) }}.
\end{equation}
Now we note that, under the condition of a suitably smooth $\lambda$, the resulting wavefunctions become
\begin{equation}\label{WKBWave}
 \Psi(x,t)_{\pm} = A_{\pm} \left(2m(E-V)\right)^{-1/4} \exp\left[\pm \frac{i}{\hbar}\int dx \sqrt{2m(E- V(x))} \right].
\end{equation}
Now we proceed to calculate the quantum potential for these wavefunctions
\begin{equation}
 V^{\pm}_q = \frac{-\hbar^2 }{2m} \left(E-V\right)^{1/4} \frac{\partial^2 }{\partial x^2}\left(E-V\right)^{-1/4}.
\end{equation}
The WKB condition, eq.\ \eqref{WKBcondition}, and the conditions on the higher derivatives then gives us that the quantum potential is negligible in comparison with the kinetic energy.
\begin{equation}
 V_q^{\pm} \ll E - V
\end{equation}
Thus, for this type of wavefunction, the quantum Hamilton-Jacobi equation, eq.\ \eqref{HamiltonJacobi} reduces to classical form, and the criterion for WKB classicality is reached.
\par One has to note however, that this notion of classicality is only restricted to states of the form  stated in eq.\ \eqref{WKBWave}. It does not in general apply to linear combinations of the WKB wavefunctions. The example given in \cite{HollandBook} is very clear. Say we take $\Psi = \frac{1}{N}\left( \frac{1}{A_+}\Psi_+ + \frac{1}{A_-}\Psi_-\right)$, with $N$ some normalisation coefficient. We can write the amplitude of the total wavefunction then as follows
\begin{equation}
 |\Psi(x,t)| = \frac{2}{N} \left(E-V\right)^{-1/4}\left| \cos\left(\int \frac{dx}{\hbar} \sqrt{2m(E- V(x))}\right)\right|.
\end{equation}
Since we assumed $E-V$ to be large, this is clearly a rapidly oscillating function. The quantum potential can be calculated to be $V_q = E-V$ and hence is very large by assumption. Thus, WKB classicality is only valid for the WKB energy eigenstates and certainly not general superpositions.
\par
For the free particle, only the first and second order in eq.\ \eqref{PhiExpansie} contribute. This is no surprise, since the de Broglie wavelength is obviously conserved. The WKB wavefunctions are the exact eigenfunctions $\exp\left( \pm i\frac{p}{\hbar}x\right)$. These are thus classical in the WKB sense.
\paragraph{WKB Classicality?}
One could ask what this `WKB classicality' actually means. We certainly have that eq.\ \eqref{HamiltonJacobi} formally resembles the classical Hamilton-Jacobi equation. What we certainly don't have in this situation is the concept of definite positions and momenta. Note that a measurement of $\hat{x}$ or $\hat{p}$ will in general be undetermined, meaning that the wavefunction is certainly not peaked. In fact, the WKB condition dictates that $E-V$ does not vary too much, thus making $|\Psi_{\pm}|^2$ an approximate uniform distribution. We conclude that the classical limit is not truly reached.
\paragraph{Wigner Function}
 To a given wavefunction $\Psi(x,t)$ in one dimension one associates a Wigner function using the relation
\begin{equation}\label{Wigner}
 P(x, p, t) = \frac{1}{\pi  \hbar} \int dy \Psi^{*}(x+y,t)\Psi(x-y, t) \exp \left(2ipy \right)
\end{equation}
Given a wavefunction on momentum space $\phi(p, t)$, an equivalent definition of the Wigner function is
 \begin{equation}\label{WignerMomentum}
 P(x, p, t) = \frac{1}{\pi  \hbar} \int dq \phi^{*}(p + q,t)\phi(p-q, t) \exp \left(-2 iqx \right).
\end{equation}
If $\Psi$ is normalised then $P(x,p,t)$ is also.
\par
The most important property of this function is the following. For any operator $F(\hat{x}, \hat{p})$ the following equality holds
\begin{align}
 \langle F(\hat{x}, \hat{p}) \rangle &= \int dxdp F(x, p) P(x,p),\\
F(x, p) &= \int dy  \Psi^*\left(x - \frac{y}{2}\right)F(\hat{x},\hat{p}) \Psi\left(x + \frac{y}{2}\right)\exp\left(ipy\right)
\end{align}
It is this property that elicits a simple interpretation of the Wigner function: one can regard it as a probability on classical phase space. However, this function is not necessarily positive everywhere; for this reason it is often called the Wigner quasi-probability distribution.
For the free particle wavefunction in eq.\ \eqref{wavefunction} one can find the explicit form of the Wigner function by solving some Gaussian integrals. It is given by
\begin{equation}
P(x,p,t) = \frac{1}{\pi \hbar} \exp\left(- \frac{(p-p_0)}{\sigma^2 \hbar^2} - \sigma^2 (pt-mx)^2\right).
\end{equation}
We see that this function is never negative, so that we can safely interpret this $P(x,p)$ as a probability distribution. In addition, we see that the distribution is peaked along the classical trajectories $x = \frac{p}{m}t$. In this way, one could say that the classical limit is reached. Note that the spreading of this classical distribution can never be zero for both $x$ and $p$! This is due to the Heisenberg uncertainty principle.
\paragraph{Wigner Classicality?} The same argument again applies: the concept of the Wigner function being peaked on classical trajectories does not imply anything for the results of measurements of $\hat{x}$ and $\hat{p}$. The wavefunction is still unaltered and the uncertainties in $\hat{x}$ and $\hat{p}$
still obey the same time evolution. Even more crudely: there is no reason \emph{a priori} to interpret the arguments $(x, p)$ of the Wigner function as classical position and momentum within the framework of quantum theory.
\section{The Inverted Oscillator}\label{InvertedOscillator}
We will now look at what the above notions have to say about the classical limit for an inverted harmonic oscillator. The Hamiltonian we wish to consider is, for real and constant $\omega$
\begin{equation}
\hat{H} = \frac{\hat{p}^2}{2m} - m\omega^2 \frac{\hat{x}^2}{2}.
\end{equation}
The classical equation of motion is easily solved:
\begin{equation}
f = Ae^{\omega t} + Be^{- \omega t},
\end{equation}
for $A$ and $B$ to be determined by the initial conditions.
\par The quantum system is a lot harder to solve. The normalisable\footnote{There are also wavefunctions for this system that are not normalisable, see eq.\ \eqref{Nonnormalisable}. The conclusions in this section will to large extent also be applicable to them, but for clarity we will only present the treatment for the normalisable solutions.} solutions are indexed by the nonnegative integers $n=0,1,2,\dots$ and are given by
\begin{equation}
  \Psi_n (x,t)  =   \sqrt{\frac{1}{  2^n n!}}\left(\frac{m}{\hbar |f|^2}\right)^{1/4} H_{n}\left(  \frac{x}{\sqrt{2}|f|} \right) \left(\frac{f}{f^{*}}\right)^{n + \frac{1}{2} }\exp \left( i\frac{f'^{*}}{f^{*}} \frac{m x^2}{2\hbar } \right),
\end{equation}
where $f$ is a solution to the classical equations of motion that satisfies a Wronskian constraint\footnote{This Wronskian constraint makes sure that Im($\frac{f}{f^{*}}$) is never zero and that the solutions are normalisable.}. For the derivation of these states, see appendix \ref{WFAppendix}.
\par For shorter notation, we will denote $\text{Re} (f) = f_{r}$ and $\text{Im}(f) = f_{i}$ and similar for other quantities.
We will take the initial conditions for $f$ to be\footnote{These initial conditions are consistent with the Wronskian condition \ref{WronskianApp}. In addition, these conditions correspond to taking the initial relations between creation and annihilation operators in the standard way.}
\begin{align}
 f(0) &= \frac{1}{\sqrt{2 \omega}}, & f'(0) = -i\sqrt{ \frac{\omega}{2}}.
\end{align}
These initial conditions determine $f$
\begin{eqnarray}\label{ConcreteF}
 f(t) = \frac{1}{\sqrt{2 \omega}}\left(\cosh \omega t - i \sinh \omega t\right).
\end{eqnarray}
\paragraph{Uncertainty}\label{InvUnc}
Using the time independent annihilation and creation operators defined in eq.\ \eqref{GeneralAnnCrea}, we can find the following expressions for $\hat{x}(t)$ and $\hat{p}(t)$
\begin{align}
\hat{x}(t)&= \sqrt{2}\left(\sqrt{\omega} f_r \hat{x}(0) - \frac{1}{m \sqrt{\omega}} f_i \hat{p}(0) \right),\\
\hat{p}(t) &=  \sqrt{2} \left( m \sqrt{\omega} f'_r \hat{x}(0) +  \sqrt{\frac{1}{\omega}}f'_i\hat{p}(0)\right).
\end{align}
This allows us to compute the time evolution of the uncertainties
\begin{align}
\begin{split}\label{Uncertainties1}
\Delta \hat{x}^2(t) &= \cosh^2(\omega t)  \Delta \hat{x}^2 (0) + \frac{1}{m^2\omega^2}\sinh^2(\omega t) \Delta \hat{p}^2(0)\\
&\quad - \frac{2}{m\omega} \cosh(\omega t) \sinh(\omega t) \left\langle \hat{x}(0)\hat{p}(0) +  \hat{p}(0)\hat{x}(0) \right\rangle \\
&\quad +  \frac{4}{m\omega} \cosh(\omega t) \sinh(\omega t)\left\langle \hat{x}(0) \right\rangle \left\langle \hat{p}(0)\right\rangle,
\end{split}\\
\begin{split}\label{Uncertainties2}
\Delta \hat{p}^2(t) &= m^2 \omega^2 \sinh^2(\omega t) \Delta \hat{x}^2 (0) + \cosh^2(\omega t)  \Delta \hat{p}^2(0) \\
&\quad- 2 m \omega \cosh(\omega t) \sinh(\omega t)\left\langle \hat{x}(0)\hat{p}(0) +  \hat{p}(0)\hat{x}(0) \right\rangle\\
&\quad + 4 m\omega \cosh(\omega t) \sinh(\omega t) \left\langle \hat{x}(0) \right\rangle \left\langle \hat{p}(0)\right\rangle.
\end{split}
\end{align}
It is very clear from these expressions that, for a generic state, the uncertainties in both variables will rise exponentially! There is no stopping the spreading of uncertainties, not even for $m$ really big. Only the limit of $\omega$ very small will slow down the spreading of the wavefunction, but we are obviously treating a free particle in that limit. The conclusion is that, for a generic state, there is no hope of maintaining small uncertainties. One can expect the classical limit to be less evident in this case.
\paragraph{Squeezing}
One can look at this problem using different variables. We define, as in \cite{Albrecht:1992kf},
\begin{align}
\hat{b}_{+} &= \frac{1}{\sqrt{2\omega}}\left(\hat{p} + m \omega \hat{x}\right), & \hat{b}_{-} = \frac{1}{\sqrt{2\omega}}\left(\frac{ \hat{p}}{m} - \omega \hat{x}\right).
\end{align}
Note that $[\hat{b}_{-}, \hat{b}_{+}] = i \hbar$, and we can regard these variables as canonical ones. For these variables, the time evolution is given by
\begin{align}
\hat{b}_{+}(t) &=   e^{-\omega t} \hat{b}_{+}(0) & \hat{b}_{-}(t)= e^{\omega t} \hat{b}_{-}(0).
\end{align}
And we can easily compute the time evolution of the uncertainties in $\hat{b}_{+}$ and $\hat{b}_{-}$
\begin{align}\label{UncertaintiesB}
\Delta \hat{b}_{+}^2 &=  e^{-2\omega t}\Delta \hat{b}_{+}^2(0), & \Delta \hat{b}_{-}^2 =  e^{2\omega t}\Delta \hat{b}_{-}^2(0).
\end{align}
The situation is thus drastically different. While the uncertainty in one variable rises exponentially, the uncertainty for the other variable vanishes.
If we start from an initial minimum uncertainty state, Heisenbergs uncertainty relations are obeyed as an equality for all time.
\par
\begin{figure}\label{SqueezingAlbrecht}
\begin{center}
\includegraphics[scale=0.75]{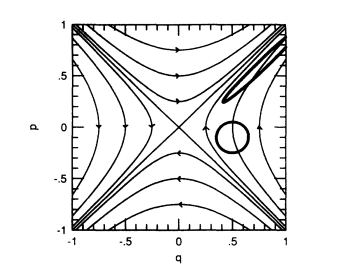}
\caption{Phase space and classical trajectories for an upside down harmonic oscillator for $\omega = m = 1$. The circular region in phase space evolves in time to produce the squeezed ellipse. Figure courtesy of Albrecht, Joyce, Ferreira and Prokopec\cite{Albrecht:1992kf}}.
\end{center}
\end{figure}

 This phenomenon is called \emph{squeezing} of the state. It get its name from the following classical picture. One can imagine an ellipse in phase space, corresponding to some spread in $x$ and $p$. The dimensions of the ellipse in both the $x$ and $p$ direction will typically grow. However, time evolution will preserve the area of the ellipse, through Liouville's theorem.  Thus, if we take $b_{+}$ and $b_{-}$ as classical quantities in the direction of the minor and major axis of the ellipse, we get that the product of the spread in $b_{+}$ and $b_{-}$ is equal to the area of the ellipse, and thus constant in time. This process is pictured in figure \ref{SqueezingAlbrecht}.
Quantum mechanically, the squeezing process results in one variable with negligible uncertainty. Thus, for large time $t$, a measurement of the observable $\hat{b}_{+}$ will result in the measured value $\langle \hat{b}_{+}\rangle$ with a probability very close to $1$. Within standard quantum theory, it is thus reasonable to think of $\hat{b}_{+}$ as having achieved the classical limit. On has to note that, for very large $t$, the conjugate variable $\hat{b}_{-}$ will have a very large uncertainty. In a way, $\hat{b}_{-}$ becomes very `quantum': a measurement of $\hat{b}_{-}$ is utterly unpredictable.
\par On a final note, we repeat that the uncertainties in $\hat{x}$ and $\hat{p}$ also rise exponentially. While we can speak about classical behavior for $\hat{b}_{+}$, we certainly cannot for the variables $\hat{x}$ and $\hat{p}$. Following this reasoning, it is important to clearly specify what the relevant quantity is.
\paragraph{WKB approximation}
We will look for WKB solutions for the inverted oscillator. More exactly we are looking for approximate energy eigenstates of energy $E > 0$.
We look at the de Broglie wavelength of this problem
\begin{equation}
 \lambda_{dB} = \frac{\hbar}{ \sqrt{2m\left[E + \frac{m\omega^2 }{2}x^2\right]}}.
\end{equation}
We know that the WKB solutions will satisfy a classical Hamilton Jacobi equation if the variation of $\lambda$ is very smooth. For example, the first derivative of $\lambda$ is
\begin{equation}
\left|\frac{d\lambda_{dB}}{dx}\right| = \frac{\hbar }{2}\frac{\sqrt{m} \omega^2 |x|}{\left(E + \frac{m\omega^2 x^2}{2}\right)^{3/2}}.
\end{equation}
This derivative is certainly small for all $x$ if $E > \frac{\hbar \omega}{2}$. Thus for energies higher than the ground state energy, we can find approximate solutions to the Schr\"odinger equation.
\par The WKB approximation leads to the following wavefunctions
\begin{equation}
\begin{split}
 \Psi^{\pm}_{WKB} &= \frac{1}{N} \left[2m \left(E + m\omega^2 x^2\right)\right]^{-1/4}\exp\left[\pm i \frac{\sqrt{m}}{\sqrt{2}\hbar} x\sqrt{E + \frac{m \omega^2}{2}x^2}\right]\\
 & \exp\left[ \pm i \frac{E}{\hbar\omega} \ln \left(\frac{\sqrt{m}\omega}{\sqrt{2}}\sqrt{E + \frac{m\omega^2}{2}x^2} + \frac{m \omega^2}{2}x\right)\right].
\end{split}
\end{equation}
Note that these functions are not normalisable. Thus, to represent normalisable physical states, we will need to make linear superpositions. However, as noted above, linear superpositions of WKB states are no longer WKB states. Classicality in the WKB sense is thus only achievable for the non-normalisable WKB energy eigenstates. We repeat that WKB wavefunctions do not truly achieve the classical limit.
\paragraph{Wigner Function}
The Wigner function for a $n$-particle state can be calculated from eq.\ \eqref{Wigner}. It is given by
\begin{equation}\label{WignerInverted}
\begin{split}
P_{n}(x,p,t) &= \frac{1}{2^n n!}\left(\frac{m}{\hbar |f|^2}\right)^{1/2} \frac{1}{\pi \hbar}\int dy H_n\left( \frac{x +y}{\sqrt{2}|f|}\right) H_n\left( \frac{x-y}{\sqrt{2}|f|}\right) \\
& \exp\left[ - \frac{m}{2\hbar|f|^2} x^2 - \frac{m}{2\hbar|f|^2}\left(y - \frac{i \hbar |f|^2}{2m}\left(p - \frac{|f|'}{|f|}mx\right)\right)^2 - \frac{\hbar |f|^2 }{4m}\left(p - \frac{|f|'}{|f|}mx\right)^2\right].
\end{split}
\end{equation}
For $n=0$ this reduces to
\begin{equation}
P_0(x,p,t) = \frac{1}{\hbar \sqrt{\pi}} \exp \left[ - \frac{m}{2 \hbar|f|^2} x^2  - \frac{\hbar|f|^2}{4m} \left(p - \omega mx \right)^2\right].
\end{equation}
Thus, for $n=0$, our semi-probability distribution is positive everywhere. We are thus free to interpret $P(x,p,t)$ as a real probability function. In addition, the Wigner function is fairly located around the classical trajectories $p = \omega m x$. In fact, as $t \rightarrow \infty$ and thus $|f| \rightarrow \infty$, the probability distribution becomes infinitely well located.
\begin{equation}
\lim_{t \to \infty} P_{0}(x,p,t) \sim \delta\left(p - \omega mx\right)
\end{equation}
However, for $n \not = 0$ the Wigner function is in general not positive definite anymore due to the oscillatory nature of the Hermite polynomials. For example, for $n = 1$, we get
\begin{equation}
\begin{split}
 P_1(x,p,t) &= \frac{(2 \pi)^{-1/2}}{\hbar m^2 |f|^2}  \left( |f|^4 \hbar^2  \left(p - \omega mx \right)^2 - 4 |f|^2\hbar m +4m^2x^2 \right)\\
& \exp \left[ -\frac{m}{2|f|^2 \hbar}x^2 - \frac{ \hbar |f|^2 }{4 m}\left(p - \omega mx \right)^2 \right]
\end{split}
\end{equation}
which is not positive for every $x$. Because of this reason, we cannot interpret the Wigner function as a probability density. We can also look at the Wigner function in terms of the classical squeezed coordinates $b_{+}$ and $b_{-}$. For $n = 0$ it becomes
\begin{equation}
P_0(b_{+},b_{-},t) = \frac{1}{ \pi \hbar} \exp \left[ - \frac{m}{2 \omega \hbar|f|^2} \left( \frac{b_{+}}{m} - b_-\right)^2   - \frac{m\hbar|f|^2}{8\omega} b_-^2  \right].
\end{equation}
Thus, for $t \rightarrow \infty$, we will retain a $\delta$ function for $b_-$, but the probability distribution will be uniform in $b_+$.  Thus, the Wigner kind of classicality is only achieved for $\hat{b}_-$ and certainly not for $\hat{b}_+$. This reflects the behavior of the quantum uncertainties, eq.\ \ref{UncertaintiesB}.
\par In conclusion, the Wigner function type of classicality will only be truly reached for the ground state, while for states with higher particle number (and in general, superpositions of different particle number states) the Wigner function is not positive definite, even though one could argue that regions where $P(x,p) < 0$ are `small'. We repeat that the Wigner classicality is not true classicality, since the properties of the Wigner function do not translate to any statement of measurements of $\hat{x}$ and $\hat{p}$.
\section{The Role of Plancks Constant, Wavelength and Mass}\label{Planck}
Let us conclude by clarifying the role of various relevant constants. We will first look at the reduced Plancks constant, $\hbar$. It is common habit for basic quantum mechanics textbooks to handwavingly invoke the classical limit as being the limit of $\hbar \rightarrow 0$. A typical example could include the commutator of two conjugate variables. By letting $\hbar \rightarrow 0$, one argues that this commutator will vanish and that the operators will behave `classically', meaning commutatively. Another consequence of this limit would be that lower limit for the Heisenberg uncertainty principle would vanish. \par Clearly, these arguments do not make sense. First of all, $\hbar$ is not a dimensionless quantity. While in SI units it can be thought of as small, this is certainly not the case other units. Secondly, it is not clear what the notion of `wavefunction' becomes in such a limit since the Schr\"odinger equation itself does not make sense any more. Thirdly, what exactly is the meaning of varying one of
natures constants?
\par A more correct formulation of this argument would involve comparing $\hbar$ to the action relevant for the problem under consideration\cite{SevenSteps}. For a particle moving in a potential $V$, the relevant length scale would be the length of variation $L$ of the potential. For such a particle the relevant dimensionless constant $C$ would be
\begin{equation}
 C = \frac{m L^2}{ \hbar t_0},
\end{equation}
where $t_0$ is the typical time scale on which we consider our system.
\par For the free particle, $L$ is very large, and thus we can regard $\hbar$ as being `small' for time scales that are not very large. This is reflected in the time behavior of the uncertainties calculated in section \ref{FreeParticle} and the fact that the Wigner function is positive definite and peaks on classical trajectories.
\par For the inverted oscillator, the scale of variation for our potential $L$ is not very clear. \cite{AlloriPHD} argues that any type of harmonic oscillator is slowly varying, meaning that we can think of $C$ as being very big.
\par These considerations are of course very much related to the WKB approximation. In the expansion of the wavefunction as $\exp \left[ \frac{i S}{\hbar}\right]$, we immediately recognize that $S$ has the dimensions of an action. Assuming that the expansion in eq.\ \eqref{PhiExpansie} is accurate to first order amounts to demanding for the second order correction $\phi_2$ (dimensions of an inverse action) that $\phi_2^{-1}$ is large compared to $\hbar$.
\par The de Broglie wavelength also enters this discussion in a natural way. It can be regarded as the length scale of the quantum mechanical problem, while $L$ can be thought of as the classical length scale of our problem. It is then natural to compare $L$ and $\lambda$ and think of the classical limit as being achieved when $\lambda$ vanishes in comparison with $L$. This is, again, very much connected to the WKB approximation. In fact, eq.\ \eqref{WKBcondition}, can be rewritten as
\begin{equation}
\frac{m}{\hbar^2} \lambda^3 \partial_x V  \ll 1.
\end{equation}
If we assume that $\left(\partial_x V\right)^{-1}$ is related to the length scale of the potential, then the WKB condition reduces to comparing $\lambda$ with $L$.
\par The role of mass is suprisingly intuitive in this discussion. The classical actions $S$ of both the free particle and the inverted oscillator are proportional to $m$. In the context of this discussion, larger $m$ will increase the ratio $\frac{S}{\hbar}$. Our intuition that really massive behave classical can thus easily be reconciled with this picture.

\chapter{Squeezing of Cosmological Perturbations}\label{Squeezing}
In this chapter we will extend the theory of squeezing to the Mukhanov-Sasaki variable. The concept was already briefly introduced in chapter \ref{QtoC} for the inverted harmonic oscillator. We will give an exposition of the theory of squeezing based principally on \cite{Polarski:1995jg}, followed by a summary of the principal arguments presented in the literature concerning classicality. This chapter will finish by arguing why the theory of squeezing is not sufficient to explain classicality for the cosmological perturbations.
\section{Theory of Squeezing}
Let us fix a certain Fourier mode of the Muhanov-Sasaki variable of wavevector $\vec{k}$. Analogous to our calculation for the inverted oscillator in section \ref{InvUnc}, we can find the operators $\nu_{ \vec{k}}(\eta)$ and $\Pi_{ \vec{k}}(\eta)$ in terms of their initial conditions
\begin{align}\label{HeisenbergPicture}
 \nu_{\vec{k}}(\eta) &= \text{Re}\left[f_k \exp\left( -ik\eta_{ini}\right) \right] \sqrt{2k} \nu_{\vec{k}}(\eta_{ini}) + \text{Im}\left[f_k \exp\left( -ik\eta_{ini}\right) \right] \sqrt{\frac{2}{k}} \Pi_{\vec{k}}(\eta_{ini}),\\
\Pi_{\vec{k}}(\eta) &= \text{Re}\left[f'_k \exp\left( -ik\eta_{ini}\right) \right] \sqrt{\frac{2}{k}} \nu_{\vec{k}}(\eta_{ini}) - \text{Im}\left[f'_k \exp\left( -ik\eta_{ini}\right) \right] \sqrt{2k} \Pi_{\vec{k}}(\eta_{ini}).
\end{align}
Using these equalities, we can find the uncertainties in the variables $\nu_{ \vec{k}}(\eta)$ and $\Pi_{ \vec{k}}(\eta)$
\begin{align}\label{OwnUncert}
\begin{split}
\Delta \nu_{\vec{k}}^2 &= \text{Re}\left[f_k \exp\left( -ik\eta_{ini}\right) \right]^2 2k \left. \Delta \nu_{\vec{k}}^2\right|_{\eta_{ini}} + \text{Im}\left[f_k \exp\left( -ik\eta_{ini}\right) \right]^2 \frac{2}{k} \left.\Delta \Pi_{ \vec{k}}^2\right|_{\eta_{ini}}\\
 & \quad + 2 \text{Re}\left[f_k \exp\left( -ik\eta_{ini}\right) \right]\text{Im}\left[f_k \exp\left( -ik\eta_{ini}\right) \right] \left\langle\nu_{ \vec{k}}(\eta_{ini})\Pi_{ \vec{k}}(\eta_{ini}) + \Pi_{ \vec{k}}(\eta_{ini})\nu_{ \vec{k}}(\eta_{ini})\right\rangle,
\end{split}\\
\begin{split}
\Delta \Pi_{ \vec{k}}^2 &= \text{Re}\left[f'_k \exp\left( -ik\eta_{ini}\right) \right]^2 \frac{2}{k} \left. \Delta \nu_{ \vec{k}}^2\right|_{\eta_{ini}} + \text{Im}\left[f'_k \exp\left( -ik\eta_{ini}\right) \right]^2 2k \left.\Delta \Pi_{ \vec{k}}^2\right|_{\eta_{ini}},\\
 & \quad- 2 \text{Re}\left[f'_k \exp\left( -ik\eta_{ini}\right) \right]\text{Im}\left[f'_k \exp\left( -ik\eta_{ini}\right) \right] \left\langle\nu_{ \vec{k}}(\eta_{ini})\Pi_{ \vec{k}}(\eta_{ini}) + \Pi_{ \vec{k}}(\eta_{ini})\nu_{ \vec{k}}(\eta_{ini})\right\rangle.
\end{split}
\end{align}
where we have omitted terms involving $\langle \nu_{ \vec{k}}(\eta_{ini})\rangle$, thus specialising to symmetric or antisymmetric wavefunctions.
These uncertainties will spread for generic states of the Mukhanov-Sasaki variable, since $|f_k|$ and $|f_k'|$ blow up. In addition, there is no limit in which these uncertainties stay bounded. Also analogous to the inverted oscillator, we can do better.
\par We define functions $u_{\vec{k}}$ and $v_{\vec{k}}$ by
\begin{align}
 u_k &= \exp \left(ik\eta_{ini}\right)\sqrt{\frac{k}{2}}f_k + i \sqrt{\frac{1}{2k}}f_k', & v_k = \exp \left(ik\eta_{ini}\right)\sqrt{\frac{k}{2}}f_k^* + i\sqrt{\frac{1}{2k}}f_k'^*.
\end{align}
Or the other way around
\begin{align}
 f_k &= \frac{1}{\sqrt{2k}}\exp \left(-ik\eta_{ini}\right)\left(u_k + v^*_k\right) & if_k' = \exp \left(-ik\eta_{ini}\right)\sqrt{\frac{k}{2}}\left(u_k - v^*_k\right)
\end{align}
The Wronskian condition, eq.\ \eqref{Wronskian} and the initial conditions on $f_k$ then imply
\begin{align}
 |u_k|^2 - |v_k|^2 &= 1, & u_k(\eta_{ini}) &= 1, & v_k(\eta_{ini}) = 0.
\end{align}
This means that there exists some functions $r_k(\eta), \theta_k(\eta)$ and $\phi_k(\eta)$ so that we can write
\begin{align}
 u_k &= e^{-i\theta_k}\cosh(r_k), & v_k = e^{i \theta + 2i \phi_k}\sinh(r_k).
\end{align}
One usually calls $r_k$ the squeezing parameter, $\theta_k$ the phase angle and $\phi_k$ the squeezing angle. %
We can now define two new variables $\hat{b}_+$ and $\hat{b}_-$ by
\begin{align}\label{SqueezeTransform}
\hat{b}_- &= \cos(\phi_k) \hat{\nu}_{ \vec{k}} + \sin(\phi_k) \hat{\Pi}_{ \vec{k}}, & \hat{b}_+ &= \sin(\phi) \hat{\nu}_{ \vec{k}} - \cos(\phi) \hat{\Pi}_{ \vec{k}}.
\end{align}
Notice that they are canonically conjugate, since $[\hat{b}_{-}(\eta_1), \hat{b}_+ (\eta_2)] = i\delta(\eta_1 - \eta_2)$. Now one can check that
\begin{align}\label{SqueezingEqs}
 \Delta \hat{b}_-^2 &= \frac{1}{2} \left. \exp\left(-2 r_k\right) \Delta \hat{b}_-^2\right|_{\eta_{ini}}, & \Delta \hat{b}_+^2 &= \frac{1}{2} \left. \exp\left(2 r_k\right) \Delta \hat{b}_+^2\right|_{\eta_{ini}}.
\end{align} As for the inverted harmonic oscillator we recover two conjugate variables that satisfy the uncertainty principle as an equality when the wavefunction is initially a minimum uncertainty wavepacket. As $r_k$ increases, the wavefunction grows more narrow in $b_-$ while it grows broader in $b_+$. We retrieve the results of the harmonic oscillator of $m = \omega = 1$ when $\phi = -\frac{\pi}{4}$ and $r_k = t$.
\par Classically, this corresponds to a volume in phase-space that gets `squeezed' into an ellipse. The surface of the ellipse stays constant in time, but the ratio of the length of the major to length of the semi-major axis grows in time as $\exp(2r_k)$. The difference with the inverted oscillator is that the transformation in eq.\ \eqref{SqueezeTransform} is time-dependent. The phase-space ellipse thus rotates in time as the squeezing angle $\phi_k$ varies.
\section{Classicality from Squeezing}\label{SqueClas}
Usually, one starts the argument by remarking that the phase of $f_k$ becomes constant as $\eta \rightarrow 0$, as proven in appendix \ref{AppLemma}. In terms of the functions $u_k$ and $v_k$ this means that,
\begin{equation}
 \lim_{\eta \rightarrow 0} \theta_k + \phi_k = \text{constant}.
\end{equation}
A typical value for $r_k$ at the end of inflation for the modes we observe in the sky would be $r_k \sim 120$ \cite{Martin:2007bw}, \cite{ Kiefer:1998qe}. This would correspond to a very squeezed state, meaning that the uncertainty in $\hat{b}_+$ would be a factor $e^{240} \sim 10^{104}$ larger than the spread in $\hat{b}_-$. A measurement of $\hat{b}_-$ would result in $\langle \hat{b}_-\rangle$ with a probability extremely close to $1$ while a measurement of $\hat{b}_+$ would be utterly unpredictable.
\par One proceeds by noting that we can make $f_k\exp(-i \eta_{ini})$ real in the late times limit by a time-independent complex rotation \cite{Lesgourgues:1998gk}. The argument is continued by noting that the Heisenberg picture equations, eqs.\ \eqref{HeisenbergPicture}, reduce to
\begin{align}\label{Approximatie}
 \hat{\nu}_{\vec{k}}(\eta) &\approx \text{Re}\left[f_k \exp\left( -ik\eta_{ini}\right) \right] \sqrt{2k} \hat{\nu}_{ \vec{k}}(\eta_{ini}),\\
 \hat{\Pi}_{ \vec{k}}(\eta) &\approx \text{Re}\left[f'_k \exp\left( -ik\eta_{ini}\right) \right] \sqrt{\frac{2}{k}} \hat{\nu}_{ \vec{k}}(\eta_{ini}),
\end{align}
since $\text{Im}(f_k\exp{-\eta_{ini}})$ vanishes in the late times limit. Since the right hand sides of these equations do not mention $\Pi_{\vec{k}}(\eta_{ini})$, the right hand sides commute. One concludes then that
\begin{equation} \label{CommutationNonsense}
 \left[ \hat{\nu}_{\vec{k}}(\eta), \hat{\Pi}_{\vec{k}}\right ] \approx 0.
\end{equation}
This is suggestively called the `pragmatical view' in \cite{ Kiefer:2008ku}. Several authors conclude that the classical limit is reached for late times \cite{Albrecht:1992kf,Lesgourgues:1998gk,Lesgourgues:1996jc,Polarski:1995jg,Kiefer:1998pb,Kiefer:2006je,Kiefer:1999gt,Kiefer:1998jk,Kiefer:1998qe} among others.
\par As extra arguments, one typically first shows that for late times the Wigner distribution of the Mukhanov-Sasaki variable becomes peaked along classical trajectories by direct computation, see e.g. \cite{Kiefer:2008ku} for the vacuum and \cite{Lesgourgues:1998gk} for generalisation to some other states. Second, one argues that the modes become indistinguishable from a `classical stochastic field'. With this, it is meant that one has at late times
\begin{align}
 \nu_{cl, \vec{k}} &= \text{Re}\left[f_k \exp\left( -ik\eta_{ini}\right) \right] \sqrt{2k} E(\vec{k}),\\
 \Pi_{cl, \vec{k}} &= \text{Re}\left[f'_k \exp\left( -ik\eta_{ini}\right) \right] \sqrt{\frac{2}{k}} E(\vec{k}),
\end{align}
where the subscript 'cl' stands for classical and $E(k)$ \cite{Lesgourgues:1998gk}
\begin{quote}
 `\textit{[\dots] is a time-independent Gaussian stochastic function of k with zero average and unit dispersion [\dots]}'
\end{quote}
and thus (from the same paper)
\begin{quote}
 `\textit{for each realization of the stochastic field $\nu_{cl, \vec{k}}$, respectively $\Pi_{cl, \vec{k}}$is the classical amplitude, respectively momentum, for the corresponding initial conditions [\dots]}'
\end{quote}
Here we have changed the names of the variables in the citations to ensure compatibility with the rest of this thesis.

\section{Classicality?}
Let us now carefully review the arguments in the previous section. A first remark can be made surrounding eqs.\ \eqref{Approximatie}. What does it mean when we throw away the terms proportional to $\hat{\Pi}_{\eta_{ini}}$? How do we compare the terms in eqs.\ \eqref{HeisenbergPicture}? This is far from clear, since we are comparing operators.
\par This problem might seem academical, but it is crucial: eq.\ \eqref{CommutationNonsense} is wrong. One can check by direct computation starting from eqs.\ \eqref{HeisenbergPicture} that
\begin{equation}
 \left[\hat{\nu}_{\vec{k}}(\eta), \hat{\Pi}_{\vec{k}}(\eta)\right] = i,
\end{equation}
as it should be. Even if one could somehow `compare' different operator terms and `throw away' some terms as done in eq.\ \eqref{Approximatie}, it is clear that any such approximation must become invalid when computing the commutator. In \cite{Albrecht:1992kf}, the author tries to explain the approximate commutativity in a more careful way in terms of expectation values instead in terms of operators. Nevertheless,
\begin{equation}
 \left\langle\left[\hat{\nu}_{\vec{k}}(\eta), \hat{\Pi}_{\vec{k}}(\eta)\right] \right\rangle \not = 0
\end{equation}
for any possible state.
\par On the subject of the Wigner function, we will repeat what was concluded in chapter \ref{QtoC}: namely that properties of the Wigner function do not imply anything about localisation of wave packets, and thus nothing on the classical limit of quantum mechanics.
\par Concerning the `classical stochastic field' argument; this is very reminiscent of a measurement. The stochastic variable $E(\vec{k})$ is exactly what would arise if we would perform a measurement of $\hat{\nu}_{\vec{k}}(\eta_{ini})$. As was already previously discussed in section \ref{PROBLEM}, the idea of measurement in this cosmological context is untenable. The implicitly assumed measurement is even more apparent in the following quote from \cite{Polarski:1995jg} (emphasis mine)
\begin{quote}
`\textit{As a result, after some \textbf{realization} of the stochastic amplitude of the field mode has occured, further evolution of the mode is deterministic and is not affected by quantum noise.}'
\end{quote}
We assume that the author means by `realization' that a measurement has been performed and that the wavefunction has collapsed to some (approximate) eigenstate of $\hat{\nu}_{\vec{k}}$. If not, it is very unclear what we should understand by the concept of `realization'.
\par As a final note, we remark that the `squeezing', represented by eqs.\ \eqref{SqueezingEqs}, is never really used in the classicality arguments. As Sudarsky puts it in his criticism \cite{Sudarsky:2009za} of a particular paper (namely \cite{Kiefer:2008ku}) on the squeezing formalism:
\begin{quote}
`\textit{But having a squeezed state has no effect on the quantum field and momentum operators which do, of course, continue to satisfy the Heisenberg uncertainties. The manuscript seems to ascribe a fundamental value to the squeezing of the state [\dots]}'
\end{quote}
The uncertainties in eqs.\ \eqref{SqueezingEqs} do not in any way present any conclusion on the uncertainties of the physical variables $\nu_{\vec{k}}$ and $\Pi_{\vec{k}}$. In fact, it can easily be seen from eqs.\ \eqref{OwnUncert} that both $\Delta \hat{\nu}_{\vec{k}}$ and $\Delta \hat{\Pi}_{\vec{k}}$ strictly grow in time with $
|f_k|$ for generic states.
\par We conclude this chapter by stating that the squeezing formalism, as presented here, is not sufficient to explain the classicality of the modes in the CMB. In fact care should be taken with several accounts of this formalism as presented in the literature as certainly not all claims are founded, with the explicit example of claims similar to the one in eq.\ \eqref{CommutationNonsense}.
\chapter{Decoherence for Cosmological Perturbations}\label{Decoherence}
In this chapter we will explore the concept of decoherence, first for the inverted oscillator and later on for the Mukhanov-Sasaki variable. It is an often invoked explanation for the classical appearance of the inhomogeneities in the CMB \cite{Barvinsky:1998cq,Burgess:2006jn,Kiefer:1998pb,Laflamme:1993zx,Martineau:2006ki}. While several authors seem to agree on the general idea, we will see that there is still substantial discussion on the specifics of the decoherence process.
\section{Classicality from Decoherence}
To start off, we will present biefly the principal aspects of the theory of decoherence. It will largely be based on \cite{Zurek:1991vd} and the introduction of \cite{Martineau:2006ki}.
\paragraph{System and Environment}Within the framework of decoherence, the problematic aspects of the quantum-to-classical transition are thought of as artefacts from the idealised notion of an `isolated system'. For a realistic experiment, one has to take into account (small) interactions of the system with its \emph{environment}. The term environment typically refers to something the observer is not interested in or has no control over. We will then need some way of reducing the degrees of freedom of the system and environment to just the degrees of the system.
\par Let us be more precise, and specify what we are talking about. Suppose we have some Hamiltonian for our system $\hat{H}_{s}$, a Hamiltonian for our environment $\hat{H}_e$ and an interaction Hamiltonian $\hat{H}_i$. The total Hamiltonian, acting on the product of the system Hilbert space and the environment Hilbert space becomes
\begin{equation}
 \hat{H} = \hat{H}_s \otimes \text{Id}_e + \text{Id}_s \otimes \hat{H}_e + \hat{H}_i.
\end{equation}
Let us further now suppose that the system and environment start out in a state that is factorisable
\begin{equation}
|\Psi(0)\rangle = \left(\sum_i \alpha_i|\phi^i_s(0)\rangle\right) \otimes \left(\sum_j \beta_j |\chi^i_e(0) \rangle\right).
\end{equation}
Without the interaction Hamiltonian $\hat{H}_i$ the time evolution of this state would guarantee it to stay factorisable for all times. However, the interaction Hamiltonian will in general act to set up entanglement. The time evolution of our state is
\begin{equation}
 |\Psi(t)\rangle = \sum_{i} \gamma_{ij}(t)|\phi^i_s\rangle \otimes |\chi^j_e\rangle.
\end{equation}
This state is in general not factorisable any more: it is in an entangled state with the environment. We now suppose that the states $\phi^{i}_s$ are an orthogonal basis in the Hilbert space of our system and compute at time $t$ the expectation value of some hermitian operator $\hat{O}$ according to the Born rule.
\begin{align}
\langle \hat{O} \rangle_{se} &= \sum_{ij} \gamma_{i}^{*}\gamma_j \left( \langle \chi^{i}_e| \otimes \langle \phi_s^{i}|\right) \left(\hat{O} \otimes \text{Id}_e \right) \left(|\phi^{j}_s(t)\rangle \otimes| \chi^{j}_e(t)\rangle \right)\\
&= \left(\sum_{i} |\gamma_i|^2 \langle \chi_e^{i}| \chi_s^{i}\rangle \langle \phi_s^{i}|\hat{O}|\phi_s^{i} \rangle \right) + \left(\sum_{ i\not = j} \gamma_i^{*}\gamma_j \langle \chi_e^{i}| \chi_s^{j}\rangle \langle \phi_s^{i}|\hat{O}|\phi_s^{j} \rangle \right)
\end{align}
Let us now assume that the states $\chi^{i}_e$ are approximately orthonormal. We can then safely neglect the second term, and we write
\begin{equation}\label{Eexpectation}
\langle \hat{O} \rangle_{se} \approx \sum_{i} |\gamma_i|^2 \langle \phi_s^{i}|\hat{O}|\phi_s^{i} \rangle.
\end{equation}
It is in this equation that we will look for the classical limit. Suppose we look at the system without coupling to the environment and then compute the expectation value of $\hat{O}$.
\begin{equation}
 \langle \hat{O} \rangle_{s} = \sum_{ij} \alpha_i^{*} \alpha_j \langle \phi_s^{i}|\hat{O}|\phi_s^{j} \rangle
\end{equation}
The difference is apparent: without the environmental interaction the expectation value $\langle \hat{O} \rangle_{s} $ contains terms that couple different basis states $\phi^{i}_s$ and $\phi^{j}_s$. They represent interference between different parts of the wavefunction. Such interference terms are absent in eq.\ \eqref{Eexpectation}. In fact, eq.\ \eqref{Eexpectation} is a pure summation of probabilities. This is reminiscent of classical mechanics, where interference is never present and probabilities behave additive.
\paragraph{Density matrix}Let us now shift our focus to the density matrix $\hat{\rho}_{se}(t)$ associated with our system state $|\Psi(t)\rangle$. On this level, it is easy to implement the fact that an observer is not interested in the environment. The best way to ignore the degrees of freedom associated with the environment is to trace them out and the result is called the reduced density matrix of the system $\hat{\rho}_s(t)$.
\begin{equation}
 \hat{\rho}_s = \text{Tr}_{e} \hat{\rho}_{se}.
\end{equation}
An important thing to note is that we have changed a pure state $\hat{\rho}_{se}$ into a state $\hat{\rho}_s$ that is generally impure by tracing out the environment. The Von Neumann entropy $S = -\text{Tr} \left(\hat{\rho} \ln \hat{\rho}\right)$ for $\hat{\rho}_s$ will be greater than the entropy of our total density matrix $\hat{\rho}_{se}$. Physically this corresponds to the loss of information we suffered by ignoring the environment. This reflects the fact that is impossible (or at least very hard) to undo the entanglement of our system with the environment.
\par We will suppose now, as before, that the states $\chi_e^{j}$ are approximately orthogonal. The reduced density matrix becomes
\begin{align}
 \hat{\rho}_s &= \sum_j \langle \chi^{j}_e| \left(\sum_{ik} \gamma_i \gamma^{*}_k |\chi_e^{i}\rangle \otimes |\phi_s^{i}\rangle \langle \phi^{k}_s| \otimes \langle \chi_e^{k}|\right) |\chi_e^{j}\rangle \approx \sum_{j} |\gamma_j|^2 |\phi_s^{j}\rangle \langle \phi^{j}_s|.
\end{align}
One sees that the reduced density matrix $\hat{\rho}_s$ is approximately diagonal in the basis of the $\phi_s^{i}$. The basis for which this (approximate) diagonalisation occurs is often called the \emph{pointer basis}. The condition of (approximate) orthonormality on the environment states is usually called the \emph{einselection condition}. The reduced density matrix now reminds us of a classical ensemble. The classical analogue would be an ensemble of many different systems, of which a fraction $|\gamma_j|^2$ would be in the state $|\phi_j\rangle$.
\par
This similarity with a classical ensemble is only formal, since $\hat{\rho}_s$ describes a single system, while a classical ensemble describes several systems. In addition, a measurement will still return values according to the Born rule. We will return to this problems in section \ref{DecoClass}.
\section{Decoherence of the Inverted Oscillator}\label{DecoInverted}
We will first apply the concept of decoherence to our toy model of the inverted harmonic oscillator, as presented in chapter \ref{QtoC}, to get some feeling for the concepts that were discussed above.
\par

We suppose our system is constituted of an inverted harmonic oscillator with Hamiltonian for real and constant $\omega_s$,
\begin{equation}\label{HamSys}
\hat{H}_S = \frac{\hat{p_s}^2}{2} - \frac{\hat{x_s}^2}{2}.
\end{equation}
As environment we will take a normal harmonic oscillator, governed by the following Hamiltonian
\begin{equation}\label{HamEnv}
\hat{H}_E = \frac{\hat{p_e}^2}{2} + \omega^2_s \frac{\hat{x_e}^2}{2},
\end{equation}
with $\omega_s$ real. Let us take a simple form of interaction between the system and the environment, with (small) interaction parameter $\lambda$. The total Hamiltonian we wish to study is then
\begin{equation}\label{TotalHam}
 \hat{H} = \hat{H}_S + \hat{H}_E + \lambda \omega_s \omega_e \hat{x_s} \hat{x_e}.
\end{equation}
One can then check that the following wavefunction satisfies the Schr\"odinger equation \cite{Kim:2000ek}
\begin{align}
 \Psi &= N(t) \exp \left[ \frac{i}{2}\left( \frac{f'^{*}_s}{f_s^{*}}x_s^2 + \frac{f'^{*}_e}{f_e^{*}}x_e^2 \right) - i B(t) x_e x_s\right],\\
N(t) &= \left( \frac{\Gamma_{\overline{x}}}{\pi |f_e|^2}\right)^{\frac{1}{4}}\left(\frac{f}{f^*}\right)^{1/2}, \\
B(t) &=\overline{\lambda}\frac{\int dt' f_e^{*} f_s^{*}}{f_e^{*} f_s^{*}}, \\
 \Gamma_{\overline{x}} &=\frac{1 - \text{Im}(B)^2 |f_e|^2 |f_s|^2}{2|f_s|^2},\\
\overline{\lambda} &= \lambda \omega_s \omega_e.
\end{align}
 The notation $\Gamma_{\overline{x}}$ will shortly be explained, while $f_s$ and $f_e$ are solutions to the following equations
\begin{align}
 \ddot{f_s} + \left[ - \omega_s^2 + B^*(t)^2\right] & = 0, & \ddot{f_e} + \left[ \omega_e^2 + B^*(t)^2\right] = 0.
\end{align}
We take solutions to these equations, obeying the following Wronskian condition, for both $f_s$ and $f_e$,
\begin{equation}
 i(\dot{f} f^{*} - \dot{f}^{*}f) = 1.
\end{equation}
It is here that lies the difference with \cite{Kim:2000ek}. Their solutions for $f_s$ do not satisfy this condition, making $\frac{\dot{f}_s^*}{f_s^*}$ real for $\lambda = 0$. Thus, their starting wavefunction is not normalisable for the noninteracting case.
\par We now consider the density matrix of this system
\begin{equation}
\begin{split}
 \rho_{se}(x_s, x_s', x_e, x_e', t) =& N^{*}(t)N(t)\exp \left[ \frac{i}{2}\left(\frac{f'^{*}_s}{f^{*}_s}x_s^2 - \frac{f'_s}{f_s}x'^2_s \right)\right],\\
& \exp \left[ \frac{i}{2}\left(\frac{f'^{*}_e}{f^{*}_e}x_e^2 - \frac{f'_e}{f_e}x'^2_e \right) - i \left(B x_e x_s - B^{*}x_e'x_s'\right) \right].
\end{split}
\end{equation}
We are now interested in the reduced density matrix of the system. Suppose we don't know anything about the harmonic oscillator, the environment, and we want to eliminate our knowledge about it. Therefore, we will trace over the environment degrees of freedom.
\begin{align}
 \rho_s (x_s, x_s') &= \int dx_e \rho_{se}(x_s, x_s', x_e, x_e),\\ \label{RedDens}
&= N^{*}(t)N(t) \sqrt{\pi |f_e|^2}\exp \left[ \frac{i}{2}\left(\frac{f'^{*}_s}{f^{*}_s}x_s^2 - \frac{f'_s}{f_s}x'^2_s \right) - \frac{|f_e|^2}{2}\left(B x_s - B^{*}x'_s\right)^2 \right].
\end{align}
Using the Wronskian condition, we can rewrite this as
\begin{align}
 \overline{x} &= \frac{1}{2}\left(x_s + x_s'\right),\\
 \delta x &= \frac{1}{2}\left(x_s + x_s'\right),\\\label{ReducedDM}
\rho_{s}\left( \overline{x}, \delta x\right) &= \sqrt{\frac{\Gamma_{\overline{x}}}{\pi}} \exp \left[ - \frac{\Gamma_{\overline{x}}}{2} \overline{x}^2 - \frac{\Gamma_{\delta x}}{2} \delta x^2 - \frac{\Gamma}{2} \delta x \overline{x}\right].
\end{align}
In these coordinates, the diagonal elements of the density matrix are given by $\delta x = 0$. If $\rho_s$ is approximately diagonal, then $\Gamma_{\delta x}$ would be very large. In fact, the ratio of $\Gamma_{\overline{x}}$ over $\Gamma_{\delta x}$ can be taken as a measure of the coherence of the density matrix. As this quantity becomes smaller, quantum coherence is lost\cite{Kim:2000ek}. Straightforward calculation gives us
\begin{align}\label{Gamma1}
 \Gamma_{\overline{x}} &= \frac{1 - \text{Im}(B)^2 |f_e|^2 |f_s|^2}{2|f_s|^2},\\
 \Gamma_{\delta x} &= \frac{1 + \text{Re}(B)^2 |f_e|^2 |f_s|^2}{2|f_s|^2}.
\end{align}
Intuitively, what we expect for $\lambda$ `small' is this: $|f_e|$ will be approximately constant, while $|f_s|$ will grow roughly exponentially. If $B$ doesn't vary much in time, we see that quantum coherence is lost.
\par In fact, this is precisely what will happen. We choose concrete realisations of $f_s$ and $f_e$ to be
\begin{align}
f_s &= \sqrt{\frac{1}{2|\Omega_s(t)|}}\left(\cosh\left(\int \Omega_s(t') dt' \right) + i \sinh\left(\int \Omega_s(t') dt'\right)\right),\\
f_e &= \sqrt{\frac{1}{2| \Omega_e|(t)}} \exp \left( - i \int \Omega_e(t') dt'\right).
\end{align}
The modified frequencies are no longer time independent. To keep the problem analytically tractable, we only look for solutions $\Omega_s, \Omega_e$ as a power series in $\lambda$ and $t$.
\begin{align}
\Omega_s(t) &\approx \omega_s - \overline{\lambda}^2 \frac{\left(\omega_e + \omega_s\right)^2}{2\left(\omega_e^2 + \omega_s^2\right)^2} + i \lambda^2 \omega_s t,\\
\Omega_e(t) &\approx \omega_s + \overline{\lambda}^2 \frac{\left(\omega_e + \omega_s\right)^2}{2\left(\omega_e^2 + \omega_s^2\right)^2} - i \lambda^2 \omega_e t.
\end{align}
To find $B^2$ to second order in $\lambda$ we only need $B$ to first order in $\lambda$
\begin{align}
 \text{Im}B &\approx - \overline{\lambda} \frac{\omega_e + \omega_s}{\omega_s^2 + \omega_e^2}, & \text{Re}B \approx 0.
\end{align}
Our measure of quantum coherence becomes, to lowest non-vanishing order in $\lambda$,
\begin{align}
 \sqrt{\frac{\Gamma_{\overline{x}}}{\Gamma_{\delta x}}} &= \sqrt{\frac{1 - \text{Im}(B)^2 |f_e|^2 |f_s|^2}{1 + \text{Re}(B)^2 |f_e|^2 |f_s|^2}},\\
&\approx 1 - \frac{\overline{\lambda}^2}{2} \left(\frac{\omega_e + \omega_s}{\omega_s^2 + \omega_e^2}\right)^2 \cosh(2 \omega_s t),
\end{align}
where we only retained the lowest non vanishing order in $\lambda$ for the last equation.
\par
As long as $\lambda t << 1$, we see that the density matrix very rapidly becomes approximately diagonal in the position basis. Quantum coherence is thus lost extremely rapidly on a time scale of $\omega_s^{-1}$, while the relevant pointer basis is the position basis. %
\par We repeat that our result is less beautiful than the result obtained in \cite{Kim:2000ek}. This is due to the fact that we impose a Wronskian constraint on our solutions for $f_e$ and $f_s$. Our wavefunctions are normalisable for all $\lambda$, while the wavefunctions in \cite{Kim:2000ek} are non-normalisable for $\lambda = 0$.
\section{System and Environment for Cosmological Perturbations}
The theory of the scalar perturbations of the Einstein equations, as explained in chapter \ref{Perturbations}, does not incorporate any decoherence in itself. The Fourier modes of the Mukhanov-Sasaki variable $\nu_{\vec{k}}$ constitute a closed system. This is a feature of the second order expansion we made of the full action of our theory in eq.\ \eqref{FullAction}. Expanding the action to third order in the perturbations will yield small corrections in the form of interaction terms: Different modes $\nu_{\vec{k}}$ will no longer evolve independently.
\par There is now an important distinction to be made. Our experiments that measure the sky have a finite angular resolution. For example, WMAP achieves a resolution often quoted as being of order $1^{\circ}$. This thus limits our ability to discern modes with very short physical wavelengths, or equivalently, modes with very high $\frac{k}{a}$. The modes we observe have all exited the Hubble radius during inflation quite early, see section \ref{ObsSpec}. We do not observe very high energy modes that only exited the Hubble radius very late (or not at all) during inflation. Since there are no observational results for these short-wavelength modes we are free to ignore their degrees of freedom. A natural choice of system and environment would thus be the long-wavelength modes and short-wavelength modes, respectively.
\par For the rest of this section, we will restrict our attention to perfect de Sitter inflation. The relevant length scale is of course the Hubble radius $H^{-1}$. A natural cut-off at any given time between the long and short wavelength modes would be the Hubble radius. The idea is thus that modes with $k \gg H$ decohere the observed modes of $k \ll H$. One thus introduces a cut-off wavelength, which separates system and environment.
\par
Of course, there are other possibilities for the choice of decohering environment. For instance, one could imagine the inflaton field as being weakly coupled to the photons present at the onset of inflation. However, the interaction between different Fourier modes of the Mukhanov-Sasaki variable will be enough to achieve decoherence. It is thus not necessary to introduce extra interactions for the sake of decoherence.
\par
It is necessary to think some more about our choice of cut-off. After all, an observer choosing a different cut-off would have a different definition of system and environment, thus seeing less or more decoherence than the observer that chose $H^{-1}$ as the cut-off wavelength. This is natural, since someone choosing a cut-off wavelength of $0$ would not see any decoherence at all, because he is keeping track of all available degrees of freedom. The choice of $H^{-1}$ as the cut-off is however the minimal one, as is argued in\cite{Martineau:2006ki}. The essential idea is that the Hubble radius represents an event horizon in perfect de Sitter spacetime.%
\section{Points of Dispute}
The ideas outlined above are generally agreed upon by the community. There are however two important aspects where the community is in disagreement. Firstly, there are two competing views of the period where the decoherence should occur. Secondly, there is still discussion of which basis is the relevant pointer basis. We will briefly discuss for both problems the dominating views.

\paragraph{When Does Decoherence Occur?}
In \cite{Martineau:2006ki,Kiefer:2006je,Barvinsky:1998cq}, among others, it is argued that decoherence occurs during the inflationary period. %
Decoherence for a mode of wavevector $\vec{k}$ is claimed to start quickly after Hubble exit, when $|k\eta| = 1$. At that moment, $k^2 - \frac{z''}{z}$ changes sign, while the short wavelength modes essentially behave as a harmonic oscillator. %
For example, in \cite{Kiefer:2008ku} the evolution of the density matrix is treated by assuming a master equation of Lindblad form. The resulting decoherence is exponential on timescales $H^{-1}ln \left(\frac{H^{-1}}{t_0}\right)$, where $t_0$ is a time scale characteristic for the considered interaction.
\par The contrasting vision is the one in \cite{Burgess:2006jn}. It is argued that decoherence is produced rapidly when the modes under investigation re-enter the Hubble radius, during the heating epoch. The short wavelength modes are then already sub-Hubble and are assumed to essentially be a thermic bath. It is calculated that, for interactions of the Mukhanov-Sasaki variable of order 3, the decoherence in this epoch is also rapid, while the decoherence time is again determined by the ratio of $H^{-1}$ to some time scale set of the interaction.
\par In \cite{Burgess:2006jn}, it is argued why the other view, decoherence during inflation, should be treated carefully. Very short wavelength modes behave essentially as they would in Minkowski space. For other experiments, unrelated to inflation, there would in general be higher-energy modes present that are excluded from observation by experimental limits. It is then concluded that such high energy modes would always decohere the long wavelength modes that are under study. If we assume this decoherence process, why do we observe any quantum coherence at all in diverse experiments?
\par This argumentation is not really founded however, since the analogy with Minkowski space time is stretched beyond its limit. For experiments in Minkowski spacetime, there is no such thing as modes that exit the Hubble radius. The long wavelength modes in our imaginary experiment also evolve in Minkowski spacetime, while during inflation we have a coupling of sub Hubble (Minkowski) modes and super Hubble (very much non-Minkowski) modes.
\paragraph{Pointer Basis}
Another point of discussion is the relevant pointer basis. In \cite{Kiefer:2006je}, \cite{Kiefer:1998jk}, \cite{Barvinsky:1998cq}, \cite{Burgess:2006jn} and many others, the relevant pointer is argued to be the field amplitude basis\footnote{To be entirely correct, the relevant pointer basis is a basis of very narrow Gaussians around some central field amplitudes. This makes the pointer basis normalisable.}. This corresponds to the position basis for the inverted harmonic oscillator. The primary argument seems to be that the observable $\hat{\nu}_{\vec{k}}$ `approximately commutes' with the Hamiltonian and is thus rather robust under time evolution. This type of argument is very reminiscent of the type of arguments considered in section \ref{SqueClas}. The system Hamiltonian was given by
\begin{equation}
 \hat{H}_{\vec{k}} = \frac{1}{2} \int d^3x \left[ \Pi_{\vec{k}}^2 + \left( k^2 - \frac{z''}{z}\right)\nu^{*}_{\vec{k}} \nu_{\vec{k}}\right].
\end{equation}
As discussed in chapter \ref{Squeezing}, there is no limit in which $\nu_{\vec{k}}$ and $\Pi_{\vec{k}}$ commute (even approximately), and thus $\nu_{\vec{k}}$ does not commute with the Hamiltonian. While this does not exclude the field amplitude basis as pointer basis, it illustrates that perhaps further motivation is necessary.
\par Several other pointer bases have been suggested. For example, in \cite{Campo:2004sz} it is suggested that the pointer basis is the basis of the coherent states, the eigenstates of the annihilation operator. The argument relies on the harmonic oscillator case. For a standard harmonic oscillator prepared in a coherent state, the expected value of both position and momentum obey the classical equations of motion. In addition, the coherent states are the states of the harmonic oscillator that entangle the least with an environment, therefore providing relatively robust under time evolution. One can doubt this argument, as \cite{Kiefer:2006je} does, since the behavior of the inverted oscillator is fundamentally different from the standard oscillator.
\section{Classicality?}\label{DecoClass}
We conclude this chapter by closely looking at what type of classicality decoherence actually achieves. What we argued in the beginning of this chapter can be loosely described as a `reduction of quantum interference' for the selected states of the pointer basis. This represents a step forward in achieving the classical limit, but it is not enough.
\par The description of the system is still using a wavefunction (or equivalently, a density matrix). Measurements will still give variable outcomes, according to the Born rule. Once measured, decoherence explains why the modes behave classically. This is best explained in terms of the paradox of Schr\"odinger's cat. The system here is the two level system comprised of $|$cat dead$\rangle$ and $|$cat alive$\rangle$. Decoherence with the remaining degrees of freedom of the cat will create a reduced density matrix that is diagonal.
\begin{equation}
 \hat{\rho}_{cat} \approx |\alpha|^2 |\text{cat alive}\rangle \langle \text{cat alive} | + |\beta|^2 |\text{cat dead}\rangle \langle \text{cat dead} |
\end{equation}
The cat is no longer in a superposition of `dead' and `alive' and the density matrix formally resembles a classical ensemble. This is only formal resemblance, since we still need a measurement to make the cat look either dead or alive to us.
\par The situation for cosmological perturbations is analogous. Decoherence, in for example the field configuration basis, will still need a measurement to select a field configuration for us to observe.\footnote{In addition, this (approximate) field configuration basis is very spread out in momentum space. For the classical limit, we need both a well determined position and a well determined momentum.} We conclude that one needs a measurement to explain the observed classicality in the night sky. But, as explained in section \ref{PROBLEM}, it is very unclear what constitutes a measurement in the cosmological constant.
\par One can consider theories which combine decoherence and an effective collapse theory. For such theories, there is no need for a measurement, but this comes at the price of introducing some kind of collapse mechanism. Examples include, but are certainly not limited to \cite{Penrose} and \cite{Perez:2005gh}.
\par Decoherence alone also does not explain why the initial symmetric state became inhomogeneous. All the modes are assumed to be in a symmetric state, and tracing over some of them will not result in a reduced density matrix that is no longer symmetric. For example, note that the interaction considered in section \ref{DecoInverted} is not symmetric in $\hat{x}_s$. Despite this fact the reduced density matrix, eq.\ \eqref{RedDens}, will still be symmetric under the change of $\left( x_s, x'_s\right) \rightarrow \left( - x_s, - x'_s \right)$.
\par Kiefer, in \cite{Kiefer:2006je}, argues that this situation is analogous to symmetry breaking. The set of minima for a certain potential is symmetric, but an individual minimum is no longer symmetric. It is then argued that for cosmological perturbations the same thing happens: the state is symmetric, but then an individual realisation is chosen that breaks the symmetry. Note first that this is exactly the procedure of a measurement. Second, in \cite{Sudarsky:2009za} it is argued that any classical realisation of spontaneous symmetry breaking, like a ferromagnet being cooled below critical temperature, involves states that are not exactly symmetric. A ferromagnet prepared in exactly zero magnetisation will conserve zero magnetisation, whatever the temperature. This process is however very much subject to perturbations, the slightest change in magnetisation will result in the magnet developing a non zero magnetisation. For the Mukhanov-Sasaki modes, there is no such initial perturbation if all modes start in the Bunch-Davies vacuum. Thirdly, for symmetry breaking in quantum field theory, the theory is equivalent for every choice of symmetry breaking vacuum. For the Mukhanov-Sasaki variable, different symmetry breaking configurations are clearly not equivalent.
\par As conclusion, we cite from \cite{Schlosshauer:2003zy} why decoherence is not sufficient to explain the classical limit:
\begin{quote}
\emph{`\dots decoherence cannot solve the problem of definite outcomes in quantum measurement: we are still left with a multitude of albeit individually well-localized quasi-classical components of the wave function, and we need to supplement or otherwise to interpret this situation in order to explain why and how single outcomes are perceived.'}
\end{quote}

\part{The Pilot-Wave Case}\label{PWCase}
\chapter{Pilot-wave Theory}\label{PilotWave}We will use this chapter to introduce Bohmian mechanics, or pilot-wave theory. It is an alternative interpretation of quantum mechanics, that postulates well-defined particles with well-defined trajectories. While it is a deterministic theory, the predictions it makes are equivalent to those made by standard quantum mechanics under the condition of quantum equilibrium.
\par First, we will explain the non-relativistic theory. It then is possible to generalize pilot-wave theory to quantum field theory: we will briefly present this theory for spin 0 fields. The possibilities that pilot-wave theory opens up in the field of quantum cosmology will also briefly be discussed. Then we will present the classical limit within Bohmian mechanics, thereby giving a strict formulation and touching upon a different characterisation and an important conjecture that is present in the literature. In conclusion, we will turn to the concrete examples of chapter \ref{QtoC}, the free particle and the inverted oscillator and develop the pilot-wave theory for them.
\section{General Theory}\label{PWTheory}
In classical mechanics a system of one particle is fully specified by the couple $(q, p)$, where $q$ is some generalised coordinate and $p$ is the momentum canonically conjugate to $q$. In Bohmian mechanics, the role of this couple is fulfilled by the couple $(q, \Psi)$, where $\Psi$ is the wavefunction of the system.
\par The wavefunction $\Psi$ is the same as the one used in standard quantum mechanics. It follows Schr\"odinger time evolution
\begin{align}
 i\frac{\partial \Psi}{\partial t}& = \hat{H}\Psi, & \hat{H} = \frac{\hat{p}^2}{2m} + V(\hat{x}).
\end{align}
 We can now, as we did with the WKB approximation, split the wavefunction as follows
\begin{equation}
 \Psi(q, t) = R(q, t) \exp \left(i S(q,t)\right)
\end{equation}
where $S$ and $R$ are both real and $R$ is positive. As before, the Schr\"odinger equation now decouples into a real and imaginary part
\begin{align}\label{Split1}
\frac{\partial S}{\partial t} &= - \frac{1}{2m} \left(\frac{\partial S}{\partial q}\right)^2 - V(x) + \frac{1}{2m R}\frac{\partial^2 R}{\partial q^2},\\\label{Split2}
\frac{\partial R^2}{\partial t} &= - \frac{\partial}{\partial q} \left( \frac{R^2}{m} \frac{\partial S}{\partial q}\right).
\end{align}
Let us now postulate an actual particle with actual velocity given by the guidance equation
\begin{equation}\label{Guidance}
 \frac{dq}{dt} = \frac{1}{m}\frac{\partial S}{\partial q}.
\end{equation}
Eq. \eqref{Split2} can then be seen as a continuity equation for an ensemble of postulated particles with density $R^2$ and velocity field given by the guidance equation eq.\ \eqref{Guidance}. Using the postulated particles interpretation, eq.\ \eqref{Split1} formally looks like the classical Hamilton-Jacobi equation\footnote{This likeness is only formal. Note that the classical Hamilton-Jacobi function $S$ is a function of various constants of motion, while the $S$ here is a function of $q$ and $t$.}. We think of it as describing a particle moving under the influence of a potential $V(q)$ and an extra quantum potential term
\begin{equation}
 V_q = -\frac{1}{2mR}\frac{\partial^2 R}{\partial q^2}.
\end{equation}
The obvious interpretation is that the pilot-wave particle moves under the influence of two forces, a classical one and a quantum one
\begin{equation}
 \frac{d^2q}{dt^2} = F_c + F_q = - \frac{dV}{dq} - \frac{dV_q}{dq}
\end{equation}
\paragraph{Quantum Equilibrium} Let us now demonstrate the equivalence between pilot-wave theory and standard quantum theory, under some condition. Let us go back to the continuity equation \eqref{Split2}. This is in fact a second continuity equation in quantum mechanics, since the Schr\"odinger equation already implies the following equation
\begin{equation}
 \frac{d |\Psi|^2}{dt} = - \frac{1}{m}\frac{\partial }{\partial q }\left(|\Psi|^2\frac{\partial S}{\partial q}\right),
\end{equation}
which is nothing but the conservation of probability within the framework of the Copenhagen interpretation. We see that both $R^2$ and $|\Psi|^2$ obey the same laws of motion. We conclude that, if $R^2(x, t_0) = |\Psi(q,t_0)|^2$ at some time $t_0$, that $R^2$ and $|\Psi|^2$ will be equal for all time. This is called the \emph{equivariance} of $R^2$ and $\Psi^2$. In general it is assumed that $R^2 = |\Psi|^2$, and this assumption goes by the name of \emph{quantum equilibrium}. Under the assumption of quantum equilibrium, the predictions of Bohmian mechanics match those of standard quantum mechanics.
To show this explicitly, consider a large amount of systems prepared in the same state $\Psi$. The expected value of some operator position dependent operator $O(\hat{q})$ in de Broglie-Bohm theory will then be
\begin{equation}
 \langle O(\hat{q})\rangle_{dBB} = \int O(\hat{q})R^2 dq= \int O(\hat{q})|\Psi|^2 dq= \langle O(\hat{q}) \rangle,
\end{equation}
where the right hand side is the expected value in standard quantum mechanics. This effectively demonstrates the equivalence of Bohmian mechanics in the equilibrium regime with standard quantum mechanics, since quantum mechanics only makes statements about large numbers of systems and most measurements only make predictions in terms of particle positions, like e.g. pointers of measurement apparati. There are however some situations, involving measuring time dependent quantities, where pilot-wave theory makes unambigous predictions while there is discussion on the Copenhagen interpretation of those measurements, as is for example mentioned in \cite{Struyve:2004xd} and \cite{AlloriPHD}.

\paragraph{Measurement and Collapse of the Wavefunction}
Collapse of the wavefunction has a natural explanation within the framework of pilot-wave theory. What one typically measures in an experiment is the position of some particle, possibly the position of some pointer of some measurement apparatus. The result of such a measurement is then the exact position of the pilot-wave particle. The collapse of the wavefunction has an effective interpretation. Consider some wavefunction that consists of two wavepackets with virtually no overlap in position space
\begin{align}
 |\Psi\rangle &= \alpha_1|\phi_1\rangle + \alpha_2|\phi_2\rangle, & \langle \phi_1|\phi_2 \rangle \approx 0.
\end{align}
Since a pilot-wave particle has a definitive position in space, we can think of it as being located inside the support of $|\phi_1\rangle$. In that region, it is clear that $\langle x |\phi_2\rangle \approx 0$. In computing the the trajectory of the particle, we can safely ignore the presence of $\phi_2$ in eq.\ \eqref{Guidance}. In this way, our description of the particle now makes use of the collapsed wavefunction $\langle x |\phi_1\rangle$. Note however, that this collapse only takes the form of a useful approximation in pilot-wave theory and that it is in no way fundamental or necessary, in contrast to the case in standard quantum mechanics.

\paragraph{Non-Locality} It is obvious that pilot-wave theory is a hidden-variables theory. As such it is necessarily non-local, as implied by Bells inequalities. To see this explicitly, consider the two-body generalisation of the previous paragraphs. We have a wavefunction $\Psi(q_1, q_2, t)$. The guidance equation for particle $1$ will depend on the \emph{instantaneous} position of particle $2$. Since this instantaneous connection is always present, no matter the distance between both particles, the theory is manifestly non-local. This is a property that will carry over to the relativistic field theory. It can be shown, see \cite{Valentini:1991zq}, that this non-locality can not be used to send faster-than-light signals under the assumption of quantum equilibrium. This places the non-locality of pilot-wave theory on the same level as the non-locality implied by entanglement in standard quantum mechanics.

\section{Quantum Field Theory}
While the pilot-wave theory presented in section \ref{PWTheory} can be generalised to particles of any spin, it is found that the interpretation in terms of particles cannot be held in a relativistic version of pilot-wave theory. In analogy with the extension of quantum theory to include special relativity by transferring from particles to fields, we will construct a pilot-wave theory for a field of spin $0$. It is possible to generalise this formalism to all bosonic fields. For spin $1/2$ one c\'an maintain the particle interpretation: the pilot-wave theory for fermionic fields is fundamentally different. It will however not be treated here. For notes on the subject see e.g. \cite{Struyve:2004xd} or \cite{HollandFieldTheory}. Our (very brief) treatment will be based on this last reference.
\par We will consider some free scalar field $\phi$. The relevant classical Lagrangian density is
\begin{equation}
 \mathcal{L} = \frac{1}{2}\partial_{\mu}\phi \partial^{\mu} \phi
\end{equation}
Moving to the Hamiltonian formalism we get
\begin{align}
 H &= \int d^3x \frac{1}{2}\left(\pi^2 + (\nabla \phi)^2\right), & \pi = \frac{\partial \phi}{\partial t}.
\end{align}
We will then quantise this Hamiltonian by imposing equal time commutation relations. In the Schr\"odinger picture we replace $\pi$ with the functional derivative $-i\frac{\partial}{\partial \phi}$. The Hamiltonian then becomes an operator $\hat{H}$ that acts on a wavefunction $\Psi(\phi(x), t)$, which is now a functional of the field configuration $\phi(x)$. We then get the functional Schr\"odinger equation
\begin{equation}
 i\frac{\partial \Psi}{\partial t} = \hat{H}\Psi.
\end{equation}
\par It is now possible to make a pilot-wave theory for the field configuration $\phi(x)$, meaning that we can formulate a guidance equation and corresponding definite trajectories for the variable $\phi(x)$. For our purposes it will be simpler to formulate the theory in terms of the Fourier components of the field. We define the Fourier transformation as follows
\begin{equation}
 \phi_{\vec{k}} = \frac{1}{(2\pi)^{3/2}}\int \phi \exp \left(-i \vec{k}\cdot \vec{x}\right)d^3x.
\end{equation}
The functional Schr\"odinger equation becomes
\begin{equation}
 i \frac{\partial \Psi }{\partial t} = \frac{1}{2}\left(\int d^3k - \frac{\partial^2}{\partial \phi_{\vec{k}} \partial \phi_{\vec{k}}^*} + k^2 \phi_{\vec{k}}\phi^*_{\vec{k}}\right) \Psi.
\end{equation}
Assuming that $\Psi$ is a product state, this equation decouples into a Schr\"odinger equation for all $\vec{k}$
\begin{equation}
 i \frac{\partial \Psi_{\vec{k}}}{\partial t} = \frac{1}{2}\left(- \frac{\partial^2}{\partial \phi_{\vec{k}} \partial \phi_{\vec{k}}^*} + k^2 \phi_{\vec{k}}\phi^*_{\vec{k}}\right)\Psi_{\vec{k}},
\end{equation}
where $\Psi_{\vec{k}}$ stands for the factor of the total wavefunction $\Psi$ associated with the variable $\phi_{\vec{k}}$. As in the non-relativistic case, we can write $\Psi_{\vec{k}} = R_{\vec{k}}\exp\left(i S_{\vec{k}}\right)$. We can then postulate that the Fourier modes follow actual trajectories, governed by the guidance equations
\begin{align}
 \frac{\partial \phi_{\vec{k}}}{dt} &= \frac{\partial S}{\partial \phi^*_{\vec{k}}}, & \frac{\partial \phi^*_{\vec{k}}}{dt} &= \frac{\partial S}{\partial \phi_{\vec{k}}}.
\end{align}
We also have an analogue for eq.\ \eqref{Split1},
\begin{equation}
 \frac{\partial S_{\vec{k}}}{\partial \eta} = - \frac{1}{2}\left(\frac{\partial S_{\vec{k}}}{\partial \phi_{\vec{k}}}\right)\left(\frac{\partial S_{\vec{k}}}{\partial \phi_{\vec{k}}^*}\right) - \frac{k^2}{2} \phi_{\vec{k}}^*\phi_{\vec{k}} - \frac{1}{2R_{ \vec{k}}} \frac{\partial^2 R_{\vec{k}}}{\partial \phi_{\vec{k}} \partial \phi_{\vec{k}}^*},
\end{equation}
and we recognize the classical Hamilton-Jacobi equation with an extra term that we identify as the quantum potential. We will use this framework (in terms of the real and imaginary parts of the Fourier modes) to construct the pilot-wave theory for the Mukhanov-Sasaki variable $\nu$ in chapter \ref{Quantifying}.
\par We conclude this brief section by remarking that this theory is not Lorentz invariant and that there exists a preferred inertial frame. For explicit argumentation see \cite{HollandFieldTheory}. This non-covariance is however not relevant, since under the condition quantum equilibrium this preferred frame can not be detected\cite{ValentiniPHD}.

\section{Pilot-wave Cosmology}\label{dBBCosm}
Pilot-wave theory opens up a whole new realm of possibilities in the context of cosmology. While we will apply it in the context of the primordial perturbations in chapter \ref{PilotWavePert}, we will briefly mention some attempts at devising a complete quantum theory of the early universe.
\par In chapter \ref{Perturbations} we only quantised the perturbations, in the form of the Mukhanov-Sasaki variable, while we took the classical solutions for the background variables $a(\eta)$ and $\varphi^{(0)}_{u}$. In this light, the theory we constructed in section \ref{QuanTheor} can be considered to be semi-classical at best. In fact, we repeatedly used the classical equations of motion for the background variables in deriving the classical action for the Mukhanov-Sasaki variable, as was indicated explicitly in section \ref{EOMSlowRoll}. However, it can be shown \cite{ Falciano:2008nk} that this action can be obtained without ever invoking the background equations\footnote{At least for an inflaton field that is not subject to any slowroll potential $V(\varphi)$.}, through an intricate scheme of canonical transformations before quantisation and unitary transformations after quantisation. In this way, we can be fairly certain that several of the features of the theory discussed in this thesis will carry over to a fully quantised theory.
\paragraph{Bouncing Cosmologies} However, as noted in \cite{Peter:2008qz}, a full quantisation can entail some very striking features and solve some problems that are not addressed by the theory of inflation. If one quantises the background variables, the result is wavefunctions for $a(\eta)$ and $\varphi^{(0)}(\eta)$. These can then be easily interpreted by pilot-wave theory. Starting from some initial condition $a(\eta_{ini})$ and $\varphi^{(0)}(\eta)$ one gets definite trajectories, meaning a prescription $a(\eta)$ and $\varphi^{(0)}(\eta)$.
\par In \cite{Peter:2008qz} these trajectories are explicitly calculated for models that exhibit a `bounce'. This bounce means that the universe is supposed to be eternal and that the scale factor in turn grows and decays with time. Bounces occur when the scale factor stops shrinking and begins to grow. It is noted that such a scenario eliminates every need for an initial singularity (a Big Bang). In addition, for some values of relevant parameters it can be shown that no physically relevant wavelength ever reaches Planck scales, thereby increasing our trust in the predictions of the model at the moment of bounces. In fact, such bouncing models do in general not need inflation to produce the observed homogeneity and the power spectrum of the perturbations. For the homogeneity case, it is enough to observe that one can evolve the critical density $\Omega$ in eq.\ \eqref{CritDensEvo} towards one by a period of accelerated expansion or, equivalently, a period of decelerated contraction. For the inhomogeneities, one can show that the same equations of motion for the inhomogeneities hold during the bounce as the ones during inflation\cite{Peter:2008qz}. Bouncing cosmologies thus don't need to introduce a hypothetical inflaton field to account for observations.
\paragraph{Inflationary Models} However, the theory also allows for models without initial singularity but with an inflaton field. Consider for example the cosmologies studied in \cite{Falciano:2007yf}. These models include a universe that has been expanding forever, starting from a universe with a finite non-zero size in the infinite past. It is even shown, that there are possible trajectories for $a(\eta)$ that start with an accelerated expansion and can be smoothly joined to a period of radiation dominated expansion. These trajectories can thus be smoothly linked to the classical theory of our universe, starting at the epoch of radiation domination.

\section{The Classical Limit}
The emergence of classicality from quantum theories is easier to discuss in the pilot-wave context. The reason for this is clear. While standard quantum mechanics has to construct ways to regain particles with well-defined positions and momenta, de Broglie-Bohm theory postulates these features. We have now a theory which unambiguously speaks about definite particles (or field configurations) for which the time evolution can be exactly computed.
\par Of course, this does not mean that we always trivially have a classical limit. The theory sketched above clearly allows for trajectories that are far from classical solutions to the equations of motion. It is the great advantage of pilot-wave theory that we can exactly see where any non-classical behaviour originates, namely in the quantum potential term $V_q$ in eq.\ \eqref{Split1}. It is easy to see that if this quantum potential is negligible compared to the kinetic energy or the classical potential of the system, that quantum effects can be ignored.
\par Since this quantum potential is dependent on the specific state the system is in, it is very non-trivial to classify the states for which the quantum potential can be neglected either in comparison with the kinetic energy or in comparison with the classical potential. To tackle this issue, we can return to the treatment of the WKB approximation in section \ref{WKBSectie}, where the quantum potential was first introduced. The pilot-wave guidance equations for the WKB energy eigenfunctions in eq.\ \eqref{WKBWave} are
\begin{equation}
\frac{dx}{dt} = \pm \frac{1}{m}\sqrt{2m (E-V(x))}.
\end{equation}
Deriving this with respect to time we immediately get the classical equations of motion
\begin{equation}
 \frac{d^2x}{dt^2} = -\frac{1}{m}\frac{dV(x)}{dx}.
\end{equation}
The WKB condition in eq.\ \eqref{WKBcondition} is thus clearly a step in the right direction, since the WKB states give rise to the classical limit. But, as we remarked before, this condition is not sufficient to guarantee the classical limit. Reconsider for instance the superposition of different WKB solutions already given before, $\Psi = \frac{1}{N}\left( \frac{1}{A_+}\Psi_+ + \frac{1}{A_-}\Psi_-\right)$. The quantum force that corresponds to this state is given by
\begin{equation}
 F_q = -\frac{\partial V}{\partial x},
\end{equation}
which is exactly equal and opposite the classical force. We recover static particles, which is certainly not classical behavior in general, unless we also consider special initial conditions which give rise to classical static trajectories. We repeat that the problem of the WKB condition is that it is state independent, while the quantum potential is state dependent.
\paragraph{A Conjecture in the Literature} One can reach more by arguments similar to the ones given in section \ref{Planck}. In \cite{SevenSteps} and \cite{AlloriPHD} it is argued that to obtain classical trajectories one needs to have that the de Broglie wavelength needs to be smaller than a scale associated with the classical potential. Specifically
\begin{equation}\label{StateInDependent}
 \lambda \ll \sqrt{\frac{\partial_x V}{\partial_x^3 V}}
\end{equation}
where the right hand side is taken to be indicative of the typical variation scale of the classical potential. For quadratic potentials, it is argued that this inequality always holds. This inequality is very reminiscent of the WKB condition, eq.\ \eqref{WKBcondition}, but a different scale of variation for the classical potential is taken to compare the de Broglie wavelength to. We imagine this condition by itself to be insufficient, since this will suffer from the same problem as the WKB condition: it will not generalise to superpositions of states. One can see this very easily in case of a quadratic potential: for an energy eigenstate of normal harmonic oscillator the condition is trivially satisfied, but the guidance equations yield static particles. Again, this is not classical movement, excepting the classical case when $x(0)= \partial_t x(0)= 0$.
\par In \cite{AlloriPHD} and \cite{SevenSteps} it is argued that this condition nees to be supplemented by a kind of scaling. We mean by this that an experimenter should fix a relevant length scale $L$ and a timescale $T$, where $L$ is the typical variation scale of the classical potential and $T$ is taken to be the time a particle needs to `feel' variation in the potential. These are defined by
\begin{align}
 L &= \sqrt{\frac{\partial_x V}{\partial_x^3 V}}, & T = \frac{L}{v}, && v = \frac{\hbar}{m \lambda}.
\end{align}
We should then look at the rescaled variables
\begin{align}\label{rescaling}
 y(t') &= \frac{x(t')}{L}, & t' = \frac{t}{T}.
\end{align}
It is not unreasonable to look at these new variables as the physical ones, since in most relevant experiments we do not have access to the (typically small) length and time scales on which the quantum particles evolve. The relevant quantity that indicates the importance of a quantum treatment of the problem is given by
\begin{align}
 \epsilon = \frac{\lambda}{L}.
\end{align}
Consider now the force acting on the macroscopic variables, due to the quantum force $F_q$
\begin{align}
 D = \frac{T^2}{L}F_q(yL, t'T).
\end{align}
The conjecture in \cite{AlloriPHD} can then be stated as
\begin{equation}
 \lim_{\epsilon \rightarrow 0} D = 0 \text{ with probability }1.
\end{equation}
The quantity $D$ is a random variable, since the initial conditions on $y(t')$ are random variables with distribution given by the norm squared of the wavefunction $|\Psi|^2$ in the case of quantum equilibrium.
\par What does this conjecture promise? It asserts that when the quantum scale $\lambda$ is negligible in comparison with the classical scale $L$, `almost all' Bohmian trajectories will follow approximately classical trajectories. Here `almost all' means that the set of initial conditions $A$, which give rise to non-classical macroscopic trajectories, has a small probability measure, meaning that
\begin{equation}
 \int_A d^3x |\Psi|^2 \ll 1.
\end{equation}
\par Let us return to our example of the harmonic oscillator. We saw that the motion was far fom classical for energy eigenstates, the Bohmian particles did not move. This is of course only classical motion when $x(0) = 0$. The macroscopic trajectories are then given by
\begin{equation}\label{Conjecture}
 y(t') = x(0)/L.
\end{equation}
The de Broglie wavelength is bounded from below for the harmonic oscillator, so we take the limit $\epsilon \rightarrow 0$ to mean that $L \rightarrow \infty$. We thus imagine $L$ to be very big, and we only see non-classical behaviour for the modes for which
\begin{equation}
 x(0) > L.
\end{equation}
Since $L$ is very big, and the wavefunction is normalised, we expect that
\begin{equation}
 \int_{|x|>L} d^3x |\Psi(x, t)|^2 \ll 1.
\end{equation}
\par The scaling in eq.\ \eqref{rescaling} is thus very essential. In the first paragraph we argued that classical limit arises when the trajectories become classical, meaning that the quantum potential is negligible compared to the classical potential (or kinetic energy). This conjecture however guarantees a less strict definition of classical limit, meaning that only the macroscopic trajectories achieve classical trajectories. Classical trajectories for the variable $x$ imply classical trajectories for $y(t')$, but the converse is certainly not true, as made clear by our example of the harmonic oscillator above.
\par In the rest of the thesis, we will not use this different characterisation of the classical limit. We will stick to our first definition for two reasons. The first one is very straightforward: the conjecture is not proven yet, and even `not precisely stated yet\cite{SevenSteps}'. Questions include the exact nature of the limit in eq.\ \eqref{Conjecture}. We will not go deeper into that instead referring the reader to \cite{AlloriPHD}, appendix B, for a summary of possible ways to interpret this limit. The second reason is that we do not want the arguments for the classicality of primordial perturbations to depend on any sort of observer. Since these perturbations couple to classical density perturbations (eq.\ \eqref{QDensity}) long before any observation by a human experimenter took place, we would like to understand this classicality independent from any human observer choosing a relevant length scale $L$.
\section{Examples}
\paragraph{The Free Particle}Let us apply the formalism of pilot-wave theory to our first example of chapter \ref{QtoC}, the quantum free particle. The energy eigenfunctions are the plane waves and the relevant normalisable states are the Gaussian wavepackets from eq.\ \eqref{wavefunction}. The wavefunctions in position space can be obtained by inverting the Fourier transform. The algebra is quite involved, but one can integrate the guidance equations for these type of wavefunctions exactly\cite{AlloriPHD}.
\begin{equation}
 x(t) = p_0 t + x(0) \sqrt{1 + \left(\frac{2\sigma^2 t}{m}\right)^2}.
\end{equation}
This reduces to the classical equations of motion on timescales $t$ for which holds that
\begin{equation}
 t \ll \frac{m}{\sigma^2},
\end{equation}
This is exactly the same criterion we found in section \ref{FreeParticle}.
\paragraph{Inverted Harmonic Oscillator}
We will now turn to the inverted oscillator. The relevant $N$-particle wavefunctions are given by eq.\ \eqref{WaveFunctionAppendix}\footnote{We will again restrict to the normalisable solutions of the Schr\"odinger equation.}. For an $N$-particle state we can explicitly integrate the Bohmian trajectories.
\begin{equation}
 x(t) = x(0)\frac{|f(t)|}{|f(0)|}
\end{equation}
For our choice of $f$, see eq.\ \eqref{ConcreteF}, we see that at $|f(t)| \sim \exp(\omega t)$ for late times $t$. The trajectories thus become classical in the late time limit. The same will happen when we discuss the primordial perturbations.
\par Another way to see is to compute the quantum force. It is given by
\begin{equation}
 F_q = \frac{x}{2|f|^4}.
\end{equation}
The classical force is given by $\omega x$ and the ratio of both forces is then
\begin{equation}
 \frac{F_q}{F_c} = \frac{\omega}{2|f|^4}.
\end{equation}
For our choice of $f$ this ratio will become exponentially smaller with time, so we can safely neglect the quantum force at large times. %
\par Pilot-wave theory thus very clearly argues that the classical limit is reached for late times.

\chapter{Classicality for Cosmological Perturbations: Pilot-Wave Theory}\label{PilotWavePert}In this chapter we will use the techniques developed in the previous chapter to formulate the pilot-wave theory for cosmological perturbations. We will first solve the Bohmian equations of motion for cosmological perturbations for the simplest cases: the $N$-particle states. We will see that pilot-wave theory offers a very natural way of achieving the classical limit.
\par An attractive feature of the standard story, presented in chapter \ref{Observation}, was that the predictions for the power spectrum were roughly the same for initial states that didn't differ too much from the Bunch-Davies vacuum. We will continue to show that for a large class of physically relevant initial states the Bohmian theory is also robust, meaning that the classical limit is always reached. This will be done by proving a theorem and then generalising it to more initial states. This will result in theorem \ref{theorem} and \ref{theorem2}. A final generalisation can be found in appendix \ref{NonProductApp}, but does not offer much more insight. In chapter \ref{Quantifying} we will investigate exactly how this classicality occurs for de Sitter and power-law inflation by presenting the results of some simulations.
\section{Bohmian Trajectories for $N$-particle States}\label{Nparticle}
We will extend the calculations done in \cite{PhysRevD.85.083506}. There, the Bohmian trajectories were calculated for the de Bunch-Davies vacuum, while we will start by considering $N$-particle states.
\paragraph{Guidance Equations}
We repeat the Hamiltonian for a Fourier mode of the Mukhanov-Sasaki variable $\nu$ that we will use
\begin{equation}
\hat{H} = \int_{\mathbb{R}^{+}} d^3k \left[ \hat{\Pi}_{\vec{k}}\hat{\Pi}_{\vec{k}}^{*} + \left(k^2 - \frac{z''}{z}\right)\hat{\nu}_{\vec{k}}^* \hat{\nu}_{\vec{k}}\right]
\end{equation}
Since $\nu$ is a real field, $\nu_{-\vec{k}} = - \nu^{*}_{\vec{k}}$. Thus we take the integral over half the momentum space and we dropped the factor $\frac{1}{2}$ present in the Hamiltonian in chapter \ref{Perturbations}.
We will develop the pilot-wave theory in terms of the real variables $\text{Re}(\nu_{\vec{k}})$ and $\text{Im}(\nu_{\vec{k}})$. We note
\begin{align}
 \nu_{+, \vec{k}} &= \frac{1}{2}\left( \nu_{\vec{k}} + \nu_{\vec{k}}^*\right), & \nu_{-, \vec{k}} &= \frac{1}{2i} \left( \nu_{\vec{k}} - \nu_{\vec{k}^*} \right).
\end{align}
The variables $\Pi_{+, \vec{k}}$ and $\Pi_{-, \vec{k}}$ are analogously defined. The Hamiltonian for these variables takes the form
\begin{equation}
 \hat{H} = \frac{1}{2}\int_{\mathbb{R}^{+}} d^3k \sum_{j = \pm} \left[ \hat{\Pi}_{j,\vec{k}}^2 + \left( k^2 - \frac{z''}{z}\right) \nu_{j, \vec{k}}^2 \right].
\end{equation}
Here the integral is again over half momentum space. We will suppose, as is often done, that the system is initially in a product state. By this we mean
\begin{equation}
 |\Psi(\eta_{ini})\rangle = \prod_{\vec{k}} \prod_{j = \pm }|\Psi_{j, \vec{k}}(\eta_{ini})\rangle,
\end{equation}
where each $|\Psi_{\pm, \vec{k}}\rangle$ is a state in the Hilbert space of mode $\nu_{\pm,\vec{k}}$. Since the Hamiltonian does not couple the different variables, time evolution will preserve this product form at later times.
We will now turn to the functional Schr\"odinger picture. The wavefunction of the Mukhanov-Sasaki variable will then factorise into wavefunctions for each mode separately. %
The wavefunction of an $N$-particle state for a Fourier mode $\nu_{\pm,\vec{k}}$ (see appendix \ref{WFAppendix}) is given by
\begin{equation}\label{Nstate}
 \Psi_{N, \vec{k}}(\nu_{\pm,\vec{k}}, \eta) = \sqrt{\frac{1}{2^N N!}}\left(\frac{1}{ 2 \pi|f_k|^2}\right)^{1/4} H_{N}\left(\frac{\nu_{\pm, \vec{k}}}{\sqrt{2}|f_k|}\right)\left(\frac{f_k}{f_k^{*}}\right)^{n+ 1/2}\exp\left[ \frac{i}{2}\left(\frac{f_k'^{*}}{f_k^{*}} \right)\nu_{\pm, \vec{k}}^2\right],
\end{equation}
where $f_k$ is a solution to the classical equation of motion, satisfying
\begin{align}
 f_k''(\eta) - \left(k^2 - \frac{z''}{z}\right) f_k(\eta) &= 0 , & f'_k(\eta_{ini}) &= -i\sqrt{\frac{k}{2}} e^{-i k \eta_{ini}} ,\\
 i(f_k' f^{*}_{k} - f'^{*}_{k} f_{k}) &= 1, & f_k(\eta_{ini}) &= \frac{1}{\sqrt{2k}} e^{-i k \eta_{ini}}.
\end{align}
All these equations were introduced in chapter \ref{Perturbations} and are repeated for the convenience of the reader. We now split the wavefunction of a single variable in its modulus and its phase, $\Psi_{\pm, \vec{k}} = R_{\pm, \vec{k}}\exp (i S_{\pm, \vec{k}})$. The real and imaginary part of the functional Schr\"odinger equation for the Hamiltonian of a single variable $H_{\pm, \vec{k}}$ then decouple in the following equations
\begin{align}\label{QHamJac}
 \frac{\partial S_{\pm, \vec{k}}}{\partial \eta} &= - \frac{1}{2}\left(\frac{\partial S_{\pm, \vec{k}}}{\partial \nu_{\pm, \vec{k}}}\right)^2 - \frac{1}{2}\left(k^2 - \frac{z''}{z}\right)\nu_{\pm,\vec{k}}^2 - \frac{1}{2R_{\pm, \vec{k}}} \frac{\partial^2 R_{\pm, \vec{k}}}{\partial \nu_{\pm,\vec{k}}^2},\\
\frac{\partial R_{\pm, \vec{k}}^2}{\partial \eta} &= - \frac{\partial}{\partial \nu_{\pm, \vec{k}}} \left( R_{\pm, \vec{k}}^2 \frac{\partial S_{\pm, \vec{k}}}{\partial \nu_{\pm,\vec{k}}}\right).
\end{align}
We will now interpret the two equations in the same way we did in chapter \ref{PilotWave}. The second equation can be seen as a continuity equation for a probability current $\frac{\partial S_{\pm, \vec{k}}}{\partial \nu_{\pm,\vec{k}}}$. We can now postulate an actual field configuration $\nu$, whose modes obey guidance equations
\begin{equation}
 \nu_{\pm, \vec{k}}' = \frac{\partial S_{\pm, \vec{k}}}{\partial \nu_{\pm, \vec{k}}}.
\end{equation}
These guidance equations then make interpretation of eq.\ \eqref{QHamJac} possible. It is formally very reminiscent of the Hamilton-Jacobi equation of the classical theory, with an added quantum potential term.
\paragraph{Trajectories}
 We proceed now to solve the guidance equations for some initial condition $\hat{x}(\eta_{ini})$. We can compute the phase $S$ from eq.\ \eqref{Nstate}.
\begin{align}
 \nu_{\pm, \vec{k}}'%
&= \text{Im} \left(\Psi_n^{-1}(\nu_{\pm, \vec{k}},\eta) \frac{\partial}{\partial \nu_{\pm, \vec{k}}} \Psi_n(\nu_{\pm, \vec{k}}, \eta)\right)
= \frac{|f_k|'}{|f_k|}\nu_{\pm, \vec{k}}
\end{align}
This equation of motion is solved by separation of variables. The solution is given by
\begin{equation}
 \nu_{\pm, \vec{k}}(\eta) = \sqrt{2k}\nu_{\pm, \vec{k}}(\eta_{ini})|f_k(\eta)|.
\end{equation}
We notice first that this result is independent of the occupation number $N$. This is not surprising, since this is also the case for the pilot-wave theories of the harmonic and inverted oscillator.
\par Secondly, we identify two regimes for these trajectories. When $|k\eta|$ is large, $f_k$ will essentially behave as a plane wave, see for example eq.\ \eqref{P1Solution} for perfect de Sitter inflation. The $N$-states are then harmonic oscillator states and the particles will stand still. This is far from classical movement, since the classical equations of motion are the equations of the harmonic oscillator.
When $|k\eta|$ is small, $f_k$ will behave as $z(\eta)$. More precisely
\begin{equation}\label{Flate}
 f_k(\eta) \approx C(k)z(\eta),
\end{equation}
where the details of $C(k)$ are dependent on the exact model of inflation under consideration. Note that $f_k$ is always a solution of the classical equations of motion. Because of eq.\ \eqref{Flate}, we see that $|f_k|$ will approximately become a solution to the classical equations of motion too. Thus, for $|k\eta|$ very small, our modes $\nu_{\pm, \vec{k}}$ will approximately obey the classical equations of motion.
\par Let us try to understand what happens a little better. The quantum potential for the state $\Psi_n(x,t)$ is
\begin{align}
 V_q &= -\frac{1}{2R}\frac{\partial^2 R}{\partial \nu_{\pm, \vec{k}}^2} = - \frac{x^2}{4|f|^4} + \frac{2n - |f|}{2|f|^3},
\end{align}
where we have used the recursion relation for the Hermite polynomials eq.\ \eqref{Hermite}. The force associated with this potential, in the Hamilton-Jacobi sense, is
\begin{align}
 F_q &= \frac{\nu_{\pm, \vec{k}}}{2|f|^4}.
\end{align}
We conclude that the ratio of the classical to the quantum force is
\begin{align}
 \frac{F_c}{F_q} &= -4|f|^4\left(k^2 - \frac{z''}{z}\right).
\end{align}
Note that when $|k\eta|$ is big, $|f_k|$ will be approximately be equal to its initial value, $(2k)^{-1/2}$. Thus, at the onset of inflation, when the modes behave as if on Minkowski space, the ratio of both forces is approximately $-1$. The quantum force exactly cancels the classical force, and the movement is very far from classical. On the other hand, after the Hubble exit when $|k\eta|$ is small, the quantum force falls as $|f_k^{-4}|$ while the classical force rises as $\frac{z''}{z}$. The quantum force thus becomes completely negligible in the late times limit, leaving the classical force to generate classical trajectories for the Bohmian particles!
\paragraph{Classical Limit}
Let us formulate what this means for the classical limit. We now have a theory in our hands that unambiguously speaks of definite field configurations and trajectories. In addition, this theory clearly demonstrates that these trajectories become `classical' for late times. `Classical' is a very well defined concept, meaning that the trajectories predicted by pilot-wave theory will become indistinguishable from solutions to the classical equations of motion of our theory. This theory thus answers our second question from section \ref{PROBLEM}.
\par
In addition, this theory provides an easy way to break the initial symmetry of the problem. While the wavefunction is symmetric, the initial condition $\nu_{\pm, \vec{k}}(\eta_{ini})$ clearly is not, and as a consequence the trajectory $\nu_{\pm, \vec{k}}(\eta)$ is no longer symmetric. Note that this is exactly analogous to classical symmetry breaking: while the laws of motion can exhibit a symmetry, individual solutions in general do not possess this symmetry.

\section{General Classicality of Trajectories}\label{Proof}
The computation of the power spectrum of the cosmological perturbations was rather robust, as noted in section \ref{Spectra}. With this we mean that corrections to the power spectrum caused by deviations from the Bunch-Davies vacuum all carry a factor $a^{-2}$. These corrections all vanish in the late times-limit. In addition, the mechanism of squeezing, as presented in chapter \ref{Squeezing}, does not distinguish much between different initial states. This section will be devoted to the proof of the analogon for pilot-wave theory.
\par We showed in the previous section that $N$-particle states give rise to classical trajectories in the late times limit. Now the question arises what kind of states also exhibit this classical behavior. We will begin by formulating an exact mathematical theorem and proceed to the proof. The consequences of this theorem will de discussed in section \ref{Explanation}.
\begin{Theorem}\label{theorem}
Let the system be in an initial state
$$|\Psi(\eta_{ini})\rangle = \prod_{\vec{k}} \prod_{j = \pm }|\Psi_{j, \vec{k}}(\eta_{ini})\rangle$$
with
$$|\Psi_{j, \vec{k}}(\eta_{ini})\rangle = \sum_{i = 0 }^{N} \alpha_i | N_i\rangle.$$
with $N \in \mathbb{N}$ and $\sum_i |\alpha_i|^2 = 1$.
Let $\nu_{\pm, \vec{k}}(\eta)$ be a solution to the guidance equation for a the real or imaginary part of a certain mode of wavevector $\vec{k}$, on the interval $I = [\eta_{ini} ,0)$. Then
$$ \lim_{\eta \rightarrow 0} \frac{\nu_{\vec{k}}(\eta)}{|f_k|}$$
exists and is finite.
\end{Theorem}
\clearpage
\begin{bewijs}
 To prove the theorem, we fix our attention to a mode $\nu_{\pm, \vec{k}}$. The wavefunction for this mode is of the form
\begin{align}
\Psi_{\pm, \vec{k}_0} &= \left(\frac{1}{2 \pi|f_k|^2}\right)^{1/4}P(\nu_{\pm, \vec{k}}, \eta)\exp\left[ \frac{i}{2}\left(\frac{f_k'^{*}}{f_k^{*}} \right)\nu_{\pm, \vec{k}}^2\right]\\\label{PDef}
P(\nu_{\pm, \vec{k}_0}, \eta) &= \sum_{i = 0}^{N} \frac{1}{\sqrt{2^n n!}}\alpha_i \left(\frac{f}{f^*}\right)^{(n_i+ 1/2)} H_{n_i}\left(\frac{\nu_{\pm, \vec{k}}}{\sqrt{2|f|^2}}\right)
\end{align}
It will be easier to change variables
\begin{align}
 q = \frac{\nu_{\pm, \vec{k}}}{\sqrt{2}|f_k|}, & & S_k = -i \ln \left(\frac{f_k}{|f_k|}\right).
\end{align}
Take $\epsilon >0$ arbitrary. The theorem is proven if we can find (with a suggestive notation) $q(0)$ and $\eta_l$ so that for all $\eta > \eta_l$ holds that $|q(\eta) - q(0)| < \epsilon$. We will first construct $\eta_n$, $n \in \mathbb{N}$ in such a way that the values $q(\eta_n)$ form a Cauchy sequence, while we note the limit of that sequence to be $q(0)$. Then we will show that a certain element of $(\eta_n)_{n\in \mathbb{N}}$ can fulfill the role of $\eta_l$.
\par The velocity field of $q$ becomes
\begin{align}
 q' &=\frac{1}{\sqrt{2}|f_k|}\frac{d \nu_{\pm, \vec{k}_0}}{d \eta} - \frac{|f_k|'}{\sqrt{2}|f_k|^2}\nu_{\pm, \vec{k}_0},\\
&= \frac{1}{\sqrt{2}|f|^2}\text{Im}\left(P(q, \eta)^{-1} \frac{\partial P(q, \eta)}{\partial q}\right),
\end{align}
where the notation $P(q, \eta)$ is possible since $P$ only depends on the combination $|f_k|^{-1}\nu_{\pm \vec{k}_0}$. Now $S_f$ has a finite limit for $\eta \rightarrow 0$, see the proof in appendix \ref{AppLemma}. We will use lemma \ref{Begrensdheid} below to continuously extend this velocity field to all of $\mathbb{R}$ and take $C \in \mathbb{R}^{+}$ so that for all $q \in \mathbb{R}$ and $\eta \in [\eta_{ini}, 0)$ holds
\begin{align}
 \frac{1}{\sqrt{2}}\left|\text{Im}\left(P(q, \eta)^{-1} \frac{\partial P(q, S_k)}{\partial q}\right)\right| < C.
\end{align}
Due to the lemma in appendix \ref{AppLemma}, $|f_k|$ is strictly increasing and unbounded on $I$. For this reason, we can take for each $n \in \mathbb{N}$, $\eta_{n'}$ so that $|f_k(\eta)| > \sqrt{nC}$. Let $\eta_n = \max \{\eta_{n'}, -1\}$. We will now prove that $q(\eta_n)$ is a Cauchy sequence. We retain from these considerations
\begin{equation}
 -\frac{C}{|f_k|^2} \leq q' \leq \frac{C}{|f_k|^2}.
\end{equation}
Integrating gives us, for $m > n$,
\begin{align}
 |q(\eta_m) - q(\eta_n)| \leq \int_{\eta_1}^{\eta_m}\frac{C}{|f_k|^2}d\eta'\leq \frac{ |\eta_m - \eta_1|}{n}\leq \frac{1}{n}.
\end{align}
And thus $q(\eta_n)$ is a Cauchy sequence. For this reason $\lim_{n \rightarrow \infty}q(\eta_n)$ exists and is finite, denoted formally by $q(0)$.
\par We will now prove that $\lim_{\eta \rightarrow 0} q(\eta) = q(0)$. Take $l > (2\epsilon)^{-1}$, $l \in \mathbb{N}$. Then, for all $\eta > \eta_l$ we can take $\eta_s > \eta$. Then the following inequalities hold
\begin{align}
 |q(\eta) - q(0)| &\leq |q(\eta_s) - q(\eta_l)| + |q(\eta_l) - q(0)| \leq \frac{\epsilon}{2} + \frac{\epsilon}{2} = \epsilon.
\end{align}
\end{bewijs}

\begin{Lemma}\label{Begrensdheid}
 Let $P(q, \eta)$ be a polynomial in $q$ of degree $N$, with time dependent complex coefficients $\alpha_i(\eta)$. Define the following quantities:
\begin{align}
I &= [\eta_{ini}, 0),\\
A &= \{(q,\eta)| p \in \mathbb{R} ,\, \eta \in I,\, P(q,\eta) = 0 \},\\\label{G}
 G(q, \eta): \left(\mathbb{R} \times [\eta_{ini},0)\right)\setminus A \rightarrow \mathbb{R}&: (q, \eta) \rightarrow \left|\text{Im}\left(P(q, \eta)^{-1} \frac{\partial P(q, \eta)}{\partial q}\right)\right|.
\end{align}
Then $G(q,\eta)$ can be continuously extended to $A$. If $\lim_{\eta \rightarrow 0} \alpha_i(\eta)$ exists for all $i = 0,1,\ldots, N$, then $\sup \{G(q, \eta)| q \in \mathbb{R}, \eta \in [\eta_{ini}, 0)\}$ exists.
\end{Lemma}
\begin{bewijs}
We will first prove that $G(q, \eta)$ can be continuously extended to A. Pick $(q_0, \eta_0) \in A$. Then we can write
\begin{equation}
 P(q, \eta_0) = (q - q_0)^r F(q, \eta_0),
\end{equation}
where $F(q_0, \eta_0) \not = 0$ and $0 < r \leq N$. We now compute
\begin{align}\label{Ontbinding}
 \lim_{(q, \eta) \rightarrow (q_0, \eta_0)}G(q, \eta) = \lim_{(q, \eta) \rightarrow (q_0, \eta_0)} \left|\text{Im} \left( \frac{r(q-q_0)^{r-1}F(q, \eta) + (q-q_0)^r \partial_q F(q, \eta)}{(q-q_0)F(q, \eta)}\right)\right| = \left|\text{Im}\left(\frac{\partial_q F(q_0, \eta_0)}{F(q_0, \eta_0)}\right)\right|
\end{align}
And this limit is finite since $F(q_0, \eta_0) \not = 0$. We conclude that $G(q, \eta)$ can be continuously extended to $A$. Let us denote this extension by $\overline{G}(q, \eta)$. Suppose now that every coefficient $\alpha_i(\eta)$ has a finite limit for $\eta \rightarrow 0$. We will first prove that
 \begin{equation}
 \lim_{\eta \rightarrow 0}\overline{G}(q, \eta)
 \end{equation}
exists and is finite for all $q$. To this end, consider $\lim_{\eta \rightarrow 0} P(q, \eta)$. Since all the coefficients $\alpha_i(\eta)$ have finite limits for $\eta \rightarrow 0$, the limit of $P(q,\eta)$ also exists. Let us formally denote this by $P(q, 0)$. We then extend $\overline{G}(q, \eta)$ by (formally) noting:
\begin{align}
 \overline{G}(q, 0) = \lim_{\eta \rightarrow 0}\overline{G}(q, \eta) = \left|\text{Im}\left(P(q, 0)^{-1} \frac{\partial P(q, 0}{\partial q}\right)\right| & &\forall q\in \mathbb{R}: P(q,0)\not = 0
\end{align}
The analogon of eq.\ \eqref{Ontbinding} applies, and $\overline{G}(q,0)$ can be continuously extended to all of $\mathbb{R}$. We will abuse notation and let $\overline{G}(q,0)$ signify this extension.
\par We will now prove that $\overline{G}(q,\eta)$ is bounded. Then clearly $G(q, \eta)$ is bounded too. We fix a time $\eta_1$.
Note that $\overline{G}(q, \eta_1)$ is a rational function, whose denumerator is a polynomial of degree $N - 1$, while the denominator is a polynomial of degree $N$. Thus $\lim_{q \rightarrow \pm \infty} \overline{G}(q, \eta_1) = 0$. Supposing that $\overline{G}(q, \eta_1)$ is not zero everywhere, fix a $q_1$ so that $\overline{G}(q_1, \eta_1) \not = 0$. We can now find $q_2$, so that for every $|q| > |q_2|$ holds
\begin{equation}
|\overline{G}(q, \eta_1)| \leq |\overline{G}(q_1, \eta)|.
\end{equation}
Now the following relation follows
\begin{align}
 \sup\{\overline{G}(q, \eta_1)| q \in \mathbb{R} \} = \sup\{\overline{G}(q, \eta_1)| q \in [-|q_2|, |q_2|]\}.
\end{align}
Since $[-|q_2|, |q_2|]$ is compact and $\overline{G}(q, \eta_1)$ is continuous, this supremum is finite.
\par It is now possible to find such a supremum for every time $\eta$, denoted by $C(\eta)$. Since `taking the supremum' is a continuous function and $\overline{G}(q,\eta)$ is continuous, $C(\eta)$ is continuous as a composition of two continuous functions. Because we could extend $\overline{G}(x,y)$ to include $\eta = 0$, we can also define $C(0)$ in the logical way. We denote the extended version of $C(\eta)$ by $\overline{C}(\eta)$.\\
The proof is completed by observing the following (in)equalities,
\begin{align}
 \sup \{G(q, \eta)| q \in \mathbb{R}, \eta \in [\eta_{ini}, 0)\} &= \sup \{\overline{G}(q, \eta)| q \in \mathbb{R}, \eta \in [\eta_{ini}, 0)\},\\
&= \sup_{\eta \in [0,1)}\left(\sup_{q \in \mathbb{R}} \overline{G(q, \eta)}\right) ,\\
&= \sup_{\eta \in [0,1)} C(\eta),\\
&= \sup_{\eta \in [0,1]}\overline{C}(\eta),
\end{align}
which is finite because $\overline{C(\eta)}$ is continuous on the compact interval $[\eta_{ini}, 0]$.
\end{bewijs}

\section{Interpretation}\label{Explanation}
This section will be devoted to understanding the theorem in the previous section. In physical terms, what was proven is
\begin{quote}
 \textbf{If the initial state of the system is a product state and each mode $\nu_{\vec{k}}$ is in a superposition of a finite number of $N$-particle states, then all trajectories generated by the guidance equation become classical in the late time limit.}
\end{quote}
\paragraph{Classicality}
Let us start with the words `become classical in the late time limit'. The meaning of this is contained in the following expression
\begin{equation}
 \lim_{\eta \rightarrow 0} \frac{\nu_{\pm, \vec{k}}(\eta)}{|f_k(\eta)|} = \text{constant}.
\end{equation}
We conclude that $\nu_{\pm, \vec{k}}(\eta) \sim D |f_k|$ at the end of inflation, where $D$ is some constant. Since we already know that $|f_k|$ itself becomes an approximate solution of the classical equation of motion, we see that $\nu_{\pm, \vec{k}}(\eta)$ becomes an approximate solution to the classical equations of motion. Thus, we can think of $\nu_{\pm, \vec{k}}$ as behaving classically for late times.
\par We can understand more of what happens when we look at the quantum potential. In terms of the polynomial $P(\nu_{\pm, \vec{k}}, \eta)$, defined in eq.\ \eqref{PDef}, we get
\begin{equation}
V_q = \frac{-\nu_{\pm, \vec{k}}^2}{4|f_k|^4} - \frac{1}{2|f_k|^2}- \frac{1}{|P|^2}\text{Re}\left[ \frac{\nu_{\pm, \vec{k}}}{|f_k|^2}P\frac{\partial P^*}{\partial \nu_{\pm, \vec{k}}} - P^*\frac{\partial^2 P}{\partial \nu_{\pm, \vec{k}}^2} + \frac{1}{2|P|} \left(P \frac{\partial P^*}{\partial \nu_{\pm, \vec{k}}} - P^*\frac{\partial P}{\partial \nu_{\pm, \vec{k}}}\right)^2\right],
\end{equation}
where we have dropped the dependencies of $P$ to shorten notation. The first term is easily interpreted. It was also present for the pure $N$-particle state and we know it gives rise to a quantum force that is negligible compared to the classical force for late times. The second term is a time dependent constant that does not give rise to any force. The term in square brackets however, can be a very complex function of $\nu_{\pm, \vec{k}}$ and time. It constitutes a very important part of the total potential for $|k\eta|$ large and it results in motion that is in general not classical. Rewriting in terms of the rescaled variable $q = \left(\sqrt{2}|f_k|\right)^{-1}\nu_{\pm, \vec{k}}$ gives
\begin{equation}
 V_q = \frac{-q^2}{2|f_k|^2} - \frac{1}{2|f_k|^2}- \frac{2}{|f_k|^2|P|^3}\text{Re}\left[ q|P|P\frac{\partial P^*}{\partial q} - P^*|P|\frac{\partial^2 P}{\partial q^2} + \frac{1}{2} \left(P \frac{\partial P^*}{\partial q} - P^*\frac{\partial P}{\partial q}\right)^2 + \text{h.c.}\right].
\end{equation}
Here $P$ is now understood to be a function of $q$ and $S_f$. Under the conditions of lemma \ref{Begrensdheid}, the phase of $P$ as a polynomial in $q$ doesn't vary too fast. The terms inside square brackets then forms a polynomial in $q$ which doesn't vary too much in time. As the factor $|f_k|^2$ becomes large, the polynomial in square brackets can't compensate. Any contribution to the quantum force will be accompanied by a factor $|f_k|^{-2}$, making sure that this part of the quantum force will be negligible compared to the classical force, which grows with a factor $\frac{z''}{z}$.
\par As a side remark, the above argument is only sufficient to sketch the picture. In itself it is not sufficient to explain the late-time classicality of the trajectories. The exact demonstration is done through the proof of theorem \ref{theorem}. For example, the quantum potential blows up for zeros of $|P|$. In contrast to the velocity field, this quantum potential will not necessarily be bounded. However, this is no problem. The zeros of $P$ correspond to zeros of the wavefunction and thus particles have zero probability of being at this nodes, under the assumption of quantum equilibrium.
\par Let us also try a dimensional analysis of what happens, analogous to the reasoning in section \ref{Planck}. We have noted in section \ref{Quantisation} that quantising the Mukhanov-Sasaki variable is equivalent to quantising a scalar field of time dependent mass $- \frac{z''}{z}$. As the effective mass becomes bigger in absolute value our trajectories become more classical. This corresponds to our intuition of `heavy' particles behaving classically. We can also look at the de Broglie wavelength for an energy eigen state of eigenvalue $E >0$\label{Note that these are NOT the particle number states $N$!}
\begin{equation}
 \lambda_{dB} = \frac{\hbar}{ \sqrt{2\left[E - \left(k^2 - \frac{z''}{z}\right)\nu_{\pm, \vec{k}}^* \nu_{\pm, \vec{k}}\right]}},
\end{equation}
in regions where the argument of the square root is positive. This wavelength is large when $k^2$ is large compared to $\frac{z''}{z}$ but rapidly decreases as $\frac{z''}{z}$ becomes the dominant term. We see that quantum effects are important when $|k\eta|\gg 1$ but the quantum scale rapidly decays when $|k\eta| \ll1$.

\paragraph{Finiteness Criterion}
What is the meaning of the condition `superposition of a finite number of $N$-particle states' in theorem \ref{theorem}? We will call this condition the `finiteness criterion'.
\par %
A physical way to realise condition this if we instate some upper boundary $N_{\max, \vec{k}}$ on the allowed occupation number for each mode. We could imagine a cut-off energy $E_l$ (with respect to the ground state) that, gives rise to such a maximum occupation number $N_{\max, \vec{k}}$. No admixtures of states above this treshold occupation number would be allowed. A logical choice for the energy scale would be the Planck energy $E_{Planck}$. Since we don't expect quantum theory to hold on that kind of energy scale, it makes sense to restrict ourselves to the lower energy states. In this way, the superpositions of a finite number of $N$-particle states are the only physically relevant states.\\
Where did we use this criterion? It was of crucial importance in the proof of lemma \ref{Begrensdheid}. The finiteness condition is needed in for two things. First, it ensures that there are no places where the velocity field $q'$ blows up in eq.\ \eqref{Ontbinding}. In effect, it states that the nodes of the velocity field (discussed in chapter \ref{PilotWave}) `behave well'. Second, it assures that the velocity field behaves well at infinity, by ensuring that the numerator of $G(q, \eta)$, eq.\ \eqref{G}, is a polynomial of lower degree than the denumerator. %

\paragraph{Product states}
The theorem presented above is valid for product states. While the proof presented is only valid for these kind of states, the condition is not necessary for classicality of the trajectories. In appendix \ref{NonProductApp} we present a further generalisation of this proof for non-product states. It does not offer more insight and has the burden of needing cumbersome notation, so we have chosen not to present it here. Also it is generally assumed in the literature that the Mukhanov-Sasaki variable started out in a product state.

\section{Getting Rid of the Finiteness Criterion}
There are states that do not satisfy the finiteness criterion, being a superposition of an infinite number of $N$-particle states. These wavefunctions $\Psi_{\infty, \vec{k}}$ have the following form
\begin{equation}
 \Psi_{\infty, \vec{k}}(\nu_{\pm, \vec{k}}, \eta) = \frac{1}{\sqrt{2|f_k|^2}}\exp\left[ \frac{i}{2}\left(\frac{f_k'^{*}}{f_k^{*}} \right)\nu_{\pm, \vec{k}}^2\right] \sum_{n = 0}^{\infty}\left[ \frac{1}{\sqrt{2^n n!}}\alpha_n \left(\frac{f}{f^*}\right)^{(n_i+ 1/2)} H_{n}\left(\frac{\nu_{\pm, \vec{k}}}{\sqrt{2|f|^2}}\right)\right].
\end{equation}
 Since the Hermite polynomials form a basis for the Hilbert space of functions with a suitable inner product (see eq.\ \eqref{HermFunctions}), we can construct almost any function at $\eta_{ini}$ by choosing the coefficients $\alpha_i$ appropriately.
Still, we want these infinite states to be normalisable. It follows that these infinite states can be approximated to arbitrary accuracy by a superposition of a finite number of $N$-particle states. The theorem \ref{theorem} would hold for every approximation. This means that the trajectories for each approximation would be classical. There is however no guarantee that the trajectories of $\Psi_{\infty, \vec{k}}$ would become classical eventually, even though the trajectories of every approximation become classical.
\par On the other hand, there are special states that do not satisfy the finiteness condition, but which do exhibit classical trajectories in the late time limit. An excellent and physically important example are the coherent states,inspired by the calculations in \cite{Ji:1997kq}. They are defined as the eigenstates of the annihilation operator. For an eigenvalue $\beta \in \mathbb{C}$ we get
\begin{equation}
 |\beta\rangle = \exp \left( \frac{-|\beta|^2}{2}\right) \sum_{n= 0}^{\infty} \frac{\beta^{n}}{n!}|n\rangle.
\end{equation}
Their wavefunctions are given by
\begin{equation}
 \Psi_{\beta, \vec{k}} = \left(\frac{1}{2\pi |f_k|^2}\right)^{1/4}\exp\left[ \frac{i}{2}\left(\frac{f_k'^{*}}{f_k^{*}} \right)\nu_{\pm, \vec{k}}^2\right] \exp \left[ - \frac{\beta^2}{2}\left(\frac{f_k}{f_k^{*}}\right) + \left(\frac{f_k}{f_k^{*}}\right)^{1/2}\frac{\beta \nu_{\pm,\vec{k}}}{\sqrt{2} |f_k|} \right],
\end{equation}
where we have used eq.\ \eqref{HermGenerating}. The guidance equation is exactly solvable for these states. The trajectories are
\begin{align}
\nu_{\pm,\vec{k}}(\eta) = \frac{\nu_{\pm,\vec{k}}(\eta_{ini})}{\sqrt{2k}}|f_k| &+ \frac{\beta}{\sqrt{2}}\int_{\eta_{ini}}^{\eta} d\eta' \frac{\sin(S_f)}{|f_k|},\\\label{Sf}
S_f &= -\frac{i}{2} \ln\left(\frac{f_k}{f_k^*}\right).
\end{align}
For late times, $\nu_{\pm,\vec{k}} \sim |f_k|$ because the integral is negligible compared to the first term. We thus recover the same classicality as the states that do satisfy the finiteness criterion.
\par The existence of states that do not satisfy the finiteness criterion but exhibit classical trajectories invites us to look for a generalisation of theorem \ref{theorem}. Careful inspection of the proof presented reveals that it is easy to prove the following generalisation.
\clearpage
\begin{Theorem}\label{theorem2}
Let the system be in an initial state
$$\Psi(\eta_{ini}) = \prod_{\vec{k}} \prod_{j = \pm } \Psi_{j, \vec{k}}(\eta_{ini}).$$
Define
$$q = \frac{\nu_{\pm, \vec{k}}}{\sqrt{2}|f_k|},$$
and
$$P_{\vec{k}}(q, \eta) = \Psi_{j, \vec{k}}(\eta) \exp \left[ \frac{-i}{2}\left(\frac{f_k'^{*}}{f_k^{*}} \right)|f|^2 q^2\right].$$
Let $\nu_{\vec{k}}(\eta)$ be a solution to the guidance equation for a certain mode of wavevector $\vec{k}$, on the interval $I = [\eta_{ini} ,0)$. Suppose
$$G(q, \eta) = \text{Im}\left(P^{-1}_{\vec{k}} (q, \eta) \frac{\partial P}{\partial q}(q, \eta) \right)$$
is bounded.
Then
$$ \lim_{\eta \rightarrow 0} \frac{\nu_{\vec{k}}(\eta)}{|f_k|}$$
exists and is finite.
\end{Theorem}
The proof is completely analogous to the proof of theorem \ref{theorem}. It is only different in the fact that it does not use lemma \ref{Begrensdheid}. We will not state the proof here.
\par This theorem is now strong enough to prove the classicality of the trajectories of the coherent states. The quantity $G(q,\eta)$ is for those states
\begin{equation}
 G(q,\eta) = \beta \frac{\sin(S_f)}{\sqrt{2}|f_k|},
\end{equation}
with $S_f$ as in eq.\ \eqref{Sf}. This is obviously bounded and the coherent states satisfy the conditions of theorem \ref{theorem2}.
\par It is very difficult to further identify states that satisfy the conditions of theorem \ref{theorem2}, due to the rather abstract mathematical conditions it poses. It should also be noted that we did not succeed in finding an explicit state that does not satisfy the conditions of theorem \ref{theorem2}. We hope that theorem \ref{theorem2} can be extended to include all possible product states of the Mukhanov-Sasaki variable. A viable possibility would maybe run along the following lines:
\begin{enumerate}[I]
 \item{An arbitrary initial wavefunction can be written as a (in)finite superposition of $N$-particle states, because the Hermite functions form a basis for the square integrable functions. So $|\Psi\rangle = \sum_i \alpha_i | N_i\rangle$}
\item{For the first $n$ terms in this sum, the quantity $G(q, \eta)$ is bounded, say by $C(n) \in \mathbb{R}$.}
\item{Normalisability of the wavefunction can be used to prove that $\lim_{n \to \infty}C(n)$ is finite.}\label{normstep}
\item{The quantity $G(q,\eta)$ is then bounded and $|\Psi\rangle$ satisfies the conditions of theorem \ref{theorem2}.}
\end{enumerate}
This sketch of a proof would work if one could find a way of implementing step \ref{normstep}. The idea of that step would be that the wavefunction would need to be normalisable, and that the `tail' of the sum over $|N_i\rangle$ states would not contribute much to $C(n_i)$ due to $|\alpha_i|$ being small in the `tail'.
However, we remind the reader that for all states of realistic interest, a finite superposition of states is sufficient. We do not claim that our theory holds once we reach energies on the Planck scale and so the accessible number of initial occupation numbers is limited.

\section{Power Spectra}\label{dBBSpectra}
We will now calculate the power spectrum for a slow-roll inflation theory in this pilot-wave formalism. Taking as initial state the Bunch-Davies vacuum and assuming quantum equilibrium, we get for the power spectrum of the rescaled Mukhanov-Sasaki variable
\begin{align}
 \frac{1}{a^2(\eta)}\langle \nu(\vec{x}, \eta)\nu(\vec{y}, \eta)\rangle &= \frac{1}{a^2}\int d\nu |\Psi(\nu, \eta)|^2 \nu(\vec{x}) \nu(\vec{y}) = \int \frac{dk}{k} \frac{\sin (k |\vec{x}-\vec{y}|)}{k|\vec{x} - \vec{y}|} \left(\frac{k^3 |f_k|^2}{2 \pi^2 a^2}\right)
\end{align}
Thus, for de Sitter or power-law inflation, we know that $a^{-1}f_k \sim k^{-3/2}$ for late times, and we regain the power spectrum for the rescaled Mukhanov-Sasaki variable as presented in section \ref{Spectra}, under the condition of quantum equilibrium. We will discuss this assumption in detail in chapter \ref{Test}.
\par Let us now look at the power spectrum of the density perturbations. We repeat the classical equation, eq.\ \eqref{ClassDens}, that couples the Fourier components of the density perturbations to the classical modes of the Mukhanov-Sasaki variable terms of the slow-roll potential $V(\varphi)$,
\begin{equation}
 \left( \frac{\delta \rho}{\rho}\right)_{\vec{k}} = \frac{\partial_{\varphi} V(\varphi)|_{\varphi_0}}{V(\phi_0)} a^{-1}\nu_{\vec{k}}.
\end{equation}
We have seen that for late times the quantum mechanical variable $\nu_{\vec{k}}$ will evolve in very good approximation according to the classical equation of motion. Thus we trust the above equation to couple the classical density perturbations to the (approximately) classical $\nu_{\vec{k}}$ at late times. Substituting our solution for the equations of motion we find
\begin{equation}
 \left( \frac{\delta \rho}{\rho}\right)_{\vec{k}} = \frac{\partial_{\varphi} V(\varphi)|_{\varphi_0}}{V(\phi_0)}\nu_{\vec{k}}(\eta_{ini})\sqrt{2k} a^{-1}|f_k|.
\end{equation}
At first sight, we get an equation that is very different from the one in section \ref{Spectra}: an extra factor of $k^{1/2}$ and the initial condition $\nu_{\vec{k}}(\eta_{ini})$ are introduced. We will explicitly show that these two contributions cancel. Under the assumption of quantum equilibrium we have the following equality
\begin{align}
 \left\langle \frac{\delta \rho}{\rho}(\vec{x}) \frac{\delta \rho}{\rho}(\vec{x} + \vec{y})\right\rangle = \frac{1}{2\pi^2}\left(\frac{\partial_{\varphi} V(\varphi)|_{\varphi_0}}{V(\phi_0)}\right)^2 \int dk \frac{\sin ky}{ky} \frac{2k|f_k|}{ a^{2}}\int d \nu_{\vec{k}}|\Psi(\nu_{\vec{k}}), \eta_{ini}|^2 |\nu_{\vec{k}}|^2.
\end{align}
The initial wavefunction is a Gaussian of variance $|f_k(\eta_{ini})| = (2k)^{-1/2}$. We get as final result
\begin{equation}
 \mathbb{P}(k) = \frac{1}{2\pi^2}\left(\frac{\partial_{\varphi} V(\varphi)|_{\varphi_0}}{V(\phi_0)}\right)^2\frac{|f_k|^2}{ a^{2}},
\end{equation}
which is the analogue of eq.\ \ref{PS}. As before, we now note that for de Sitter and power law inflation the ratio $a^{-1}|f_k|$ is approximately constant for late times and proportional to $k^{-3/2}$. We regain thus a scale invariant power spectrum, in agreement with observation.

\chapter{Quantifying Classicality}\label{Quantifying} In the previous chapter, we have proven that pilot-wave trajectories for a large class of initial states will always resemble the classical trajectories in the large time limit. In this chapter we will deepen our understanding of this phenomenon by presenting some simulations for de Sitter and power-law inflation. We will primarily determine how much these trajectories will deviate from the classical trajectories for a given amount of inflation and find that for realistic amounts of inflation this process is fast enough to create classicality for the relevant modes.
\par As a starting remark, note that theorem \ref{theorem} will be sufficient for the rest of this chapter. Computers can only handle a superposition of a finite number of $N$-particle states numerically. For any simulation executed, classicality will eventually set in due to this calculational constraint. In addition, we will only study the case for product states, since the case for entangled modes is computationally harder and expected to be analogous.
\section{Two Sources of Deviation from Classicality}\label{Sources}
We will fix a mode $\nu_{\pm, \vec{k}}$. We showed that the trajectories for finite superpositions will all resemble the classical ones in the late times limit. This was implemented by proving that for every trajectory
\begin{equation}
 \nu_{\pm, \vec{k}}(\eta) \approx D |f_k|,
\end{equation}
for some constant $D$, when $\eta \rightarrow 0$. There are two important sources here that make $\nu_{\pm,\vec{y}}(\eta)$ deviate from the classical trajectories. First, the trajectory of $\nu_{\pm, \vec{k}}(\eta)$ will deviate from $D|f_k(\eta)|$. Second, $|f_k(\eta)|$ will never solve the classical equation of motion exactly, only approximately.
We will start by addressing the first issue. In the proof of theorem \ref{theorem} we stated that the velocity field of the rescaled variable $q$ was
\begin{equation}
 |q'| \leq \frac{C}{|f_k|^2}.
\end{equation}
Integration gives us
\begin{equation}
\frac{|\nu_{\pm, \vec{k}}(\eta)|}{|f_k(\eta)|} \leq \left|\frac{\nu_{\pm, \vec{k}}(\eta_{ini})}{|f_k(\eta_{ini})|}\right| + \int_{\eta_{ini}}^{\eta} d\eta'\frac{C}{|f_k|^2}.
\end{equation}
The trajectory will not deviate much from $|f_k|$ time at time $\eta_{c}$ when
\begin{equation}\label{EpsilonC}
\epsilon_{C} = \int_{\eta_c}^{0}\frac{C}{|f_k|^2} \ll \left|\frac{\nu_{\pm, \vec{k}}(\eta_c)}{f_k(\eta_c)}\right|.
\end{equation}
This condition is always fulfilled in the case of pure $N$-particle states, since $C = 0$ in that case. For other initial states, this happens when $|f_k(\eta_c)|$ becomes very large. This is the case when the mode has exited the Hubble radius, i.e. when $\frac{z''}{z} = k^2$ and the classical potential switches sign. We will take the lefthand side of this equation as a measure of the deviation from classicality from this source.
\par When does $|f_k|$ become an approximate solution to the classical equations of motion? Note that $|f_k|$ satisfies the following equation
\begin{equation}
|f_k|'' = -\left(k^2 - \frac{z''}{z}\right)|f_k| + \frac{1}{2|f_k|^3}.
\end{equation}Since
The term with $|f_k|^{-3}$ will lose its importance when
\begin{equation}
\epsilon_f = \left| k^2 - \frac{z''}{z}\right|^{-1}|f_k|^{-4} \ll 2
\end{equation}
This happens soon after the Hubble exit of the mode as both $|f_k|$ and $\frac{z''}{z}$ increase rapidly. We will take the left hand side of this equation as a measure of the deviations from classicality due to this source $\epsilon_f$.
\par The conclusion of this section is thus that we expect the trajectories to be indistinguishable soon after the Hubble exit of the mode. To be able to say more, we have to specify a model of inflation, thereby giving us a formula for $\frac{z''}{z}$.

\section{De Sitter and Power-Law Inflation}
We will further focus on the special cases of de Sitter and power-law inflation. Both are slow-roll models and we can thus safely approximate $\frac{z''}{z}$ by $\frac{a''}{a}$ (see eq.\ \eqref{ZalsA}). The scale factor is then given by
\begin{equation}
 a \sim |\eta|^{-p},
\end{equation}
where $p=1$ corresponds to de Sitter inflation and $p>1$ corresponds to power-law inflation. So, in terms of the cosmic time $t$ the scale factor is
\begin{align}
a &= \exp \left(H t\right) && \text{If } p = 1,\\
a &\sim t^{\frac{p}{p-1}} && \text{If } p > 1.
\end{align}
Lucchin shows \cite{Lucchin} that this type of inflation is compatible with observations when $p = 1, \frac{p}{p-1} = 2$ or $\frac{p}{p-1} \geq 10$. Thus we get as set of possible values $1 \leq p < \frac{10}{9}$ or $p = 2$. These are the values we will use in simulations.
These types of inflation have the additional bonus that we can solve for the function $f_k$ exactly. We repeat eq.\ \eqref{exactF}
\begin{equation}
f_{k}(\eta) = \sqrt{\frac{\pi}{4k}}e^{- i k \eta_{i}}e^{\pi \left(\frac{1+p}{2}\right)}\sqrt{-k\eta}H^{(1)}_{p+ \frac{1}{2}}(-k\eta).
\end{equation}
\par Let us compare the two sources of deviations from classicality in the previous section for de Sitter and power-law inflation. Taking $a \sim \eta^{-p}$ we compute the ratio of the deviations some time after Hubble exit, when $|f_k| \sim a$
\begin{equation}
 \frac{\epsilon_C}{\epsilon_f} \sim C\eta^{-2p-1}.
\end{equation}
For $C \not = 0$ the dominant source for late times will always be $\epsilon_C$. In what follows, we will neglect $\epsilon_f$ as source of non-classicality.
\section{Technicalities of the Simulations}
Before we move on to the simulations themselves, we will briefly comment on some technical details.
\par The variable $\nu_{\pm,\vec{k}}$ has a velocity field that becomes extremely large for $\eta \rightarrow 0$. The differential equations for this variable present an extremely stiff problem for $\eta \rightarrow 0$ and have proven to be very hard to solve reliably. Instead we simulated the rescaled variable
\begin{equation}
 q = \frac{\nu_{\pm,\vec{k}}}{|f_k|}.
\end{equation}
The variable $q$, by virtue of lemma \ref{Begrensdheid}, has a bounded velocity field and represents a computational problem that is far better conditioned. Since the functions $|f_k|$ are exactly know for de Sitter and power-law inflation this change of variables does not entail any kind of extra uncertainty. In addition, graphs for this rescaled variable are far more clear, since trajectories of $\nu_{\vec{k}}$ will become unbounded in the $\eta \rightarrow 0$ limit.
\par The guidance equations for the trajectories were solved using a fourth-order Runge Kutta-method of adaptive step size, implemented by the module ode45 in Octave. Accuracy was checked by calculating the time-inverted trajectories, starting from their end position, and comparing the time-reversed end position with the initial condition. Errors were estimated using this method to be of order $10^{-6}$ to $10^{-8}$ for superpositions of around $4$ different quantum numbers up to $N \sim 20$. Higher accuracies, higher quantum numbers and superpositions of more different quantum numbers are achievable for increased computer time, but this was deemed unnecessary.
\par Concerning the subject of simulational parameters, it is important to realise that the choice of $\eta_{ini}$ is not irrelevant. Different choices of $\eta_{ini}$ will imply different initial conditions for $f_k$. In general, $\eta_{ini}$ was chosen in such a way that the mode under consideration starts out in very nearly Minkowski space $|k\eta| \gg 1$. Practically we chose the parameters so that at least $k \eta_{ini} > 10$ was guaranteed. In terms of the equations of motion, this guarantees that the term $\frac{z''}{z}$ is at least two orders of magnitude smaller than the $k^2$ term. Computationally, this places some restrictions on the parameters used. One doesn't want to choose $\eta_{ini}$ to be too big, since that would involve taking many time steps. Furthermore taking $k$ big results in problems in the Minkowski regime. In that regime, the system will essentially behave as a harmonic oscillator. The typical frequency of the system is determined by energy differences, proportional to $k$. High $k$ means high frequency which needs a lot of time steps. We thus need to strike a balance between high $k$ and early times $\eta_{ini}$.
\par To finalise this section, let us introduce the notation for the parameters of the simulation we will adopt. `Quantum numbers $[0, 1, 2]$' with `weights $[\alpha, \beta, \gamma]$' means that the initial state of the mode $\nu_{\vec{k}}$ was prepared in
\begin{equation}
 |\Psi_{\vec{k}}(\eta_{ini})\rangle = \left(|\alpha|^2 + |\gamma|^2 + |\delta|^2\right)^{-1}\left(\alpha |0\rangle + \beta |1\rangle + \gamma |2\rangle\right),
\end{equation}
where the numbers $\alpha, \beta, \gamma$ are in general complex and were often taken randomly. Initial non-product states were not simulated. Their behaviour is thought not to be very different from the product states, by virtue of the theorem in appendix \ref{NonProductApp}. To keep in contact with the physically relevant situation, $k$ was taken to be bigger than $H$. Relativistic units were chosen.
\clearpage
\section{Typical Trajectories}
\begin{figure}[h]
\begin{center}
 \includegraphics[scale = 0.3]{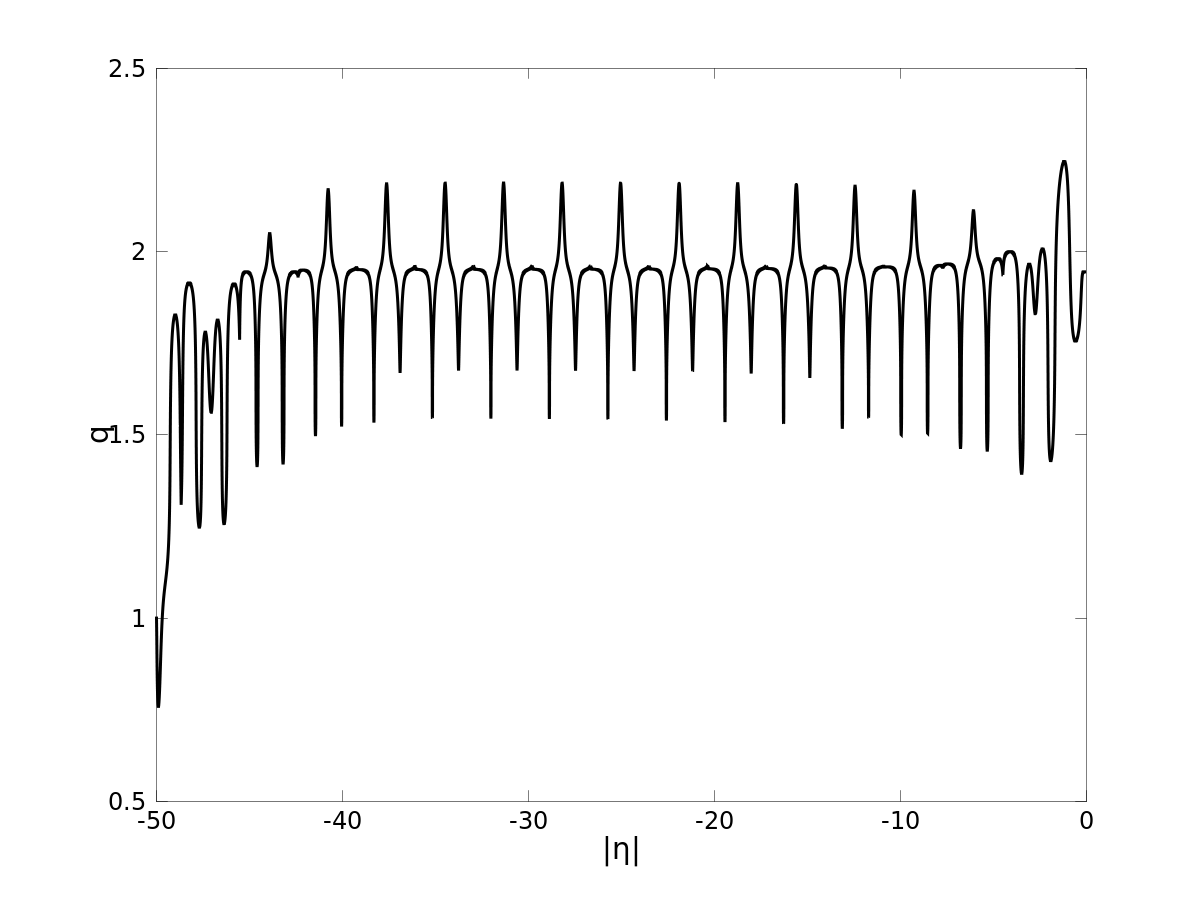}
 \caption{Superposition of quantum numbers $[1,6,9]$ with weights of equal amplitude but random phase. Parameters used were $H=0.01$, $k = 1$ and $q(\eta_{ini})= 1$ for de Sitter inflation.}
\label{EasyP1}
\end{center}
\end{figure}

We will begin by analysing some trajectories of the variable $q$. The case for pure $N$-particle states is not interesting, the velocity field of $q$ will be zero. We will take as first example a superposition of $[1,6,9]$ weights of equal amplitude but random phases. The trajectory of such a mode is shown in fig. \ref{EasyP1}. The general process is clear. $q$ starts off by moving in a seemingly oscillatory way, which corresponds to the field evolving on Minkowski vacuum. At that time $|f_k|$ is approximately and the variable $q$ is directly proportional to $\nu_{\pm, \vec{k}}$.
\par
\begin{figure}[p]
\begin{center}
 \includegraphics[scale = 0.3]{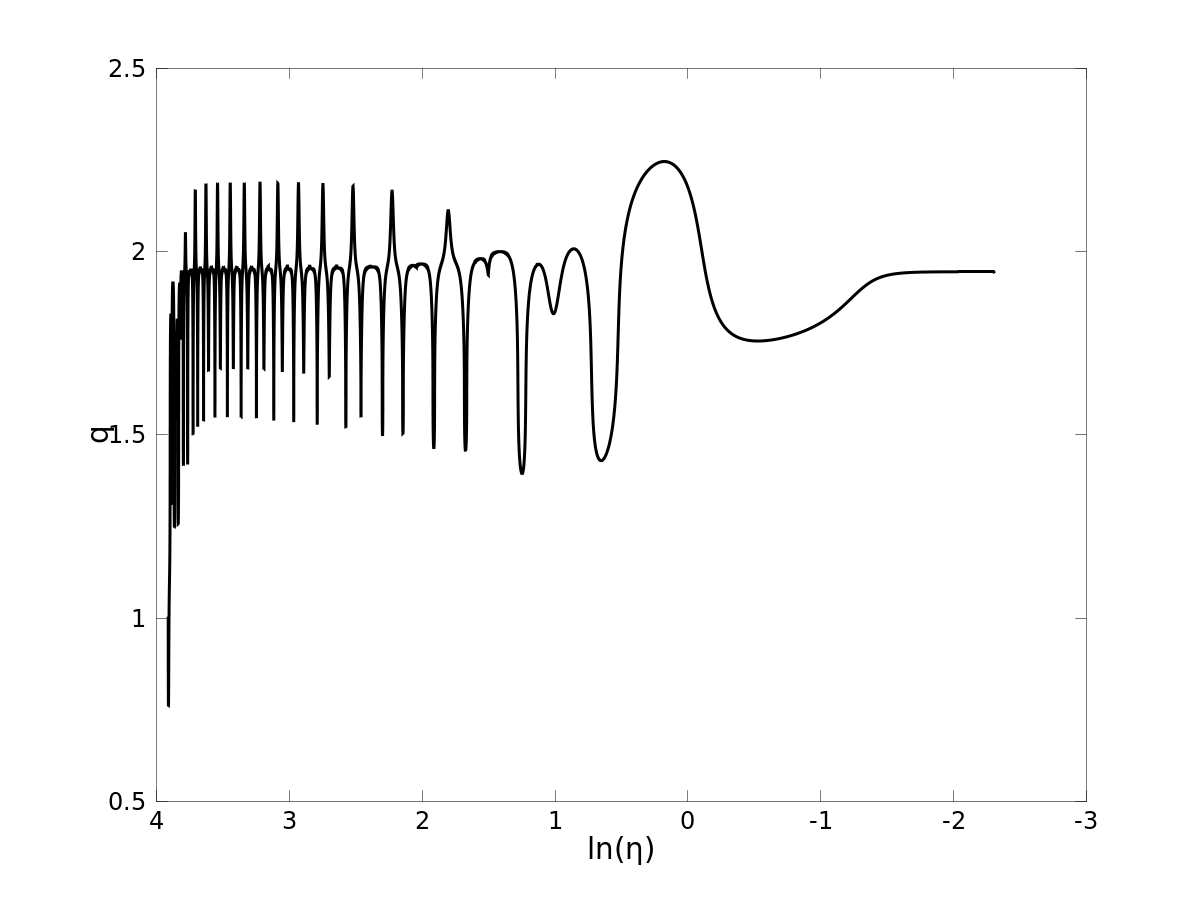}
 \caption{Superposition of quantum numbers $[1, 6,9]$ with weights of equal amplitude but random phase. Parameters used were $H=0.01$, $k = 1$ and $q(\eta_{ini})= 1$ for de Sitter inflation.}
\label{EasyP1log}
\end{center}
\end{figure}

 As $\frac{a''}{a}$ increases, the effective frequency of our oscillator drops and the oscillations become slower and slower. Their period rises and their amplitude gets damped as $|f_k|$ increases. In the limit of $k\eta \rightarrow 0$ the velocity field of $q$ becomes negligible and $q(\eta)$ settles to a constant value. This means that $\nu_{\vec{k}} \sim |f_k|$ in that limit, as was proven in theorem \ref{theorem}. This can be seen more clearly on fig. \ref{EasyP1log}, which describes the same simulation but with a logarithmic horizontal and restricted to later times. The picture is more clear here. One can see that the period of the oscillations increases and that the freezing of the mode clearly takes place at $\ln(|k\eta|) \approx 0$, meaning at Hubble exit $|k \eta| = 1$.
\par
\begin{figure}[p]
\begin{center}
 \includegraphics[scale = 0.3]{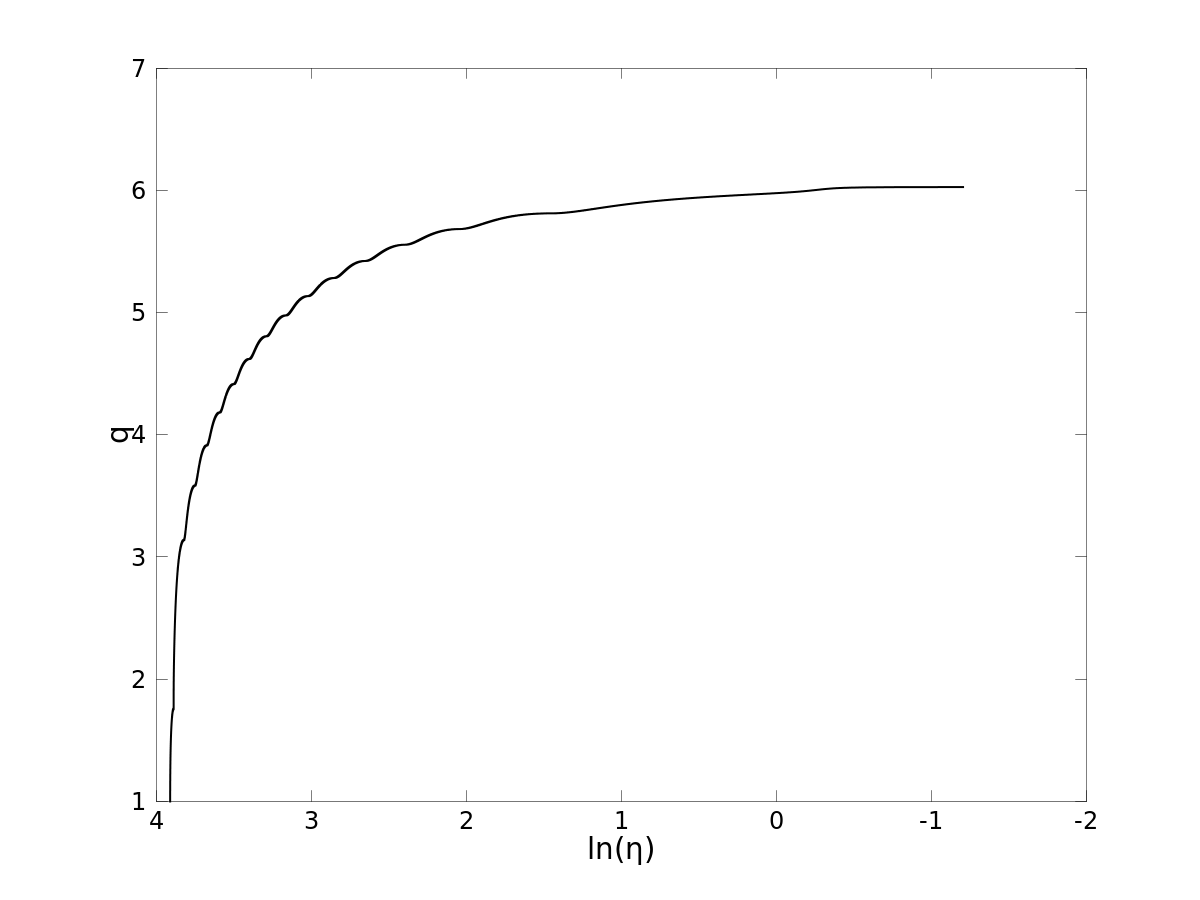}
 \caption{Superposition of quantum numbers $[2,3]$ with random weights of equal amplitude. Parameters used were $H=0.01$, $k = 1$ and $q(\eta_{ini})= 1$ for $p=2$ power-law inflation.}
\label{EasyP2}
\end{center}
\end{figure}

 It turns out that this behaviour is quite generic for different superpositions and different values of $p$. For example, in fig. \ref{EasyP2} a trajectory is plotted for $p=2$ power-law inflation. The behaviour of $q$ does not differ quantitatively from the previous figure. This type of trajectories was reliably confirmed for $p \in \{1, 10/9, 2\}$ and superpositions of up to $4$ different $N$ with $N$ between $0$ and $20$. Qualitatively the same behaviour was observed for $N$ up to $50$ with superpositions of up to $6$ different $N$. These results are not presented since the error estimates proved to be too big for systematic quantitative analysis.

\section{Freezing Time}
We will now come to a more systematic study of the time when $\nu_{\pm, \vec{k}}$ becomes classical. $\epsilon_f$ can be easily computed since we can solve for $f_k$ exactly. The only unknown quantity in section \ref{Sources} is the $C$. We will thus inquire about the moment when eq. \eqref{EpsilonC} holds.
\par The first problem to for doing this systematically is that eq.\ \eqref{EpsilonC} is a worst-case scenario, since $C|f_k|^{-2}$ is an upper boundary on the initial velocity field of the variable $q$. Given a certain initial state, one can in principle compute $C$. But it is very much possible that a trajectory $q(\eta)$ will have derivatives that are a lot smaller than $\frac{C}{|f_k|^2}$. Second, while one could compute $C$ for \emph{one} given initial state, it is very hard to do this more generally. Let's say we have an initial state
\begin{equation}
 |\Psi_{\pm, \vec{k}} \rangle = \sum_{i = 0}^{N_{max}}\alpha_i |n_i\rangle.
\end{equation}
$C$ will then be a very complex function of the $\alpha_i$ and $N_{max}$, since the wavefunctions of states $|n_i\rangle$ involve the Hermite polynomials and time dependent phases.                                                                        While we have tried to get an analytical estimate of the `average velocity' of $q$ along some trajectory, the two problems mentioned are the reasons we failed. It is here that simulations bring solace.
\par The method we used to extract information from the solutions is simple. We take the value of $q$ at the end time $\eta_f$ of our simulation, and then select the time $\eta_c$ so that for all $\eta > \eta_c$ we get that $|q(\eta) - q(\eta_f)| < \epsilon$. We effectively measured the time at which deviations from the limit value were smaller than $\epsilon$. The problem with this approach is that we have to choose a `hard' value of $\epsilon$. Taking $\epsilon$ to be too small will result in $\eta_c$ to be very susceptible to numerical errors, either because of numerical instabilities or because of our finite simulation time. Taking $\epsilon$ too large will place $\eta_c$ in the regime where the trajectories still oscillate.
\par
\begin{figure}[p]
\begin{center}
\includegraphics[scale = 0.3]{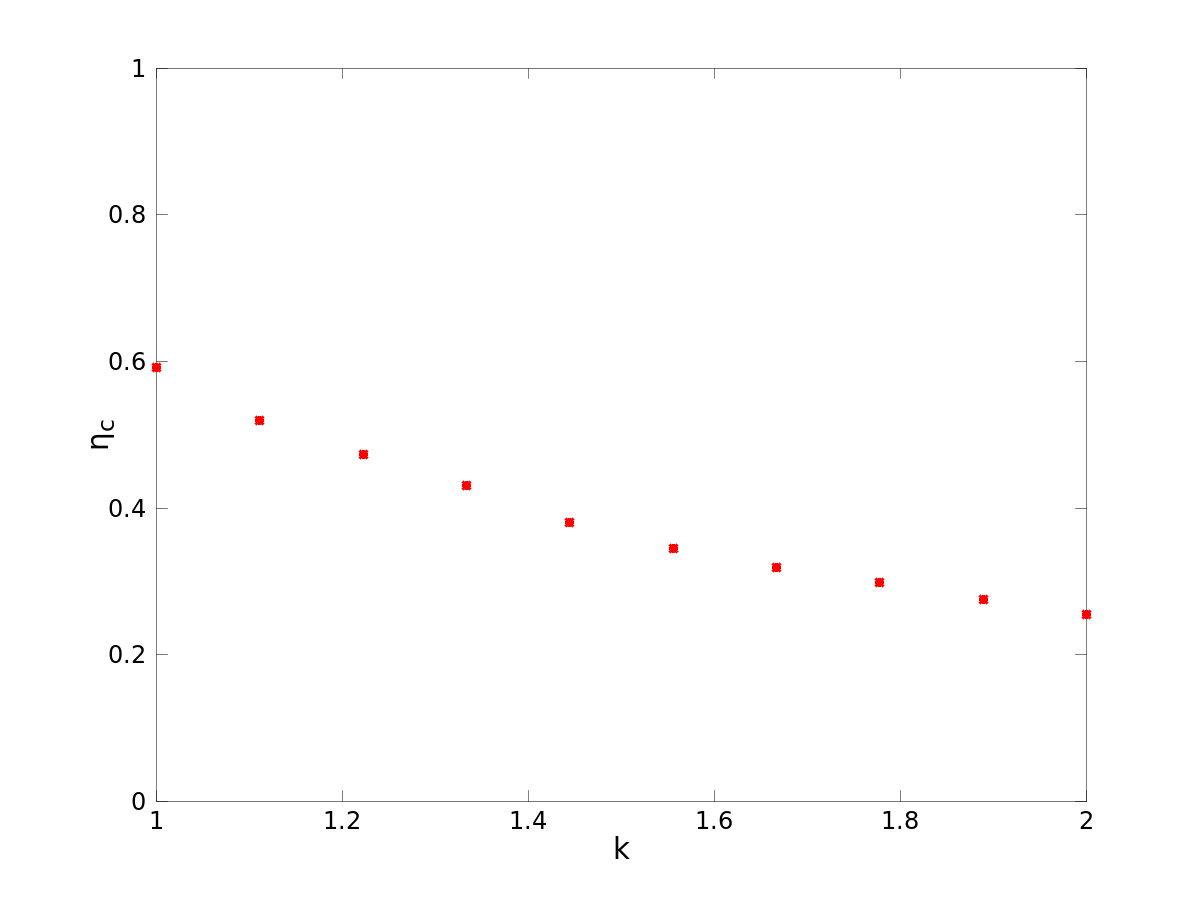}
\caption{The $k$ dependence of $\eta_c$ for $\epsilon = 10^{-3}$ and a superposition of quantum number $[0,1,2]$ with weights $[10, \exp(3i), \exp(i)]$, with $H = 0.01$, $\eta_{ini} = -50$, $q(\eta_{ini}) = 1$ and de Sitter inflation.}
\label{KDep1}
\end{center}
\end{figure}

\begin{figure}[p]
\begin{center}
\includegraphics[scale = 0.3]{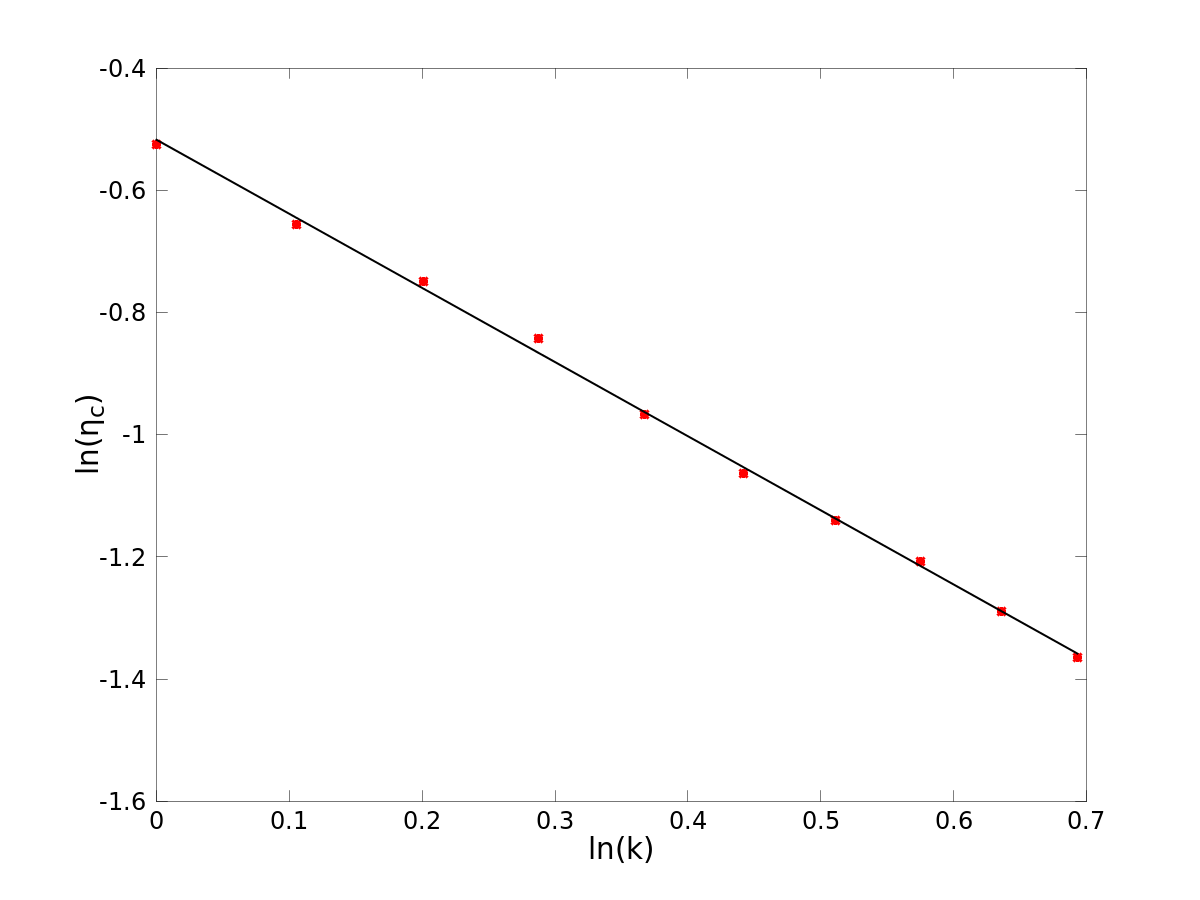}
\caption{Log-log plot of the $k$ dependence of $\eta_c$ for $\epsilon = 10^{-3}$ and a superposition of quantum number $[0,1,2]$ with weights $[10, \exp(3i), \exp(i)]$, with $H = 0.01$, $\eta_{ini} = -50$, $q(\eta_{ini}) = 1$ and de Sitter inflation. Red stars indicate simulation values, black line indicates first order polynomial fit.}
\label{KDep1Log}
\end{center}
\end{figure}

 A good value of $\epsilon$ would be between $10^{-2}$ and $10^{-4}$. Looking back at fig. \ref{EasyP1} we see that the typical scale on which is $q(\eta)$ oscillates is of order $1$ to $10^{-1}$. This scale of oscillations was generally confirmed, for example by fig. \ref{EasyP2} where we see qualitatively the same behaviour. In addition, a value for $\epsilon$ larger than $10^{-4}$ will make $\eta_c$ rather robust to numerical effects, since those are expected to be at most of order $10^{-6}$.
\paragraph{Dependence on $k$} Let us first look at the $k$-dependence of $\eta_c$. We can see the dependence clearly in fig. \ref{KDep1} for a simple superposition. A log-log version of this plot is shown in fig. \ref{KDep1Log}, together with a first order fit. On that figure it is even more clear that $\eta_c \sim \frac{1}{k}$ is a constant. A first order polynomial fit in the least squares sense gives
\begin{equation}
 \log(|\eta_c|) = \left(-1.21 \pm 0.001\right) \log(k) - 0.5\pm 0.01.
\end{equation}
We can safely assume that $\eta_c \approx k^{-1}$. %
\clearpage
\section{Is Classicality reached?}
While the dependence on other parameters was not very clear, we demonstrated in the previous section that $|k\eta_c|$ was approximately a constant for a given $\epsilon$. Let us use this information to make an estimate of how much realistic modes can deviate from the classical trajectories.
\par Let us go back to eq.\ \eqref{EpsilonC}. For $a \sim \eta^{-p}$, $\epsilon_C \sim \eta^{2p+1}$. Of course, eq.\ \eqref{EpsilonC} was a worst case scenario. We can however expect some variant of eq.\ \eqref{EpsilonC} to hold, meaning that we will replace the maximum of the velocity field $C$ by some suitable constant $A$ that could be thought of as representing the average velocity of $q$. We thus look for a dependence
\begin{equation}\label{Gedrag}
 \epsilon_C \sim A\eta^{2p+1}.
\end{equation}
long after the Hubble exit.
\par
\begin{figure}[p]
\begin{center}
\includegraphics[scale = 0.3]{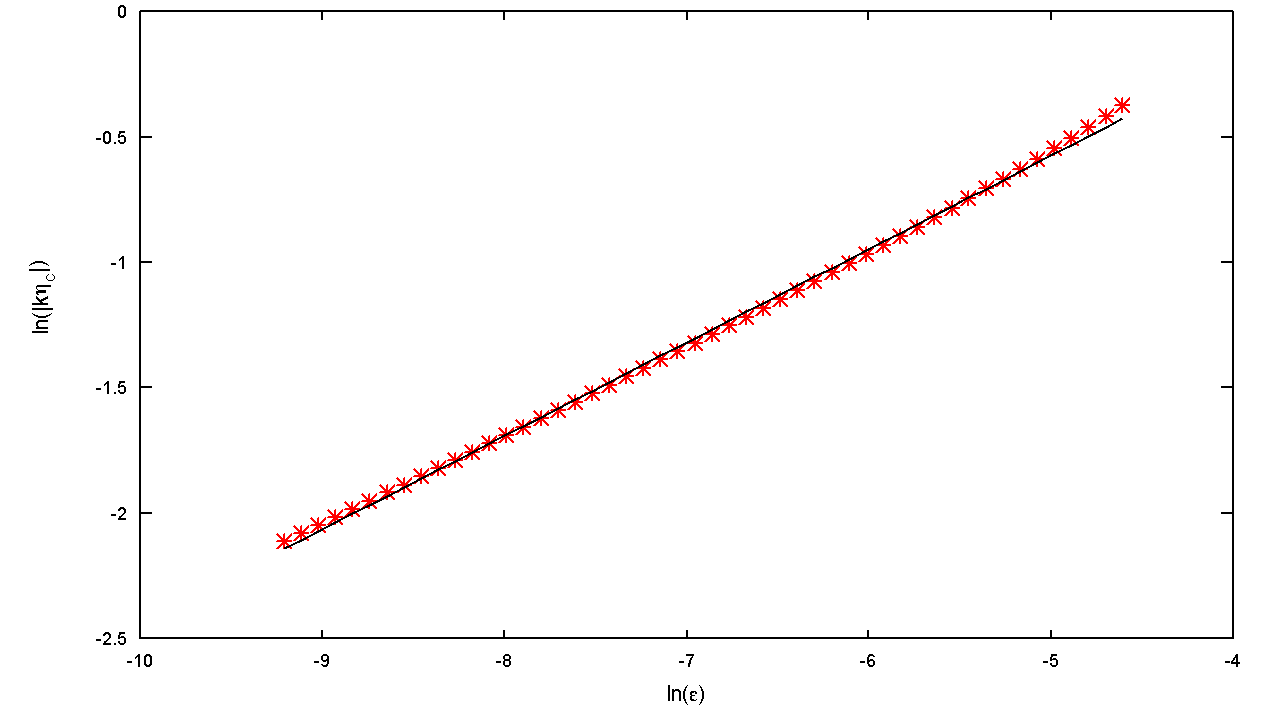}
\caption{Log-log plot of the average of $|k\eta_c|$ as a function of $\epsilon$. Superposition of $[0, 1, 2]$ with weights $[10, \exp(3i), \exp(i)]$ with $H = 0.01$, $\eta_{ini} = -50$, $q(\eta_{ini}) = 1$ and de Sitter inflation. The red stars are simulated values, the black line is a first order polynomial fit in the least squares sense.}
\label{Classicality10MeiLogLog}
\end{center}
\end{figure}

 Let us take again $\eta_c(\epsilon)$ as in the previous section. Consider fig. \ref{Classicality10MeiLogLog} for de Sitter inflation. Plotted is the behaviour of $|k\eta_c|$ as a function of our choice of $\epsilon$ between $10^{-4}$ and $10^{-2}$. The values of $|k\eta_c|$ were obtained by averaging over the simulations for different $k$ presented in fig. \ref{KDep1}. The polynomial fit is given by
\begin{equation}
 \ln(|k\eta_c|) = \left(0.37 \pm 0.001\right) \ln (\epsilon) + 1.28 \pm 0.001 .
\end{equation}
This is consistent with eq.\ \eqref{Gedrag}
\begin{equation}
 |k\eta_c| =  \left(A\epsilon\right)^{1/3},
\end{equation}
where the constant coefficient $A$ is of order $1$.
\par
\begin{figure}[p]
\begin{center}
\includegraphics[scale = 0.3]{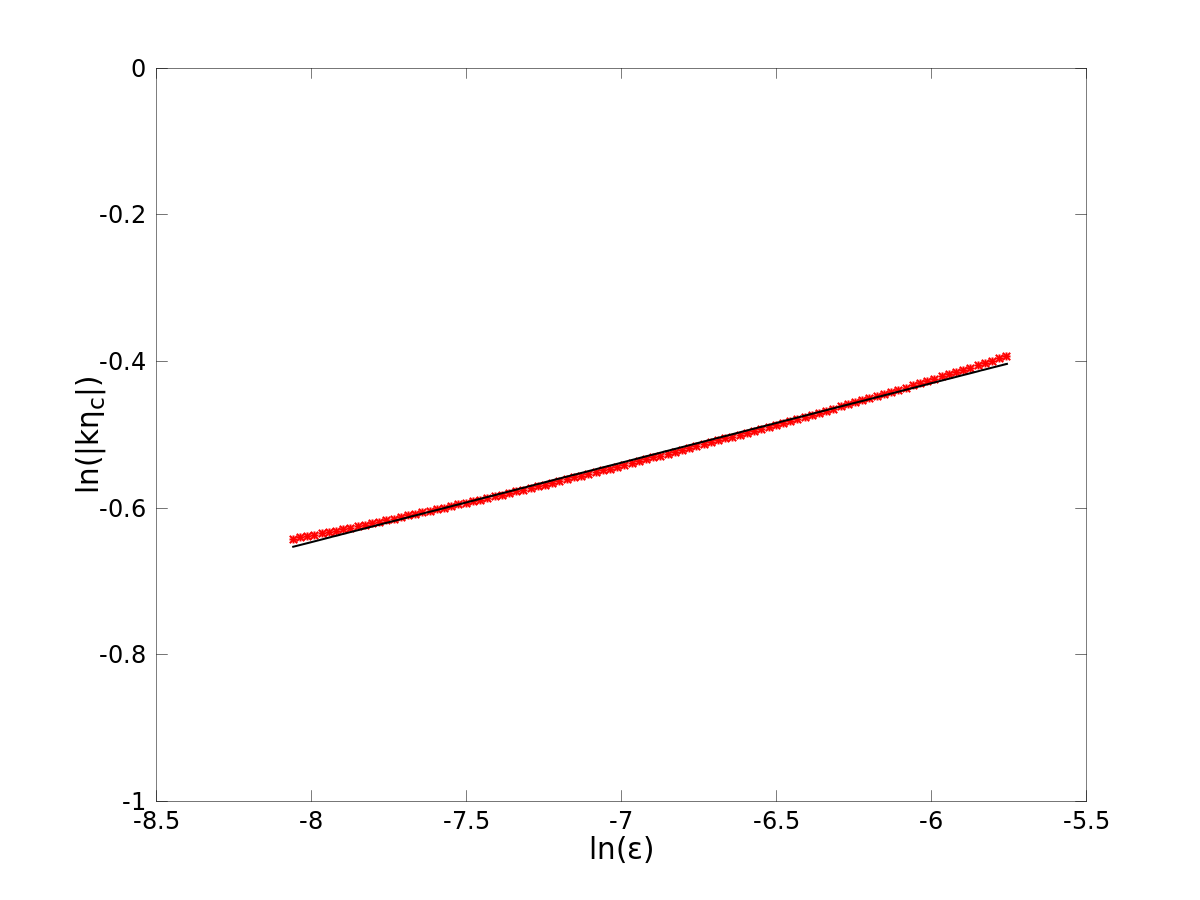}
\caption{Log-log plot of the average of $|k\eta_c|$ as a function of $\epsilon$. Superposition of $[0, 1, 2]$ with random weight of equal amplitude. $H = 0.01$, $\eta_{ini} = -50$, $q(\eta_{ini}) = 0.1$ and power-law inflation for $p = 2$. The red stars are simulated values, the black line is a first order polynomial fit in the least squares sense.}
\label{P2Epsilon}
\end{center}
\end{figure}

 We observe essentially the same behaviour for $p=2$ power law inflation. In figure $\ref{P2Epsilon}$ we again see a log-log plot of classicality time $|k\eta_c|$ as a function of the required accuracy $\epsilon$. The black line is the first order polynomial fit given by
\begin{equation}
 \ln(|k\eta_c|) = \left(0.217 \pm 0.001 \right) \ln (\epsilon) + 3.26 \pm 0.01 .
\end{equation}
which is again fully consistent with the relation
\begin{equation}
 |k\eta_c| =  \left(A\epsilon\right)^{1/5},
\end{equation}
where $A$ is now of order $10^7$.

For de Sitter inflation, the results of the previous section allow us to estimate the classicality of the modes at the end of inflation by extrapolating our results. The non-classicality is dominated by $\epsilon_C$. For simplicity, we take $60$ e-folds of inflation. This means that $\frac{\eta_{ini}}{\eta_f} = e^{60}$. Let us take a mode $\nu_{\vec{k}}$ that left the Hubble radius early, so $|k\eta_{ini}| \approx 1$. Then we get
\begin{equation}
 \epsilon_C(\eta_F) = A k^3 \eta_{F}^3 \approx A \left(\frac{k \eta_{F}}{k \eta_{ini}}\right)^{3} \approx A e^{-180}.
\end{equation}
Thus the process we described is certainly fast enough to achieve classicality de Sitter inflation, if we take reasonable estimates for $A$. For the example in the previous section, $A$ was of order $1$.
\par
The same result applies for power-law inflation. Taking again $60$ e-folds of inflation, we have
\begin{equation}
 \epsilon_C(\eta_F) \approx A e^{-120p - 60}.
\end{equation}
Again, taking reasonable assumptions for $A$ we get extremely small deviation from classicality. As an illustration, we take the example of fig. \ref{P2Epsilon}, meaning $p= 2$ power-law inflation and $A$ of order $10^7$. We get an upper bound on $ \epsilon \leq 10^{-98}$.
\section{Conclusion}
We can conclude from this chapter that the classical limit for the primordial modes is expected to be reached. We repeat that there are two sources of deviation, as presented in section \ref{Sources}. For pure $N$-particle states, the only contribution arises from $\epsilon_f$, which we can calculate exactly. For finite superpositions, we can neglect $\epsilon_f$ compared to $\epsilon_c$ for late times. Since this contribution is very hard to calculate in general, we resorted to numerical simulations. The deviation $\epsilon$ is seen to follow the estimate $A|k\eta|^{2p+1}$, where $A$ is some state dependent parameter. This relation suggests that for reasonable $A$ the classical limit is reached sufficiently fast.

\chapter{Cosmological Perturbations: a Test for Quantum Mechanics?}\label{Test}
In chapter \ref{PilotWavePert} and chapter \ref{Quantifying} we have shown that pilot-wave theory presents a solution to the problems stated in section \ref{PROBLEM}. Firstly, it explains why the cosmological perturbations look classical to us. Secondly, it presents a very natural way of breaking the initial symmetry present in the de Bunch-Davies vacuum. This picture is not very dependent on the choice of initial state. Thirdly pilot-wave theory leads to the same scale invariant power spectrum as the standard account does under the assumption of quantum equilibrium.
\par This chapter will be devoted to exploring the possibilities of non-equilibrium in the early universe. We will first discuss the concept of non-equilibrium and relaxation to equilibrium for pilot-wave theory in general. Then we will look at what opportunities inflation presents us with to detect possible non-equilibrium in the early universe, as they are displayed in the literature.
\section{Non-Equilibrium and Relaxation}
In the Copenhagen interpretation of quantum mechanics the Born rule is simply postulated. In pilot wave theory, the Born rule arises automatically when the density of the postulated particle $\rho$ is equal to the modulus squared of the wavefunction $|\Psi|^2$. The continuity equation, eq.\ \eqref{Split2}, guarantees that the equality $\rho = |\Psi|^2$ will be preserved by time evolution.
\par It does not have to be this way. While it is completely possible to regard quantum equilibrium as an additional postulate in pilot-wave theory, the formalism itself gives no reason why $\rho$ needs to be equal to $|\Psi|^2$. Taking $\rho = |\Psi|^2$ makes sure that pilot-wave theory cannot be differentiated from standard quantum theory, in this regime the predictions of both theories are the same. While one could prefer the picture pilot-wave theory paints, there is no reason in that case to prefer one theory over the other. Only when $\rho \not = |\Psi|^2$ does pilot-wave theory differ from standard quantum theory. The search for systems that are in non-equilibrium is thus an important test of quantum mechanics.
\par So far, no quantum non-equilibrium has been observed. If it exists, how do we explain its apparent absence? There are indications\cite{Valentini:1991zq} that systems prepared with initial non-equilibrium distributions tend to relax to equilibrium. Note that the continuity equation forbids any non-equilibrium distribution to evolve to the exact equilibrium distribution, but non-equilibrium distributions can come arbitrarily close. Both distributions are then essentially the same on a coarse grained level. Using numerical simulations it is often found that relaxation is very efficient\cite{Relax, ColinStruyve}. Relaxation typically happens on the same time scale as the considered wavefunction varies, so
\begin{equation}\label{TimeScale}
 \tau \sim \frac{1}{\Delta E},
\end{equation}
where $\Delta E$ is the spreading in energy of the considered wavefunction. Notice that relaxation occurs on very large time scales when $\Delta E$ is small. In the case of a pure $N$-particle state of an harmonic oscillator, the particles do not move. Every distribution is a stationary one and no relaxation occurs. This is also the case for modes of the Mukhanov-Sasaki variable, long before Hubble exit, see section \ref{Nparticle}. For complex superpositions of modes, the relaxation time scale will generally be small.
\section{Non-Equilibrium in the Early Universe}\label{NonEqEarly}
Since relaxation is expected to be very efficient in realistic situations this would explain why we only observe quantum equilibrium. If there was initially some quantum equilibrium in the early universe, it would very rapidly have relaxed.\footnote{It has even been suggested that non-equilibrium may have been produced in the early universe due to quantum gravitational effects, see \cite{Valentini:2006yj}.}
\par Without falling into pure speculation we can think about what would have happened to early non-equilibrium. In standard Hot Big Bang theory, it would have relaxed on a very short time scale and would be well outside present experimental limits. In inflationary models of our universe, there is another possibility. We have proven in chapter \ref{PilotWavePert} that trajectories in pilot-wave theory typically get `frozen'. We mean by this that the physical variable $a^{-1}\nu_{\pm, \vec{k}}$ does not move in the late times limit. It is clear that any distribution $\rho$ present at that time will also be frozen and thus not evolve any more in time. When such a frozen mode then re-enters the Hubble radius, it will again behave as a harmonic oscillator and relaxation can commence.
\paragraph{Freezing Inequality} The scenario explained in the previous paragraph would mean that modes with wavelengths $\lambda$ that are larger than the Hubble scale \emph{now} can still carry evidence of initial quantum non-equilibrium. They would have exited the Hubble radius very early during inflation, effectively `freezing in' any non-equilibrium that would have been present. This non-equilibrium could, in principle, be observed by us as soon as the modes re-enter the Hubble radius. In \cite{Valentini:2008rg}, Valentini derives the \emph{freezing inequality}. Quantum non-equilibrium will not relax on an time interval $[\eta_i, \eta_f]$ for a mode of some scalar field $\phi$ of wavevector $\vec{k}$ when
\begin{equation}
 4 a(\eta_f) \sqrt{a(\eta_f)\langle \hat{H}_{\pm, \vec{k}} \rangle|_{\eta_f}}\int_{\eta_i}^{\eta_f}d\eta' \frac{1}{a(\eta')}\sqrt{a(\eta') \langle \hat{H}_{\pm, \vec{k}}\rangle|_{\eta'}} < 1.
\end{equation}
Here $\hat{H}_{\pm, \vec{k}}$ is the Hamiltonian associated with the real or imaginary part of the mode $\phi_{\vec{k}}$, given by
\begin{equation}\label{HamPie}
 \hat{H}_{\pm, \vec{k}} = \frac{\hat{\Pi}^2_{\pm, \vec{k}}}{2a} + \frac{a}{2}k^2\phi_{\pm, \vec{k}}^2,
\end{equation}
where $\Pi_{\pm, \vec{k}} = a \phi'$. %
 These equations were derived for a free massless scalar field without potential that lives on an expanding, but not necessarily inflationary, background.
This is of course a harmonic oscillator for every mode $\phi_{\pm, \vec{k}}$ with time dependent mass. One can do something in close analogy with appendix \ref{WFAppendix} and we can find a number operator $\hat{N}_{\pm, \vec{k}}$ with positive integer eigenvalues so that\cite{Valentini:2008rg}
\begin{equation}
 \hat{H}_{\pm,\vec{k}} = \frac{k}{a}\left(\hat{N}_{\pm, \vec{k}} + \frac{1}{2}\right).
\end{equation}
The freezing inequality becomes
\begin{equation}
 \frac{1}{k} > 4 a(\eta_f) \sqrt{\langle \hat{N}_{\pm,\vec{k}} \rangle|_{\eta_f} + \frac{1}{2}}\int_{\eta_i}^{\eta_f}d\eta' \frac{1}{a(\eta')}\sqrt{ \langle \hat{N}_{\pm, \vec{k}} \rangle|_{\eta'} + \frac{1}{2}}.
\end{equation}
This freezing inequality was derived for general expanding spacetimes. The author argues that this inequality predicts freezing of non-equilibrium for super Hubble modes for pre-inflationary eras, say between Planck time and the start of inflation. Valentini suggest then, in another article \cite{Valentini:2008dq}, that non-equilibrium will be transferred intact troughout some inflationary period. One can see this as follows. Starting from the de Bunch-Davies vacuum, the modes of the Mukhanov-Sasaki variable will evolve in time as (see section \ref{Nparticle})
\begin{equation}
 \nu_{\pm, \vec{k}}(\eta) = \sqrt{2k} \nu_{\pm, \vec{k}}(\eta_{ini}) |f_k(\eta)|.
\end{equation}
Since this time evolution is so simple, any initial distribution will essentially spread out in time but retain its form. More exactly
\begin{equation}
 \rho(\nu_{\pm, \vec{k}},\eta) = \rho\left(\frac{\nu_{\pm, \vec{k}}}{\sqrt{2k}|f_k(\eta)|}, 0\right)
\end{equation}
This of course corresponds to the physical variables being frozen in the late times limit.
\par There are however some problems with this picture. Firstly, pre-inflationary periods are very poorly understood. We are not even entirely sure whether inflation happened and which of the many possible models is closest to reality. We thus should be rather distrustful about the validity of the action from which the Hamiltonian in eq.\ \eqref{HamPie} was derived. Second, the picture painted in \cite{Valentini:2008dq} is slightly too simplistic. For an exact Bunch-Davies vacuum the trajectories are indeed that simple, but this is not longer true for states that deviate a little bit from the Bunch-Davies vacuum. We have explicitly shown in chapter \ref{Quantifying} that the trajectories for superpositions of states are far from simple. Before Hubble exit the system is a harmonic oscillator and we would expect relaxation on the time scale in eq.\ \eqref{TimeScale}.
\par One could for instance try to remedy this by applying the freezing inequality to inflationary periods. A best-case scenario will give us some insight. The inequality is best satisfied for $\langle \hat{N}_{\pm,\vec{k}}\rangle = 0$. We take a de Sitter inflationary period of $60$ e-folds of inflation for simplicity. The modes that satisfy the freezing inequality then satisfy
\begin{equation}
H > k e^{120}.
\end{equation}
This is clearly a very extreme result. This would mean that the wavelengths of modes that retain quantum non-equilibrium would be $e^{120} \sim 10^{52}$
times bigger than the Hubble radius at the start of inflation. This would correspond to wavelengths of $10^{78}H^{-1}$ at the end of inflation. It is out of the question that we would observe any such modes.
\par The problem lies in some of the inequalities used in the derivation of the freezing inequality. While certainly valid, they seem to be too crude to give useful results. %

\paragraph{Sharpening the Freezing Inequality} For actual inflation, we will sketch how better estimates could be made. We focus on de Sitter inflation of $60$ e-folds for $p = 1$. The criterion used to look for the freezing of quantum non-equilibrium in \cite{Valentini:2008rg} reads for our rescaled variable $q = \frac{\nu_{\pm, \vec{k}}}{|f_k|}$
\begin{equation}\label{InEq}
 \langle| \delta (q)|(\eta_f) |\rangle \ll \Delta q
\end{equation}
The left hand side is the `distance travelled' between $\eta_{ini}$ and $\eta_f$. To be precise: $\delta q = q(\eta_f) - q(\eta_{ini})$ and the expected value should be taken with respect to an equilibrium distribution. The right hand side is the uncertainty in $q$ of the considered wavefunction. The idea is now the following: if the particles do not move over scales comparable to the spread of the wavefunction, there can be no relaxation on the level of the entire wavefunction\footnote{The inequality is of course more rigorously argumented in \cite{Valentini:2008rg}.}.
\par Analogously to the treatment in \cite{Valentini:2008rg}, we will estimate an upper bound for the left hand side for wavefunctions that satisfy the conditions of theorem \ref{theorem2}. We thus suppose an initial product state with a bounded velocity field for $q$. Notice that the following inequalities hold
\begin{equation}
 \langle| \delta (q)|(\eta_f) |\rangle \leq \left\langle \int_{\eta_{ini}}^{\eta_f}d\eta' |q'|\right\rangle \leq \int_{\eta_{ini}}^{\eta_f} \frac{C}{|f_k|^2},
\end{equation}
where the constant $C$ is defined as in chapter \ref{PilotWavePert}, being an upper bound on the quantity $G$. Let us first focus on modes that start inside the Hubble radius at the onset of inflation. The last integral is then comprised of two different parts: before and after Hubble exit of the mode. In the first regime $|f_k|$ is essentially constant, and in the second it roughly behaves as $\eta^{-1}$. We estimate the integral as
\begin{equation}\label{SplitIntegral}
 \int_{\eta_{ini}}^{\eta_f} \frac{C}{|f_k|^2} \approx \int_{\eta_{ini}}^{-1/k} 2kC + \int_{-1/k}^{\eta_f} \frac{k^3}{H^2}\eta^{2} C = C\left(\frac{-1}{k} - \eta_{ini} \right) + C \frac{k^3}{H^2}\left(\eta_{f}^{3} + \frac{1}{k^3}\right),
\end{equation}
We have seen in chapter \ref{Quantifying} that typical trajectories do not move appreciably after Hubble exit. We will thus neglect the second term in this equation.
\par We will now turn to $\Delta q$ in the right hand side of eq.\ \eqref{InEq}. For the vacuum, $\Delta \nu_{\vec{k}}$ is given by $|f_k|$. We will thus estimate, as a lower bound, $\Delta q \approx 1$.
\par Collecting everything, we estimate that non-equilibrium will survive during a period of de Sitter inflation is
\begin{equation}
C \left|\frac{1}{k} + \eta_{ini}\right| \leq 1
\end{equation}
For a given initial state, and thus given $C$, this is a condition on the wavelength of the mode under consideration. We see that non-equilibrium will be maintained if the mode exits the Hubble radius soon enough. This is could have been anticipated: before Hubble exit the modes behave as harmonic oscillators and relaxation can occur.
\par Let us now focus on modes that start far outside the Hubble radius at the onset of inflation. The velocity integral is then
\begin{equation}
 \int_{\eta_{ini}}^{\eta_f} \frac{C}{|f_k|^2} \approx C \frac{k^3}{H^2}\left(\eta_{f}^{3} -\eta_{ini}^3\right)
\end{equation}
The freezing inequality becomes
\begin{equation}
 C\frac{k^3}{H^2}\left(\eta_{f}^{3} -\eta_{ini}^3\right) \leq 1
\end{equation}
Since the mode is initially outside of the Hubble radius, we know that $|k\eta_f| < |k\eta_{ini}| \ll 1$. We thus see that this inequality is very easily satisfied for modes that start outside the Hubble radius. For such modes, we do not expect appreciable relaxation to equilibrium.
\par This is only a sketch of an argument, mainly for two reasons. First, the splitting of the integral in eq.\ \eqref{SplitIntegral} is very rough. Second, we have ignored any role that $C$ plays in the discussion. Relaxation is obviously heavily dependent on the initial states of the Mukhanov-Sasaki variable. For example, if it started in the Bunch-Davies vacuum, $C$ would be zero and no relaxation at all could occur. On the other side, if some initial super-Hubble mode would be in a very complex superposition of $N$-particle states, then we could imagine $C$ being very large and thus permitting relaxation for that mode. A complete theory of relaxation during inflation would thus include some estimates for $C$.
\section{Detection}
Let us now come to the detection of possible non-equilibrium. Let us assume for simplicity that the Mukhanov-Sasaki variable $\nu$ started out in the de Bunch-Davies vacuum.
Every Fourier mode $\nu_{\vec{k}}$ would then have a Gaussian wavefunction. As first proposed test \cite{Valentini:2008dq} lies in the power spectrum itself. The calculation done in section \ref{dBBSpectra} was obviously dependent on the assumption of quantum equilibrium. We conclude that non-equilibrium would lead to a different power spectrum.
\par There is however an ongoing search for \emph{non-Gaussianity}, on which pilot-wave theory also has an interesting take. Let us to that end compute the correlator of three modes
\begin{equation}
 \langle \hat{\nu}_{\vec{k}_1} \hat{\nu}_{\vec{k}_2} \hat{\nu}_{\vec{k}_3}\rangle = \int d\nu_{\vec{k}_1} d\nu_{\vec{k}_2} d\nu_{\vec{k}_3} \rho_{\vec{k}_1} \rho_{\vec{k}_2} \rho_{\vec{k}_3} \nu_{\vec{k}_1} \nu_{\vec{k}_2} \nu_{\vec{k}_3}.
\end{equation}
If we assume quantum equilibrium, then $\rho_{\vec{k}_i}$ is a Gaussian. It is then easy to see that this integral will always be zero, for symmetry reasons. This means that the three-point-correlator $\langle \nu(\vec{x}) \nu(\vec{y}) \nu(\vec{z})\rangle$ should be zero under the assumption of quantum equilibrium. We conclude that in the case of non-equilibrium being present in some of the modes that are discernible across the sky that the three-point-correlator will in general not be zero. Since the results of the WMAP experiment were not sensitive enough to give any definite information on the three-point-correlators, it is one of the aims of the Planck mission to provide us with more data \cite{PlanckMissie}.
\par As mentioned before, this idea fits in the broader scheme of the search for non-Gaussianity. All the correlators of arbitrary order can be trivially specified in terms of the correlator of second order if the distribution is Gaussian. The odd orders are zero (as shown above for degree three) while the even powers are just the second order to some appropriate power. Deviations from this Gaussian distribution can be easily inferred from the behaviour of the different correlators and are generally termed non-Gaussianity. It is one of the primary goals of the Planck mission to test for non-Gaussianity\cite{PlanckMissie}. If the Planck mission would find non-trivial behaviour of the correlators, we have shown that quantum non-equilibrium offers a possible explanation. It should be noted that non-Gaussianity is a very generic prediction. For example, taking into account non-linear effects in inflation (meaning taking a higher order expansion in section \ref{Overview}) already yields a (small) amount of non-Gaussianity. Multi-field inflationary models can also easily account for any non-Gaussianity, as well as several non-inflationary models\cite{PlanckMissie}.
\section{Non-Inflationary Tests}
To conclude this chapter, we will briefly mention some other tests for non-equilibrium in the early universe, taken from \cite{Valentini:2006yj}. Based on the freezing inequality presented in section \ref{NonEqEarly}, one can expect suppression of relaxation in other expanding regimes. More precisely, particles that decoupled very early from the rest of the energy density in the universe could be expected to retain some degree of non-equilibrium. Such particles have been given the suggestive name of `relic particles'. Gravitinos, for which decoupling at Planck time is expected, are suggested as a possible candidate. One however has to keep in mind that inflation, if it occurred, has probably diluted their presence to unobservable densities. The search for relic particles can thus only expected to be fruitful in non-inflationary models.
\par As a final possibility, Hawking radiation from black holes might be in non-equilibrium. Roughly speaking, because of the no-hair theorem it seems that black holes can effectively destroy information. While completely thermal Hawking radiation does not solve this, it has been suggested that out-of-equilibrium Hawking radiation can carry more information than normal radiation, thus possibly providing an answer to the black hole information paradox \cite{Valentini:2006yj}.
\addcontentsline{toc}{chapter}{Conclusion}
\chapter*{Conclusion and Outlook}\label{Conclusion}
 The aim of this thesis was to try and understand the problem of classicality in the context of primordial perturbations. The proposed solutions are ample, but we have seen that the scientific community is in general divided on the subject. The important questions are
\begin{enumerate}[I]
 \item How does one go from an initial homogeneous and isotropic state to a state with inhomogeneities?
 \item Why can we consider the density perturbations to behave essentially classical at the end of inflation?
\end{enumerate}
Already in section \ref{PROBLEM} we argued that the answer quantum mechanics provides, a measurement, is not enough to explain the classicality of the primordial perturbations. In chapter \ref{QtoC} we explicitly reviewed some traditional answers to the question and found many to be insufficient. Two proposed solutions were examined, the theory of squeezing and the theory of decoherence. The first one was found to make invalid approximations, in stating that two conjugate quantum mechanical operators commute. The second is physically extremely feasible and explains the absence of quantum interference. However one has to supplement decoherence with a measurement or some effective collapse mechanism to achieve the classical limit.
\par We examined the answers that pilot-wave theory offers. The classical limit in this framework is more easily attained since the theory treats definite particles (or field modes) with definite trajectories. The answer to the second question above is also easy: the symmetry is broken by the initial conditions. The answer to the first question is more involved: we found in chapter \ref{PilotWavePert} that the trajectories for a large class of initial states, including the vacuum, become classical in the late times limit. We explicitly showed that pilot-wave theory reproduces the predictions of standard quantum mechanics in this situation, under the assumption of quantum equilibrium. In chapter \ref{Quantifying} we provided the reader with explicitly calculated trajectories to help intuition. We argued, using numerical simulations, that this mechanism is capable of reaching classical trajectories in realistic periods of inflation.
\par Further investigations could possibly be made to extend our picture to arbitrary initial states, and mathematically prove that they all give rise to classical trajectories in the late times limit. The possibility of quantum non-equilibrium in the early universe also needs more work. We gave some sketches in chapter \ref{Test}, but the details are yet to be understood. The data the Planck mission will provide should bring more insight into the discussion, possibly discerning between different inflationary models and putting upper limits on the amount of quantum non-equilibrium in the early universe.
\chapter*{Nederlandstalige samenvatting}\addcontentsline{toc}{chapter}{Nederlandstalige Samenvatting}
\markboth{Nederlandstalige Samenvatting}{Nederlandstalige Samenvatting}
Als we naar de hemel kijken, merken we dat we vanuit alle richtingen een merkwaardig homogeen signaal opvangen, vooral in het micrometer regime. Dit is straling vanuit de verste delen van het observeerbare universum. Bovendien is deze \emph{kosmische achtergrondstraling} een van onze beste manieren om informatie over het vroege universum te krijgen. De eerste observatie is al meteen de belangrijkste: de achtergrondstraling is uiterst homogeen. De temperatuur van de straling die we opvangen is zo goed als onafhankelijk van de richting waarin we kijken. Dit is een belangrijke experimentele bevestiging van het \emph{kosmologische principe}: het universum is homogeen op zeer grote schaal.
\par Maar het is duidelijk dat het universum op kleine schaal verre van homogeen is. Dit is ook te zien in de achtergrondstraling: deze is niet perfect homogeen maar bevat kleine inhomogeniteiten. Als we uitgaan van het kosmologisch principe moeten we natuurlijk een manier vinden om die kleine storingen te verklaren: waar komen ze vandaan en waarom zijn ze zo verdeeld over de hemel? Daarover gaat deel \ref{Exploring}.
\par In hoofdstuk \ref{Inflation} wordt de relativistische theorie van \emph{inflatie} uit de doeken gedaan. Het idee is dat het vroege universum een zeer korte periode van versnelde expansie doormaakte. Wij presenteren het simpelste model, namelijk \emph{slow-roll inflatie}. Hierin wordt de expansie aangedreven door een re\"eel scalair veld, het \emph{inflaton}veld, dat leeft op een achtergrond van een homogeen universum met FLRW-metriek. Voor een universum waarin de energiedichtheid wordt gedomineerd door zo'n veld, zal de schaalfactor $a(t)$ exponentieel snel stijgen, met $H$ de Hubble parameter
\begin{equation}
 a = \exp(Ht).
\end{equation}
\par In hoofdstuk \ref{Perturbations} veranderen we het homogene universum uit het vorige hoofdstuk in een inhomogeen universum. Dat wil zeggen dat we de kleine storingen in zowel de metriek als het inflatonveld beschouwen. Het zal voldoende zijn om de variabelen te splitsen in homogene achtergrond en \emph{scalaire} storingen. Vector- en tensorstoringen zijn van ondergeschikt belang. Deze splitsing is natuurlijk niet invariant onder veralgemeende co\"ordinaattransformaties en we gaan dus op zoek naar ijkinvariante combinaties. Uiteindelijk komt het erop neer dat de scalaire perturbaties maar \'e\'en relevante vrijheidsgraad hebben, de Mukhanov-Sasaki variable $\nu$ met Hamiltoniaan in momentumruimte
\begin{equation}
 H = \frac{1}{2}\int d^3k \left[ \Pi_{\vec{k}}\Pi_{\vec{k}}^* + \left(k^2 - \frac{a''}{a}\right)\nu^*_{\vec{k}} \nu_{\vec{k}}\right],
\end{equation}
waar $\Pi_{\vec{k}}^*$ het momentum geconjugeerd aan $\nu_{\vec{k}}$ is en apostrofes wijzen op afgeleides naar een andere tijdsco\"ordinaat, de conforme tijd $\eta$ . Deze Hamiltoniaan beschrijft Fourier modes die zich eerst gedragen als harmonische oscillatoren. Vanaf $|k\eta| = 1$ gedragen ze zich echter heel anders, als omgekeerde oscillatoren. Op het einde van dit hoofdstuk kwantiseren we deze Hamiltoniaan. We vinden annihilatie en creatie operatoren en de bijbehorende $N$-deeltjestoestanden.
\par In hoofdstuk \ref{Observation} beginnen we met de inhomogeniteiten aan de hemel beter te beschrijven: we karakteriseren ze in termen van hun vermogensspectrum. Vervolgens leggen we uit hoe de kwantumtheorie uit het vorige hoofdstuk aanleiding geeft tot een \emph{schaalinvariant vermogensspectrum} dat de observaties verklaart. Hier blijken een aantal onduidelijkheden te bestaan. In sectie \ref{PROBLEM} stellen we de volgende vragen:
\begin{enumerate}
 \item{Hoe geeft een kwantummechanische $\hat{\nu}$ aanleiding tot klassieke fluctuaties van de energie?}
 \item{De initi\"ele golffunctie voor $\nu$ is symmetrisch. De verdeling van de inhomogeneiteiten aan de hemel is dat niet. Hoe is dit mogelijk?}
\end{enumerate}
Een uitleg gebaseerd op een \emph{meting} is niet mogelijk in deze kosmologische context.
\newline
\par In deel \ref{Usual} onderzoeken we de mogelijke antwoorden die de literatuur aanbiedt. Hoofdstuk \ref{QtoC} houdt zich bezig met het onderzoeken van allerlei criteria voor de klassieke limiet. We bespreken de rol van onzekerheden, de WKB benadering, de Wigner functie en de rol van de natuurconstanten met als voorbeelden het vrije deeltje en de omgekeerde oscillator. In hoofdstuk \ref{Squeezing} bekijken we het formalisme van de zogenaamde \emph{squeezed states}. Dit formalisme is gebaseerd op een benadering, namelijk
\begin{equation}
 \left[\nu_{\vec{k}}, \Pi_{\vec{k}}\right] \approx 0.
\end{equation}
Deze benadering is duidelijk verkeerd, omdat deze commutator altijd $i$ is.
\par In hoofdstuk \ref{Decoherence} bekijken we een andere prominente verklaring. \emph{Decoherentie} veronderstelt extra interacties tussen verschillende modes van de Mukhanov-Sasaki variabele of interacties met andere velden. Deze resulteren in een (bij goede benadering) diagonale dichtheidsmatrix en verklaart dus het afwezig zijn van interferentie tussen verschillende toestanden. We merken echter op dat verscheidene details in de literatuur nog niet uitgeklaard zijn, zoals bijvoorbeeld het moment waarop decoherentie zijn werk deed. Helaas heeft decoherentie geen antwoord op onze tweede vraag: er is een extra meting (of een effectief collapsemechanisme) nodig om de overgang van een symmetrische naar een assymetrische toestand te verklaren.
\newline
\par In deel \ref{PWCase} bekijken we welke antwoorden de Broglie-Bohm mechanica kan aanreiken. In hoofdstuk \ref{PilotWave} leggen we de theorie uit. Het verschil met standaard kwantummechanica ligt er voornamelijk in dat deze theorie het heeft over goed gedefinieerde deeltjes (of Fourier modes in veldentheorie) met goed gedefinieerde trajecten, die worden bepaald door eerste orde differentiaalvergelijkingen. Meer concreet schrijven we een golffunctie $\Psi$ als
\begin{equation}
 \Psi = R \exp (i S),
\end{equation}
met $R>0$ en $S$ re\"eel. De vergelijking wordt dan
\begin{equation}
 \frac{dx}{dt} = \frac{1}{m}\frac{\partial S}{\partial x}.
\end{equation}
Dit komt overeen met de beweging van een deeltje onder de invloed van de klassieke potentiaal $V$ en een extra potentiaal, de kwantumpotentiaal $V_q$
\begin{equation}
 \frac{d^2 x}{dt^2} = -\frac{\partial V}{\partial x} - \frac{\partial V_q}{\partial x}.
\end{equation}
Onder een bepaalde voorwaarde, die van \emph{kwantumevenwicht}, zijn de voorspellingen van deze theorie niet onderscheidbaar van de voorspellingen van de standaard kwantummechanica. Vervolgens bekijken we hoe precies de klassieke limiet moet gezien worden in dit kader. Het is duidelijk dat de klassieke limimet bereikt wordt wanneer de trajecten oplossingen zijn van de klassieke bewegingsvergelijkingen. Dit gebeurt als de kwantumpotentiaal niet te hard varieert. We bekijken dit ook weer kort aan de hand van de voorbeelden van het vrije deeltje en de omgekeerde oscillator.
\par In hoofdstuk \ref{PilotWavePert} lossen we de differentiaalvergelijkingen op voor de Mukhanov-Sasaki variabele. Voor de $N$-deeltjestoestanden vinden we de klassieke limiet zeer natuurlijk terug voor late tijden: de modes van $\nu$ volgen in die limiet de klassieke bewegingsvergelijkingen. Vervolgens zoeken dan verder naar de oplossingen voor superposities van een eindig aantal $N$-deeltjestoestanden. Deze zijn niet meer analytisch op te lossen, maar we bewijzen rigoureus dat deze klassiek worden voor late tijden.
\par In hoofdstuk \ref{Quantifying} werken we aan onze intu\"itie: we lossen de bewegingsvergergelijkingen numeriek op en bevestigen dat ze zich klassiek gedragen voor late tijden. Daarbij zoeken we uit op welk moment we eigenlijk kunnen spreken van de klassieke limiet. We vinden door numerieke simulaties en theoretische schattingen dat de afwijking van een klassiek traject $\epsilon$ op het einde van een de Sitter inflationaire periode aan de volgende ongelijkheid voldoet
\begin{equation}
 \epsilon \leq Ae^{-180},
\end{equation}
waar $A$ een constante is van orde $1$.
\par Tenslotte be\"eindigen we deze thesis in hoofdstuk \ref{Test} met ons af te vragen wat er gebeurt als er niet voldaan is aan de voorwaarde van kwantumevenwicht. In de literatuur wordt geargumenteerd dat we dan eventueel kwantumonevenwicht zouden kunnen detecteren in de achtergrondstraling. We geven aan waar deze idee\"en wat te snel door de bocht gaan, maar we schetsen hoe een volledige theorie er zou moeten uitzien en hoe dit onevenwicht precies zou kunnen gedetecteerd worden.
\addcontentsline{toc}{chapter}{Acknowledgement}
\chapter*{Dankwoord}
Veel van ons hebben zich er al sinds de jaren '20 mee verzoend dat de wereld niet deterministisch is. Ik weiger echter te aanvaarden dat de volgende mensen een of ander golfpakket vormden tot ik ze omeeting op ze uitvoerdde.
\par Aan mama en papa, die toch wel voor zowat alles gezorgd hebben. Ook al kregen gasten de opmerking `Vraag maar niet naar zijn thesis' en ondanks dat papa de indruk had dat `ze weten toch eigenlijk niet wat aanvangen met jullie, die laatste twee jaar', hebben ze zich altijd achter mij geschaard.
\par Aan Ward, die me zowel voor mijn thesis als mijn bachelorproject begeleid heeft. Hij is een van de weinige personen op deze wereld die het belang van eenvoud en duidelijkheid volledig naar waarde schatten. De vraag `maar wat betekent dat eigenlijk' is een vaak genegeerde vraag in veel artikels en boeken, en ik bewonder hem enorm dat hij toch van die vraag zijn werk heeft gemaakt.
\par Aan Wina en aan mijn jaargenoten in het bijzonder, aan wie ik veel indirect te danken heb. Met sommigen van hen heb ik (te) veel gedronken, met heel veel heb ik gelachen en van allemaal heb ik ongelofelijk veel bijgeleerd. Ik ga altijd met veel plezier terugkijken op allerlei discussies: `many-worlds interpretaties', `staat er een overaftelbaar aantal sterren aan de hemel' en `proffen die collapsen naar overal, behalve waar ze moeten zijn' zijn er slechts enkele. Speciale dank aan de kerels van het bureau: Hans, Ruben en Alberto voor nog veel meer discussies over zo mogelijk nog gekkere onderwerpen.
\par Aan Hanne, want ze verdient toch een speciale vermelding, omdat ze al die jaren voor eten heeft gezorgd.\footnote{En omdat ze het zelf niet zo subtiel heeft gevraagd. Ook wel een beetje als verontschuldiging: we hebben vijf jaar lang plezier gehad aan haar pijn.}
\par Aan Kaat, die maar ongeveer de helft van het maken van deze thesis heeft meegemaakt, maar wel het grootste deel van het gevloek heeft opgevangen. Zij zorgde voor vele `topdagen'.
\par Aan Rowie, mijn aantal leuke herinneringen aan jou overtreft het aantal pinten dat we samen uitdronken veruit. Ik denk dat het zo duidelijk genoeg is.
\par Aan het USO, ik hoop dat ik een fractie heb teruggegeven van dat wat jullie mij gegeven hebben. Jullie hebben mij leren luisteren: naar Mahler, naar stemming\footnote{Wat is dat moeilijk!} en toonafstanden, maar vooral naar anderen.
\par En omdat ik toch ergens een krachtige one-liner moet hebben staan:
\begin{quote}
 `\dots but criticizing [\dots] is not the job. The job is to find the theory that's right.'
 \begin{flushright}
L. Smolin, The Trouble With Physics
 \end{flushright}
\end{quote}

\begin{appendices}
\appendixpage
\openany
\addtocontents{toc}{\protect\setcounter{tocdepth}{-1}}
\chapter{Wavefunctions of Time Dependent Harmonic Oscillators}\label{WFAppendix}
In this appendix we will construct the wavefunctions for time dependent harmonic oscillators. We prefer using the method of Lewis-Riesenfeldt invariants as it was originally written down in \cite{LewisRiesen}. The flaw of this method is that it will only allow us to construct normalisable wavefunctions. For inverted harmonic oscillator systems there clearly are unbound states with non-normalisable wavefunctions: these cannot be constructed in this way and we will merely quote the more general result from \cite{CaldirolaKanai}.
\par The type of Hamiltonian that we will discuss is given by
\begin{equation}\label{AppendixHamiltonian}
 \hat{H} = \frac{\hat{p}^2}{2m} + \frac{1}{2}m \omega^2(t)\hat{x}^2,
\end{equation}
where the frequency is allowed to depend on time. We do not restrict to real frequencies, thus allowing $\omega^2$ to be negative. The classical equation of motion for this Hamiltonian is
\begin{equation}\label{ClassicalEOMCaldKanai}
 f'' + \omega^2(t) f = 0,
\end{equation}
where apostrophes indicate derivatives with respect to time.
In classical physics, one would look for a constant of motion and then do a canonical transformation to find the solutions. Quantum mechanically, we look for an invariant $\hat{I}$, meaning that $\frac{d\hat{I}}{dt} = 0$. We propose, for $f$ an arbitrary solution to the classical equation of motion,
\begin{equation}
 \hat{I} = \frac{1}{2}\left[ |f|^{-2}\hat{x}^2 + \left(|f|\frac{\hat{p}}{m} - \frac{d|f|}{dt}\hat{x}\right)^2\right].
\end{equation}
Using Heisenbergs equations of motion for $\frac{d\hat{x}}{dt}$ and $\frac{d\hat{p}}{dt}$ one can readily check that $\hat{I}$ is indeed an invariant of the Hamiltonian. This means that the eigenvalues of $\hat{I}$ will be independent of time.
We are now free to impose a Wronskian constraint on the functions $f, f'$ and their conjugates, the equivalent of eq.\ \eqref{Wronskian}.
\begin{equation}\label{WronskianApp}
 i(f' f^{*} - f'^{*}f) = 1.
\end{equation}
We now introduce operators $\hat{a}$ and $\hat{a}^{\dagger}$, defined as
\begin{align}\label{GeneralAnnCrea}
 \hat{a} &= \frac{1}{\sqrt{2}} \left(-if'^{*}\hat{x} + i\frac{f^{*}}{m}\hat{p} \right)& \hat{a}^{\dagger} = \frac{1}{\sqrt{2}}\left(if' \hat{x} -i \frac{f}{m}\hat{p}\right)
\end{align}
One can check that these operators satisfy the following relations
\begin{align}
[\hat{a}, \hat{a}^{\dagger}] &= 1 & \hat{a}\hat{a}^{\dagger} = \hat{I} + \frac{1}{2}\\
\frac{d\hat{a}}{dt} &= 0 & \frac{d\hat{a}^{\dagger}}{dt} = 0
\end{align}
Thus $\hat{a}$ and $\hat{a}^{\dagger}$ respectively annihilate and create quanta of $\hat{I}$. The operator $\hat{I}$ is for the time dependent system what the number operator is for the time independent system.
\par We will now limit ourselves to the normalisable eigenstates of the operator $\hat{I}$. When $\omega^2 >0$ all the eigenstates are normalisable. When $\omega^2 < 0$ it is clear that the energy is not bounded from below and the Schr\"odinger equation allows unbound states with non-normalisable wavefunctions.
\par Diracs algebraic treatment\footnote{Note that this treatment doesn't work for non-normalisable states.} of the harmonic oscillator can now be used for the normalised eigenstates of $\hat{I}$, and we find that these eigenstates of the operator are the non-negative integers $n = 0,1,2, \dots$.
There thus exists a state $|0\rangle$ which gets annihilated by $\hat{a}$; this gives us the differential equation
\begin{equation}
\left[ i mf'^{*} x - f^{*}\frac{\partial}{\partial x}\right]\phi_{0}(x, t) = 0,
\end{equation}
which can be solved by guessing a Gaussian form. The result is, up to some time dependent normalisation factor
\begin{equation}
 \phi_0(x,t) \sim \exp\left[i \frac{m}{2}\left(\frac{f'^{*}}{f^{*}} \right)x^2\right].
\end{equation}
Anticipating the results for the standard harmonic oscillator, we expect that Hermite polynomials will play a role in constructing the eigenfunctions for higher eigenvalues of $\hat{I}$. Notice that the groundstate already incorporates the Hermite polynomial of degree 0. The recursion relation for Hermite polynomials is
\begin{equation}\label{Hermite}
 H_{n+1}(q) = 2qH_{n} - 2\frac{\partial H_{n}(q)}{\partial q}.
\end{equation}
Using this relation and normalising, one can check that the wavefunctions of the states $|n\rangle \sim \left(\hat{a}^{\dagger}\right)^n |0\rangle$ are given by
\begin{equation}\label{WaveFunctionAppendix}
 \phi_{n}(x, t) = \sqrt{\frac{1}{2^n n!}}\left(\frac{m}{2 \pi \hbar |f|^2}\right)^{1/4} H_{n}\left(\frac{x}{\sqrt{2}|f|}\right)\left(\frac{f}{f^{*}}\right)^{n+ 1/2}\exp\left[i \frac{m}{2}\left(\frac{f'^{*}}{f^{*}} \right)x^2\right].
\end{equation}

where we have reinstated $\hbar$ for future reference. For a standard oscillator, these states are also energy eigenstates: it is the factor $\left(\frac{f}{f^*} \right)^{n+ 1/2}$ that gives rise to the Schr\"odinger time evolution, $\exp\left(-i E t\right)$. For completeness sake, the most general solution to the Schr\"odinger equation for Hamiltonian \eqref{AppendixHamiltonian}is, indexed by a real $\lambda$ \cite{CaldirolaKanai}
\begin{align}\label{Nonnormalisable}
 \Psi_{\lambda}(x,t) &= \frac{1}{\sqrt{|f|}} \left(\frac{f}{f^{*}}\right)^{\frac{\lambda}{\hbar} } \exp \left( i\frac{f'^{*}}{f^{*}} \frac{m x^2}{2\hbar } \right) \Phi_{\frac{\lambda}{\hbar}}\left(\sqrt{\frac{2}{\hbar}}\frac{x}{|f|}\right) \exp\left(i \alpha_{\lambda}\right),\\
\alpha_{\lambda} &= \frac{\lambda}{\hbar}\int_0^t dt' \frac{1}{m|f|^2}.
\end{align}
These are eigenfunctions of $\hat{I}$ with eigenvalue $\lambda$. For each $\lambda$ there are two linearly independent $\Phi_{\frac{\lambda}{\hbar}}$. They are the odd and even parabolic cylinder functions. For $\lambda = \hbar \left(n + \frac{1}{2}\right) $, there is a linear combination of the even and odd solution that is normalisable. This combination then reduces to eq.\ \eqref{WaveFunctionAppendix}.

\chapter{An Alternative Action}\label{AppAlt}
Eq. \eqref{QHamiltonian} is where we started the quantum description of cosmological perturbations. There is however an alternative approach commonly used by several authors. In this approach, the action of the Mukhanov-Sasaki variable is derived by considering a minimally coupled real massless scalar field $\chi$ on a flat FLRW background. The action is, in terms of the conformal time
\begin{equation}\label{AltAction}
 S^{alt} = \frac{1}{2}\int d \eta\: d^3x \: a^2 \: \eta^{\mu \nu}\partial_{\mu}\chi^{*} \partial_{\nu} \chi.
\end{equation}
And the momentum conjugate to $\chi$ is $P_{\chi} = a^2 \chi'$. We will write the dynamics in terms of a new variable $y = a \chi$. The conjugate momentum is then $P_{y} = y' - \frac{a'}{a}y$. Immediately moving into momentum space, we write the action in terms of the Fourier modes $y_{\vec{k}}$ \cite{Martin:2004um}
\begin{equation} \label{AltActionFourier}
 S^{alt} = \frac{1}{2}\int d\eta \int_{\mathbb{R}^{+3}} d^3k \left[ y_{\vec{k}}'y_{\vec{k}}'^{*} + y_{\vec{k}}'^{*}y_{\vec{k}}'- 2\frac{a'}{a} \left( y_{\vec{k}}'y_{\vec{k}}^{*} + y_{\vec{k}}' y_{\vec{k}}^{*}\right) - \left( k^2 - \frac{a'}{a}\right) \left(y_{\vec{k}}^{*}y_{\vec{k}} + y_{\vec{k}} y_{\vec{k}}^{*}\right)\right],
\end{equation}
 where the $k$-integral is restricted, since we are dealing with a real field and thus $y_{- \vec{k}} = -y_{\vec{k}}^{*}$. We see that $P^{*}_{\vec{k}} = y_{\vec{k}}' - \frac{a'}{a} y_{\vec{k}}$. The Hamiltonian for a single Fourier mode then reads
\begin{equation}\label{AltHamiltonian}
 H^{alt}_{\vec{k}} = P_{\vec{k}} P_{-\vec{k}} + k^2 y_{\vec{k}} y_{-\vec{k}} + \frac{a'}{a}\left( P_{-\vec{k}} y_{\vec{k}} + P_{\vec{k}} y_{-\vec{k}}\right)
\end{equation}
This is the Hamiltonian\footnote{Notice that this Hamiltonian couples modes with wavevector $\vec{k}$ and $-\vec{k}$. This interaction originates in the reality of the field and is similar to the fact that we restrict the integral in eq.\ \ref{AltActionFourier}.} that is used to quantise the theory by some authors; \cite{PhysRevD.85.083506}, \cite{Kiefer:1998pb} and \cite{Albrecht:1992kf} among others.
The Lagrangian density in eq.\ \eqref{AltAction} differs from the one in eq.\ \eqref{actie} by a total derivative. They are thus equivalent on the classical level and we assume that they also are equivalent in their quantised forms.
\par As before we will write the theory in terms of annihilation and creation operators. These relations still hold
\begin{align}
\hat{a}_{\vec{k}} &= g_k^{*}\hat{\nu}_{\vec{k}} + i f_{k}^{*}\hat{\Pi}_{\vec{k}}, &
\hat{a}^{\dagger}_{\vec{k}} = g_k\hat{\nu}_{\vec{k}} - i f_{k}\hat{\Pi}_{\vec{k}},
\end{align}
if we enforce the Wronskian condition
\begin{equation}
 (g_{k} f^{*}_{\vec{k}} + g^{*}_{k} f_{k}) = 1.
\end{equation}
The main difference is that the function $g_k$ is now
\begin{equation}
g_{k} = i \left( f_{k}' - \frac{a'}{a}f_{\vec{k}}\right).
\end{equation}
In terms of $a^{\dagger}_{\vec{k}}$ and $a_{\vec{k}}$, the Hamiltonian becomes
\begin{equation}\label{AnnCreHamiltonian}
\hat{H}^{alt} = \int_{\mathbb{R}^{+3}}d^3k \left[k(\hat{a}^{\dagger}_{\vec{k}}\hat{a}_{\vec{k}} + \hat{a}^{\dagger}_{-\vec{k}}\hat{a}_{-\vec{k}}+ 1) + \frac{z'}{z}\left( \hat{a}^{\dagger}_{\vec{k}} \hat{a}^{\dagger}_{-\vec{k}} + \hat{a}_{\vec{k}} \hat{a}_{-\vec{k}}\right)\right]
\end{equation}
We get the same interpretation as before, when $|k\eta|$ is big, this is essentially the Hamiltonian of a harmonic oscillator. But it is in the last two terms of this Hamiltonian that we get a more intuitive picture of what happens when $|k\eta|$ is small. When $\frac{z'}{z}$ is large compared to $k$, particle number eigenstates $|n\rangle$\footnote{Note that the states $|n\rangle_{\vec{k}}$ are not the same anymore, since $a^{\dagger}_{\vec{k}}$ is dependent on the Hamiltonian used.} are no longer energy eigenstates and the time evolution induces pair creation. \footnote{This is in fact very reminiscent of the Schwinger effect, as shown in \cite{Martin:2007bw}. In that case, the energy recquired for pair creation is supplied by an external electric field, while the source in this case is the gravitational field. }
\par This could have been guessed, since for this action
\begin{equation}
 \frac{d \hat{a}}{dt} \not= 0,
\end{equation}
and the operator $\hat{a}^{\dagger}_{\vec{k}}\hat{a}_{\vec{k}}$ is no longer an invariant. This means that particle number is not conserved by time evolution in this case. This is the most important reason why we chose to work with Hamiltonian in eq.\ \eqref{QHamiltonian}, since for that action the particle number is a good quantum number. This has the disadvantage that the interpretation of the regime after Hubble exit is severely harder.

\chapter{Properties of the Hermite polynomials}\label{HermApp}
The Hermite polynomials $H_n(x)$, $n \in \mathbb{N}$ can be recursively defined by
\begin{align}
H_0(x) &= 1,\\
H_{n+1}(x) &= 2xH_{n}(x) - \frac{\partial H_n(x)}{\partial x}.
\end{align}
They satisfy the following differential equation
\begin{equation}
 \frac{\partial^2 H_n}{\partial x^2} - 2 x H_n + 2 n H_n = 0.
\end{equation}
The Hermite functions $\phi_n(x)$ are determined by
\begin{equation}\label{HermFunctions}
 \phi_n(x) = (-1)^n \sqrt{\frac{1}{2n! \sqrt{\pi}}}H_n(x)\exp \left(-\frac{x^2}{2}\right)
\end{equation}
And they form a complete orthonormal basis of the Hilbert space of square integrable functions.
The Hermite polynomials have a generating function, meaning that
\begin{equation}\label{HermGenerating}
 \exp \left[ - z^2 + 2xz \right] = \sum_{i = 0}^{\infty}\frac{z^n}{n}H_i(x).
\end{equation}
\chapter{Properties of $f_k$}\label{AppLemma}
We will use this appendix to collect the properties of $f_k$, as discussed in the text. Some properties that may not be found explicitly in the literature, but that prove useful in the text (particularly section \ref{Proof}) are also presented.
The basic properties of $f_k$ are
\begin{align}\label{FVergelijkingen}
 f_k''(\eta) - \left(k^2 - \frac{z''}{z}\right) f_k(\eta)&= 0, & f'_k(\eta_{ini}) &= -i\sqrt{\frac{k}{2}} e^{-i k \eta_{ini}} ,\\
 i(f_k' f^{*}_{k} - f'^{*}_{k} f_{k}) &= 1, & f_k(\eta_{ini}) &= \frac{1}{\sqrt{2k}} e^{-i k \eta_{ini}}.
\end{align}
\begin{Theorem}
If $\frac{z''}{z}$ is a positive increasing function on $I = [\eta_{ini},0 )$ that is unbounded, then $|f_k|$ is an unbounded, increasing function on $I$ .\end{Theorem}
\begin{bewijs}
We first prove that $|f_k|$ is increasing on $I$, by proving that $|f_k|'$ is positive on $I$.
Note that
\begin{equation}
 |f_k|'(\eta_{ini}) = 0
\end{equation}
due to the initial conditions of $f_k$. In addition the equations \ref{FVergelijkingen} imply that
\begin{equation}
|f_k|'' = -\left(k^2 - \frac{z''}{z}\right)|f_k| + \frac{1}{2|f_k|^3}.
\end{equation}
Filling in for the initial conditions, we notice $|f_k(\eta_{ini})|'' = \frac{1}{\sqrt{2k}}\frac{z''}{z}$. Thus $|f_k|''$ is positive on $I$, since $\frac{z''}{z}$ is increasing. We thus conclude that $|f|'$ is increasing on $I$. Since $|f_k (\eta_{ini})|' = 0$, it follows that $|f_k|'$ is positive on $I$. In addition, it follows that $|f_k| \geq 1$, for all $\eta \in I$.
\par
To prove that $|f_k|$ is unbounded, it is enough to observe that $|f|''$ is unbounded on $I$. Since $|f_k|'' - \frac{1}{2|f_k|^3}$ is unbounded, and $|f_k| \geq 1$, it follows that $|f_k|''$ is unbounded.
\end{bewijs}
\begin{Theorem}
 We write $f_k(\eta) = \exp(i S_f(\eta)) |f_k(\eta)|$. Then $ \lim_{\eta \rightarrow 0}\exp(i S_f(\eta))$ exists and is finite if $\frac{z''}{z}$ grows faster than $\eta^{-1/2}$.
\end{Theorem}
\begin{bewijs}
Notice that the following equalities hold
\begin{equation}
S_f' = \frac{1}{2i}\left[\ln \left( \frac{f_k}{f^*_k}\right)\right]' = \frac{1}{2|f_k|^2}.
\end{equation}
Note that this is integrable when $|f_k|$ grows faster than $\eta^{-1/2}$ close to $0$. This is the case for every model of inflation studied in this thesis.
\end{bewijs}

\chapter{Generalisation of Theorem \ref{theorem2} to Non-Product States}\label{NonProductApp}
We will generalise theorem \ref{theorem2} in chapter \ref{PilotWavePert} to non-product states.
\begin{Theorem}
Suppose the system is in an initial state
$$
\Psi = \sum_{i} \alpha_i \prod_{\vec{k}} \prod_{j = \pm }\Psi_{i,j, \vec{k}}(\eta_{ini})
$$
Define
$$q_{\pm, \vec{k}} = \frac{\nu_{\pm, \vec{k}}}{\sqrt{2}|f_k|},$$
and
$$P_{i,j,\vec{k}} = \Psi_{i, j, \vec{k}}(\eta) \exp \left[ \frac{-i}{2}\left(\frac{f_k'^{*}}{f_k^{*}} \right)|f|^2 q_{j,\vec{k}}^2\right].$$
Finally, define
$$
P = \sum_{i}\prod_{\vec{k}}\prod_{j} \alpha_i P_{i,j,\vec{k}}
$$
Let $\nu_{\vec{k}}(\eta)$ be a solution to the guidance equation for a certain mode of wavevector $\vec{k}$, on the interval $I = [\eta_{ini} ,0)$. Suppose
$$G_{j, \vec{k}} = \text{Im}\left(P^{-1} \frac{\partial P}{\partial q_{j, \vec{k}}} \right)$$
is bounded for all $j$ and $\vec{k}$.
Then
$$ \lim_{\eta \rightarrow 0} \frac{\nu_{\pm, \vec{k}}(\eta)}{|f_k|}$$
exists and is finite.
\end{Theorem}
\begin{bewijs}
 For clearness and ease of notation, we will only prove the case for entanglement between the real parts of two modes, as the case for entanglements of more states is completely analogous. Thus we suppose
\begin{equation}
 \Psi = \left( \alpha \Psi_{+,\vec{k}_1}\Psi_{+,\vec{k}_2} + \beta \chi_{+,\vec{k}_1}\chi_{+,\vec{k}_2} \right) \Psi_{-,\vec{k_1}}\Psi_{-,\vec{k}_2} \prod_{\vec{k}\not = \vec{k}_1, \vec{k}_2} \prod_{j = \pm}\Psi_{i,j, \vec{k}}(\eta_{ini}),
\end{equation}
where $|\alpha|^2 + |\beta|^2 = 1$.
Proving the theorem for the non-entangled degrees of freedom is analogous to the proof of theorem \ref{theorem2}. For the real parts of the modes of wavevector $\vec{k}_1$ and $\vec{k}_2$ the guidance equation for $\nu_{+,\vec{k}_1}$ becomes
\begin{align}
\nu'_{+, \vec{k_1}} &= \text{Im} \left[\alpha\frac{1}{\Psi_{+,\vec{k}_1}\Psi_{+,\vec{k}_2} + \beta\chi_{+,\vec{k}_1}\chi_{+,\vec{k}_2}} \left( \alpha\Psi_{+,\vec{k}_2}\frac{\partial \Psi_{\vec{k}_1}}{\partial \nu_{\vec{k}_1}}+ \beta\chi_{+,\vec{k}_2} \frac{\partial \chi_{\vec{k}_1}}{\partial \nu_{\vec{k}_1}}\right)\right],
\end{align}
and analogous for $\nu_{+,\vec{k}_2}$. This becomes in terms of the $P_{i,j, \vec{k}}$ and the $q_{j,\vec{k}_1}$
\begin{align}
 q'_{+, \vec{k_1}} = \frac{1}{\sqrt{2}|f_{k_1}|^2}\text{Im}\left[\left(\alpha P_{\Psi,+, \vec{k}_2} P_{\Psi, +, \vec{k}_1} + \beta P_{\chi,+, \vec{k}_2}P_{\chi, +, \vec{k}_1} \right)^{-1}\left(\alpha P_{\Psi,+, \vec{k}_2} \frac{\partial P_{\Psi, +, \vec{k}_1}}{\partial \nu_{+, \vec{k}_1}} + \beta P_{\chi,+, \vec{k}_2} \frac{\partial P_{\chi, +, \vec{k}_1}}{\partial \nu_{+, \vec{k}_1}}\right)\right],
\end{align} and analogous for $q_{+, \vec{k}_2}$. Now the imaginary part of the quantity in square brackets is exactly $G_{+, \vec{k}_1}$ and is bounded by assumption. The proof can continue in the same way as the proof of theorem \ref{theorem} in chapter \ref{PilotWavePert}.
\end{bewijs}
\end{appendices}
\printbibliography
\end{document}